\documentclass[12pt,epsfig]{article}
\usepackage[usenames]{color}
\usepackage{epsfig}

\newcommand{\be}{\begin{eqnarray}}
\newcommand{\ee}{\end{eqnarray}}

\newcommand{\beq}{\begin{equation}}
\newcommand{\eeq}{\end{equation}}

\newcommand{\nn}{\nonumber}

\def\la{\mathrel{\mathpalette\fun <}}
\def\ga{\mathrel{\mathpalette\fun >}}
\def\fun#1#2{\lower3.6pt\vbox{\baselineskip0pt\lineskip.9pt
\ialign{$\mathsurround=0pt#1\hfil##\hfil$\crcr#2\crcr\sim\crcr}}}

\begin{document}

\newcommand{\bea}{\begin{eqnarray}}
\newcommand{\eea}{\end{eqnarray}}
\def\la{\mathrel{\mathpalette\fun <}}
\def\ga{\mathrel{\mathpalette\fun >}}
\def\fun#1#2{\lower3.6pt\vbox{\baselineskip0pt\lineskip.9pt
\ialign{$\mathsurround=0pt#1\hfil##\hfil$\crcr#2\crcr\sim\crcr}}}
\def\wh{\widehat}
\newcommand{\gmp}{g_\mu^\top}
\newcommand{\gnp}{g_\nu^\top}
\newcommand{\gmn}{g_{\mu\nu}}
\newcommand{\gmnp}{g_{\mu\nu}^\top}
\newcommand{\gmnpp}{g_{\mu\nu}^{\top\top}}
\newcommand{\gabpp}{g_{\alpha\beta}^{\top\top}}
\newcommand{\gmmpri}{g_{\mu\mu'}}
\newcommand{\gmmprip}{g_{\mu\mu'}^{\top}}
\newcommand{\gnnprip}{g_{\nu\nu'}^{\top}}
\newcommand{\gmmpripp}{g_{\mu\mu'}^{\top\top}}
\newcommand{\gnnpripp}{g_{\nu\nu'}^{\top\top}}
\newcommand{\mfi}{m_\phi}
\newcommand{\bk}{{\bf k}}
\newcommand{\gampp}{\gamma_\mu^{\top\top}}
\newcommand{\ganpp}{\gamma_\nu^{\top\top}}
\newcommand{\kompp}{k_{1\mu}^{\top\top}}
\newcommand{\ktmpp}{k_{2\mu}^{\top\top}}
\newcommand{\konpp}{k_{1\nu}^{\top\top}}
\newcommand{\kpp}{k^2_{\top\top}}
\newcommand{\kompripp}{k^{'\top\top}_{1\mu}}
\newcommand{\konpripp}{k^{'\top\top}_{1\nu}}
\newcommand{\koapp}{k^{\top\top}_{1\alpha}}
\newcommand{\kobpp}{k^{\top\top}_{1\beta}}
\newcommand{\aaa}{\alpha}
\newcommand{\dd}{\delta}
\newcommand{\vx}{{\bf x}}
\newcommand{\vy}{{\bf y}}
\newcommand{\vp}{{\bf p}}
\newcommand{\vq}{{\rm {\bf q}}}
\newcommand{\vF}{{\bf F}}
\newcommand{\vE}{{\bf E}}
\newcommand{\vB}{{\bf B}}
\newcommand{\mT}{{\rm\bf T}}
\newcommand{\mS}{{\rm\bf S}}
\newcommand{\vV}{{\bf V}}
\newcommand{\tilphi}{\tilde{\varphi}}
\newcommand{\ka}{\mbox{\rm \ae}}
\newcommand{\hm}{\hspace{-1.5ex}}
\newcommand{\ep}{\epsilon}
\newcommand{\pp}{\partial}
\newcommand{\bp}{\mbox{\boldmath$p$}}
\newcommand{\bg}{\mbox{\boldmath$\gamma$}}
\newcommand{\bs}{\mbox{\boldmath$\sigma$}}
\newcommand{\bta}{\mbox{\boldmath$\tau$}}
\newcommand{\bq}{\mbox{\boldmath$q$}}
\newcommand{\bb}{\mbox{\boldmath$b$}}
\newcommand{\al}{\alpha}
\newcommand{\bet}{\beta}
\newcommand{\bkp}{\bf k_\perp}

\newcommand{\KL}{\rm\Lambda K^+}
\newcommand{\KS}{\rm\Sigma K}
\newcommand{\er}{$\pm$}
\newcommand{\widt}{\rm\Gamma_{tot}}
\newcommand{\wadd}{$\rm\Gamma_{miss}$}
\newcommand{\gpiN}{$\rm\Gamma_{\pi N}$}
\newcommand{\getN}{$\rm\Gamma_{\eta N}$}
\newcommand{\gkla}{$\rm\Gamma_{K \Lambda}$}
\newcommand{\gksi}{$\rm\Gamma_{K \Sigma}$}
\newcommand{\gNpi}{$\rm\Gamma_{P_{11} \pi}$}
\newcommand{\gDpf}{$\rm\Gamma_{\Delta\pi(L\!<\!J)}$}
\newcommand{\gDps}{$\rm\Gamma_{\Delta\pi(L\!>\!J)}$}
\newcommand{\sqgDpf}{$\rm\sqrt\Gamma_{\Delta\pi(L\!<\!J)}$}
\newcommand{\sqgDps}{$\rm\sqrt\Gamma_{\Delta\pi(L\!>\!J)}$}
\newcommand{\gNpf}{$\rm\Gamma_{D_{13}\pi}$}
\newcommand{\gNps}{$\Gamma_{D_{13}\pi(L\!>\!J)}$}
\newcommand{\gnsi}{$\rm N\sigma$}
\newcommand{\roper}{$ N(1440)P_{11}$}
\newcommand{\srma}{$  N(1535)S_{11}$}
\newcommand{\trma}{$ N(1520)D_{13}$}
\newcommand{\srmb}{$ N(1650)S_{11}$}
\newcommand{\trmb}{$ N(1700)D_{13}$}
\newcommand{\trmc}{$ N(1875)D_{13}$}
\newcommand{\trmd}{$ N(2170)D_{13}$}
\newcommand{\fvma}{$ N(1675)D_{15}$}
\newcommand{\fvmb}{$ N(2070)D_{15}$}
\newcommand{\fvpa}{$ N(1680)F_{15}$}
\newcommand{\srpb}{$ N(1710)P_{11}$}
\newcommand{\trpa}{$ N(1720)P_{13}$}
\newcommand{\trpb}{$ N(2200)P_{13}$}
\newcommand{\trpd}{$ N(2170)D_{13}$}
\newcommand{\dtpa}{$\Delta(1232)P_{33}$}
\newcommand{\doma}{$\Delta(1620)S_{31}$}
\newcommand{\dtma}{$\Delta(1700)D_{33}$}
\newcommand{\dtmb}{$\Delta(1940)D_{33}$}
\newcommand{\bc}{\begin{center}}
\newcommand{\ec}{\end{center}}

\title{
The analysis of  reactions $\pi N\to two\, mesons + N$ within reggeon exchanges.\\
1. Fit and results.
}

\author{V.V. Anisovich and A.V. Sarantsev \\
Petersburg Nuclear Physics Institute, Gatchina, 188300, Russia}

\maketitle

\begin{abstract}

The novel point of this analysis is a direct use of reggeon exchange
technique for the description of the reactions
$\pi N\to two\, mesons + N$ at
large energies of the initial pion. This approach allows us to
describe simultaneously distributions over $M$ (invariant
mass of two mesons) and $t$ (momentum transfer squared to nucleons).
Making use of this technique, the following resonances (as well as
corresponding bare states), produced in the $\pi N\to \pi^0\pi^0 N$
reaction are studied: $f_0(980)$, $f_0(1300)$
($f_0(1370)$ in PDG notation), $f_0(1200-1600)$,
$f_0(1500)$, $f_0(1750)$, $f_2(1270)$, $f_2(1525)$, $f_2(1565)$,
$f_2(2020)$, $f_4(2025)$. Adding data for the reactions
  $p\bar p({\rm at\, rest,\, from\, liquid\,H_2})\to \pi^0\pi^0\pi^0$,
 $\pi^0\pi^0\eta$, $\pi^0\eta\eta$
 and
$p\bar p({\rm at\, rest,\, from\, gaseous\, H_2})\to \pi^0\pi^0\pi^0$,
$\pi^0\pi^0\eta$, $\pi^0\eta\eta$, we have
performed simultaneous $K$-matrix fit of two-meson spectra in
all these reactions.
The results of combined fits to the above-listed
isoscalar $f_J$-states and to isovector ones, $a_0(980)$,
$a_0(1475)$, $a_2(1320)$, are presented.

\end{abstract}
 \vspace{0.5cm}

\section{Introduction}

The study of the mass spectrum of hadrons and their properties is
the key point for the understanding  of colour particle interactions
at large distances. But even the meson sector, though less
complicated than the baryon one, is far from being
completely understood. We mean that \\
({\it i}) there is no sufficient
information about states above 2 GeV, \\
({\it ii}) certain quark--antiquark states below 2 GeV ({\it e.g.}
$2^{--}$ states) are still missing, \\
({\it iii}) there is no clear understanding of the glueball spectrum
(although strong candidates in  the $0^{++}$ and $2^{++}$ sectors
exist, we have no definite information about the $0^{-+}$ sector), \\
({\it iv}) some analyses reported the observation of other exotics
({\it e.g.} hybrid) states, \\
({\it v}) in the scalar sector not only the properties but also the
existence of states like $\sigma$, $\kappa$, $f_0(1300)$
($f_0(1370)$ in PDG notation) is under discussion.

So, there is indeed a strong demand for new data which can help us
to identify
the meson states in a more definite way. However, the situation is
only partly connected with the lack of data. In the lower mass
region there is a lot of data taken from the proton--antiproton
annihilation at rest (Crystal Barrel, Obelix), from the
$\gamma\gamma$ interaction (L3), from the proton--proton central
collisions (WA102), from $J/\Psi$ decay (Mark III, BES), from
 $D$- and $B$-meson decays (Focus, D0, BaBar, Belle, Cleo C) and
from $\pi N\to two\, mesons+N$ reactions with high energy pion beams
(GAMS, VES, E852).
Most of these data are of high statistics, thus allowing us to
determine resonance properties with a high accuracy (though, let us
emphasize, in the reactions $\pi N\to two\, mesons+N$
polarized-target data are lacking).

Nevertheless, in many cases there are significant contradictions between
analyses performed by different groups. The ambiguities originate from
two circumstances.

First, in the discussed sectors the analyses of data taken from a
single experiment cannot provide us with a unique solution. A unique
solution can be obtained only from the combined analysis of a large
set of data taken in different experiments.

Second, there are some simplifications inherent in many analyses.
The unitarity was neglected frequently even when the amplitudes were
close to the unitarity limit. A striking example is that up to now
there is no proper $K$-matrix parametrization of the $1^{--}$ and
$2^{++}$ waves which are considered by many physicists as mostly
understood ones. As to multipartical final states, only a few
analyses have ever considered the contributions of triangle or box
singularities to the measured cross sections. However, these
contributions can simulate the resonant behavior of the studied
distributions, especially in the threshold region (for more detail,
see \cite{book3} and references therein).

In the analysis of meson spectra in high energy reactions $\pi N\to
two\, mesons+N$, many results are related to the decomposition of
the cross sections into natural and unnatural amplitudes that is
based on certain models developed for the two-pion production at
small momenta transferred,  ({\it e.g.}, see
\cite{Hyams:1973zf,etkin,Chung:1997qd}). However, as was discussed
by the cited authors, a direct application of these methods at large
momenta transferred  to the analysis of data may lead to a wrong
result. In addition, the $\pi N\to two\, mesons+N$ data were
discussed mostly in terms of  $t$-channel particle exchange, though
without proper analysis of the $t$-channel exchange amplitudes.

A decade ago our group  performed a combined analysis of data on
proton--antiproton annihilation at rest into three pseudoscalar
mesons, together with the data on two-meson $S$-waves extracted form
the $\pi N\to \pi\pi N$, $\eta\eta N$, $K\bar K$ and $\eta\eta' N$
reactions \cite{Tkm,TkmBugg,TYF}. The analysis has been carried out
in the framework of the $K$-matrix approach which preserves
unitarity and analyticity of the amplitude in the two-meson physical
region. Although the two-meson data extracted from the
 reaction $\pi N\to two\, mesons+N$ at small
momentum transfer  appeared to be
highly compatible with those found in proton--antiproton
annihilation, we have faced a set of problems,  describing  the
$\pi N\to two\, mesons+N$  data at large momentum transfer. As we have
seen now, the problems  were owing to the use of partial wave
decomposition which was performed by the E852 Collaboration and showed
a huge signal at 1300 MeV in the $S$-wave.

The strategy of our present approach is  as follows. The analysis of
a large set of experimental data on proton--antiproton annihilation
at rest is carried out together with the analysis of the $\pi N\to
two\, mesons+N$ data based on the $t$-channel reggeized exchanges.
For the $\pi N\to two\, mesons+N$ reactions, the data at small and
large momentum transfers are included. Here, as the first step, we
perform the analysis in the framework of the $K$-matrix
parametrization for all fitting channels ($K$-matrix approach
insures the unitarity and analyticity in the physical region). At
the next stage, we plan to use the $N/D$ method for two-meson
amplitudes satisfying these requirements in the whole complex plane.

In this paper, we present the method for the analysis of the $\pi N$
interactions based on the $t$-channel reggeized exchanges
supplemented by a study of the proton--antiproton annihilation at
rest. The method is applied to a combined analysis of the
$\pi N\to  \pi^0\pi^0 N$ data taken by E852 at small and large momentum
transfers and Crystal Barrel data on the proton--antiproton
annihilation at rest into three neutral pseudoscalar mesons.
The even waves, which
contributed to this set of data, are parametrized within the
$K$-matrix approach.
To check a strong $S$-wave signal around
1300 MeV, which has been reported by E852 Collaboration from the
analysis of data at large momentum transfers, is a subject of a
particular interest in the present analysis.

We present the results of the new $K$-matrix analysis of two-meson
spectra in the scalar, $J^P=0^+$, and tensor, $J^P=2^+$, sectors:
these sectors need a particular attention because just here we meet
with the low-lying glueballs, $f_0(1200-1600)$ and $f_2(2000)$. The
situation with the tensor glueball is rather transparent allowing us
to make a definite conclusion about the gluonium structure of
$f_2(2000)$, while the status of the broad state $f_0(1200-1600)$
requires a special discussion: this state is nearly flavour-blind
but the corresponding pole of the amplitude dives deeply into the
complex-$M$ plane. It is definitely seen only in  the analysis of a
large number of different reactions in broad intervals of mass
spectra (for example, see \cite{book3} and references therein).

So, here we consider the following reactions: \\
({\it i}) $\pi p \to \pi^0\pi^0 n$ at high energies of initial pion
and small and large momenta transferred to nucleon, and \\
({\it ii})
$p\bar p({\rm at\, rest \,H_2})\to \pi^0\pi^0\pi^0$,
 $\pi^0\pi^0\eta$, $\pi^0\eta\eta$ in liquid and gaseous $H_2$ ---
the data on these reactions give us the most reliable information about
 scalar and tensor sectors.\\
As was stressed above, the novel point of the performed $K$-matrix
analysis is the use of reggeon exchange technique for the
description of $\pi p \to \pi\pi n$ at high energies that allows us
to analyze the two-meson invariant mass spectra and nucleon momentum
transfer distributions simultaneously.

The paper is organized as follows.

In Section 2, we consider meson--nucleon collisions at high energies
and present formulas for peripheral two-meson production amplitudes
in terms of reggeon exchanges. Amplitudes for the description of
low-energy three-meson production in the $K$-matrix approach are
given in Section 3. The fitting procedure is described in Section 4.
In Conclusion we summarize the results. Technical aspects of the
fitting procedure are discussed in \cite{km08A}.

\section{ Meson--Nucleon Collisions at High Energies:
Peripheral Two-Meson Production in Terms of Reggeon Exchanges}

The two-meson production reactions $\pi p\to\pi\pi n$, $K\bar Kn$,
$\eta\eta n$, $\eta\eta'n$ at high energies and small momentum
transfers to the nucleon are used for obtaining  the $S$-wave
amplitudes $\pi\pi\to\pi\pi$, $K\bar K$, $\eta\eta$, $\eta\eta'$ at
$|t|<0.2\,(\rm GeV/c)^2$ because, as commonly believed, the $\pi$
exchange dominates this wave at such momentum transferred. At larger
momentum transfers, $|t|\ga 0.2\,(\rm GeV/c)^2$, we observe
definitely a change of the regime in the $S$-wave production ---  a
significant contribution of other reggeons is possible
($a_1$-exchange, daughter-$\pi$ and daughter-$a_1$ exchanges).
Nevertheless, the study of the two-meson production processes at
$|t|\sim0.5-1.5\,(\rm GeV/c)^2$ looks promising, for at such
momentum transfers the contribution of the broad resonance (the
scalar glueball $f_0(1200-1600)$) vanishes. Therefore, the
production of other resonances (such as the $f_0(980)$ and
$f_0(1300)$) appears practically without background -- this is
important for finding out their characteristics as well as a
mechanism of their production.

What we know about the reactions $\pi p\to\pi\pi n$, $K\bar Kn$,
$\eta\eta n$, $\eta\eta'n$ allows us to suggest that a consistent
analysis of the peripheral two-meson production in terms of reggeon
exchanges may be a good tool for studying meson resonances. Note
that investigation of two-meson scattering amplitudes by means of
the reggeon exchange expansion of the peripheral two-meson
production amplitudes was proposed long ago \cite{Tan-sh} but was
not used because of the lack of data until now.

\begin{figure}
%Fig. 1
\centerline{\epsfig{file=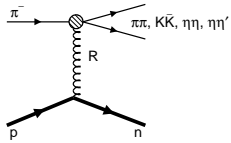,width=3cm}}
\caption{Example of a reaction with the production of two mesons
(here $\pi\pi$, $K\bar K$, $\eta\eta $, $\eta\eta'$  in $\pi^- p$
collision) due to reggeon
($R$) exchange. }
\label{T6km-20}
\end{figure}

The $K$-matrix amplitude of the peripheral production of two mesons with
total angular momentum $J$ reads:
\be
\bigg(\bar\psi_N(k_3)\hat G_R\psi_N(p_2)\bigg)R(s_{\pi N},t)
\widehat K_{\pi R(t)}(s) \bigg[1-\hat\rho(s)\widehat K(s)\bigg]^{-1}
Q^{(J)}(k_1,k_2) \ ,
\label{Tperiph}
\ee
This formula is illustrated by  Fig. \ref{T6km-20} for the production of
$\pi\pi$,
 $K\bar K$, $\eta\eta $, $\eta\eta'$  systems. Here the factor
$(\bar\psi_N(k_3)\hat G_R\psi_N(p_2))$ stands for the
reggeon--nucleon vertex, and $\hat G_R$ is the spin operator;
$R(s_{\pi N},t)$ is the reggeon propagator depending on the total
energy squared of colliding particles, $s_{\pi N}=(p_1+p_2)^2$, and
the momentum transfer squared $t=(p_2-k_3)^2$, while the factor
$\widehat K_{\pi R(t)}[1-i\hat\rho (s)\widehat K(s)]^{-1}$ is
related to the block of two-meson production; $s\equiv
M^2=(k_1+k_2)^2$, and $\hat\rho (s)$ is the phase space matrix . In
the reactions $\pi p\to\pi\pi n$, $K\bar Kn$, $\eta\eta n$,
$\eta\eta'n$, the factor $\widehat K_{\pi
R(t)}(s)[1-i\hat\rho(s)\widehat K(s)]^{-1}$ describes transitions
$\pi R(t)\to\pi\pi$, $K\bar K$, $\eta\eta$, $\eta\eta'$: in this way
the block $\widehat K_{\pi R(t)}$ is associated with the prompt
meson production, and $[1-i\hat\rho(s)\widehat K(s)]^{-1}$ is the
$K$-matrix factor for meson rescattering (of the type of
$\pi\pi\to\pi\pi$, $\pi\pi\to K\bar K$, $K\bar K\to\eta\eta$, and so
on). The prompt-production block for transition $\pi R\to b$ (where
$b=\pi\pi$, $K\bar K$, $\eta\eta$, $\eta\eta'$,  $4\pi$, ...) is
parameterized with singular (pole) and smooth terms
\cite{Tkm,TYF,Tepja}:
\beq\label{Tperiph1}
 \left(\widehat K_{\pi R(t)}\right)_{\pi R,b}\ =\ \sum_n
\frac{G^{(n)}_{\pi R}(t)g^{(n)}_b}{\mu^2_n-s}+f_{\pi R,b}(t,s)\ .
\eeq
 The pole singular term, $G^{(n)}_{\pi R}(t)g^{(n)}_b/(\mu^2_n-s)$,
determines the bare state: here $G^{(n)}_{\pi R}(t)$ is the bare
state production vertex while the parameters $g^{(n)}_b$ and $\mu_n$
are the coupling and the mass of the bare state -- they are the same
as in the partial wave transition amplitudes $\pi\pi\to\pi\pi$,
$K\bar K$, $\eta\eta$, $\eta\eta'$, $4\pi$, $...$. The smooth term
$f_{\pi R,b}$ stands for the background production of mesons. The
$G^{(n)}_{\pi R}(t)$, $f_{\pi R,b}$, $g^{(n)}_b$, $\mu_n$ are free
parameters of the fitting procedure, while the characteristics of
resonances are determined by poles of the $K$-matrix amplitude
(remind that the position of poles is given by zeros of the
amplitude denominator, $det | 1-\hat\rho(s)\widehat K(s) |=0$).

Below we  explain in detail the method of analysis of meson spectra
 using as an example the reactions $\pi N\to\pi\pi N$, $K\bar K N$,
$\eta\eta N$, $\eta\eta' N$, $\pi\pi\pi\pi N$.

\subsection{Reggeon exchange technique and the
$K$-matrix analysis of meson spectra in the waves $J^{PC}=0^{++}$,
$1^{--}$, $2^{++}$, $3^{--}$, $4^{++}$ in high energy reactions
$\pi N\to two\; mesons+ N$}

Here we present  the technique of the analysis of high-energy
reaction  $\pi^- p\to mesons+n$, with the production of mesons
in the $J^{PC}=0^{++}$, $1^{--}$,  $2^{++}$, $3^{--}$, $4^{++}$ states
at small and moderate momenta transferred to the nucleon.

The following points are to be emphasized:\\
(1) The technique can be used for performing the $K$-matrix analysis
not only for $0^{++}$ and
$2^{++}$ wave, as in \cite{Tkm,TYF,Tepja}, but simultaneously in
$1^{--}$, $3^{--}$, $4^{++}$ waves as well.\\
(2) We use the reggeon exchange technique for the
description of the $t$-dependence in all analyzed amplitudes. This
allows us to perform a partial wave decomposition of the produced
meson states directly on the basis of the measured cross
sections without using the published moment expansions (which
were done under some simplifying assumptions -- it is discussed below
in more detail). \\
(3) The mass interval of the analyzed spectra is extended
up to 2500 MeV thus overlapping with the mass region studied in
reactions $p\bar p$(in flight)$\to\, mesons$ \cite{Tpnpi-ral}.

We discuss in detail the reactions at incident pion momenta 20--50
GeV/c, such as measured in
\cite{Tgams1,Tgams2,Tbnl,TC-M,Tamelin,TE852}: ({\it i}) $\pi^- p\to
\pi^+\pi^- +n$, ({\it ii}) $\pi^- p\to \pi^0\pi^0 +n$, ({\it iii})
$\pi^- p\to  K_S K_S +n$, ({\it iv}) $\pi^- p\to \eta\eta +n$. At
these energies, the mesons in the states $J^{PC}=0^{++}$, $1^{--}$,
$2^{++}$, $3^{--}$, $4^{++}$ are produced via $t$-channel exchange
by  reggeized mesons belonging to the leading and daughter $\pi $,
$a_1$ and $\rho$ trajectories.

But, first, let us present notations used below.

\subsubsection{Cross sections for the reactions ${\pi}N\to\pi\pi N$,
$KKN$, $\eta\eta N$}

We consider the process of the  Fig. \ref{T6km-20} type, that is,
$\pi N$ interaction at large momenta of the incoming pion with the
production of a two-meson system with a large momentum in the beam
direction. This is a peripheral production of two mesons.

The cross section is defined as follows:
 \bea
&& \hspace{-1cm}
d\sigma=\frac{(2\pi)^4|A|^2}{8\sqrt{s_{{\pi}N}}|\vec{p}_2|_{cm(\pi
p)}}\,
d\phi(p_1+p_2; k_1,k_2,k_3), \nn \\
&&  \hspace{-1cm} d\phi(p_1+p_2;
k_1,k_2,k_3)=(2\pi)^3d\Phi(P;k_1,k_2)\, d\Phi(p_1+p_2;P, k_3)\,ds\,,
\eea
 where $|\vec{p_2}|_{cm(\pi p)}$ is the pion momentum in the c.m.
frame of the incoming hadrons. Taking into account that invariant
variables $s$ and $t$ are inherent in the meson peripheral
amplitude, we rewrite the phase space in a more convenient form:
\bea \label{Tperiph2}
 && d\Phi(p_1+p_2; P,
k_3)=\frac{1}{(2\pi)^5}\frac{dt}{8|\vec{p_2}|_{cm(\pi p)}
\sqrt{s_{{\pi}N}}},\qquad t=(k_3-p_2)^2 ,\nn \\ &&
d\Phi(P;k_1,k_2)=\frac{1}{(2\pi)^5}\rho(s)d\Omega\, , \qquad
\rho(s)=\frac{1}{16\pi}\frac{2|\vec k_1|_{cm(12)}}{\sqrt{s}}.
\eea
Momentum $|\vec k_1|_{cm(12)}$ is calculated in the c.m. frame of
the outgoing mesons: in this system one has $P=(M, 0,0,0,)\equiv
(\sqrt {s},0,0,0)$ and $g_{\mu\nu}^{\perp
P}k_{1\nu}=-g_{\mu\nu}^{\perp P}k_{2\nu} = (0, k\,\sin \Theta
\sin\varphi ,\,k\,\cos \Theta \sin\varphi$, $ k\,\cos \Theta \, k)$
while
 $d\Omega =d(\cos \Theta)d\varphi$.
We have:
\bea \label{Tperiph3}
d\sigma=\frac{(2\pi)^4|A|^2(2\pi)^3}{8|\vec{p_2}|_{cm(\pi p)}
\sqrt{s_{{\pi}N}}}\frac{1}{(2\pi)^5}
\frac{dt\,dM^2\,d\Phi(P,k_1,k_2)}{8|\vec p_2|_{cm(\pi
p)}\sqrt{s_{\pi N}}}=\frac{|A|^2\rho(M^2)\, MdM\, dt\,
d\Omega}{32(2\pi)^3 |\vec p_2|^2_{cm(\pi p)}\,s_{\pi N}}.~~
\eea
The cross section can be expressed in terms of the spherical
functions:
\be \label{Tperiph4}
\frac{d^4\sigma}{dt{d\Omega}dM}=N(M,t)
\sum\limits_l\bigg({\langle{Y}^0_l\rangle} Y^0_l(\Theta
,\varphi)+2\sum \limits_{m=1}^{l}{\langle{Y}^m_l\rangle}{\rm
Re}\,Y^m_l(\Theta ,\varphi) \bigg).
\ee
 The coefficients $N (M,t)$, $\langle{Y}^0_l\rangle$,
$\langle{Y}^m_l\rangle$ are subjects of study in the determination
of meson resonances.

 Before describing the results of analysis based on the reggeon
exchange technique, let us comment methods used in other approaches.

\subsubsection{The CERN-Munich approach}

The CERN-Munich model \cite{TC-M} was developed for the analysis of
the data on the $\pi^- p\to \pi^+\pi^- n$ reaction. It is based
partly on the absorption model but mainly on phenomenological
observations. The amplitude squared is written as
\be \label{Tcm-1}
|A|^2=\bigg|\sum\limits_{J=0} A^0_J Y^0_J(\Theta ,\varphi)+
\sum\limits_{J=1} A^{(-)}_JRe Y^1_J(\Theta ,\varphi)\bigg|^2 +
\bigg|\sum\limits_{J=1} A^{(+)}_{J}Re Y^1_J(\Theta ,\varphi)\bigg|^2
\ , \ee
 and additional assumptions are made:\\
1) The helicity-1 amplitudes are equal for natural and unnatural
exchanges $ A^{(-)}_{J}=A^{(+)}_{J}$;\\
2) The ratio of the $A^{(-)}_{J}$ and the $A^0_{J}$  amplitudes is
a  polynomial over the mass of the two-pion system which does not
depend on $J$ up to the total normalization,
$A^{(-)}_{J}=A^0_{J}\bigg(C_J\sum\limits^3_{n=0}b_n M^n\bigg)^{-1}$.\\
Then, in \cite{TC-M},  the amplitude squared was rewritten using
density matrices $\rho^{nm}_{00}=A^0_{n}A^{0*}_{m}$,
$\rho^{nm}_{01}=A^0_{n} A^{(-)*}_{m}$,
$\rho^{nm}_{11}=2A^{(-)}_{n}A^{(-)}_{m}$ as follows:
\bea  \label{Tcm}
|A|^2&=&\sum\limits_{J=0}Y^0_J(\Theta ,\varphi) \left
(\sum\limits_{n,m}
d^{0,0,0}_{n,m,J}\rho^{nm}_{00}+d^{1,1,0}_{n,m,J}\rho^{nm}_{11}\right
)
\nn \\
& +&\sum\limits_{J=0} Re Y^1_J(\Theta ,\varphi) \left
(\sum\limits_{n,m}
d^{1,0,1}_{n,m,J}\rho^{nm}_{10}+d^{0,1,1}_{n,m,J}\rho^{mn}_{11}\right
) , \nn \\
d^{i,k,l}_{n,m,J}&=&\frac{ \int d\Omega \,Re
Y^i_n(\Theta,\varphi)\,Re Y^k_m(\Theta,\varphi)\, Re
Y^l_J(\Theta,\varphi) }{ \int d\Omega \, Re
Y^l_J(\Theta,\varphi)\,Re Y^l_J(\Theta,\varphi) } .
\label{dcoef}
\eea
Using this amplitude for the cross section,
the fitting to the moments $<Y^m_J>$ has been carried out.

The CERN--Munich approach cannot be applied to large $t$, it does
not work for many other final states either.

\subsubsection{GAMS, VES, and BNL approaches}

In GAMS \cite{Tgams1,Tgams2}, VES \cite{Tamelin}, and BNL
\cite{TE852} approaches, the $\pi N$ data are described  by a sum of
amplitudes squared with an angular dependence defined by spherical
functions:
\be
|A^2|\!=\!\bigg|\sum\limits_{J=0} A^{0}_J Y^0_J(\Theta
,\varphi)\!+\! \sum\limits_{J=1} A^{(-)}_{J}\sqrt 2\,
Re\,Y^1_J(\Theta ,\varphi)\bigg|^2 \!\!+\! \bigg|\sum\limits_{J=1}
A^{(+)}_{J} \sqrt 2 \, Im\, Y^1_J(\Theta ,\varphi)\bigg|^2~~
\label{Tves} \ee
The $A^0_J$ functions are denoted as
$S_0,P_0,D_0,F_0\ldots$, the $A^{(-)}_J$ functions are defined as
$P_-,D_-,F_-,\ldots$ and the $A^{(+)}_J$ functions as
$P_+,D_+,F_+,\ldots$. The equality of the helicity-1 amplitudes with
natural and unnatural exchanges is not assumed in these approaches.

However, the discussed approaches are not free from other assumptions like
the coherence of some amplitudes or the dominance of the one-pion
exchange. In reality the interference of the amplitudes being
determined by $t$-channel exchanges of different particles leads to
a more complicated picture than that given by (\ref{Tves}), this latter
may lead (especially at large $t$) to a misidentification of
quantum numbers for the produced resonances.

For example, in \cite{TE852} the S-wave appears in an unnatural set
of amplitudes only. Natural exchanges have moments with m=1,2,3....
However, the a1-exchange is a natural one, therefore it contributes
into the S-wave and does not interfere with unnatural exchanges --
in this point the moment expansion \cite{TE852} does not coincide
with formula with reggeon exchanges.

\subsection{The $t$-channel exchanges of pion trajectories  in the
reaction $\pi^- p \to \pi\pi \, n$}

Consider now in more detail the production amplitude for the $\pi\pi$
system with $I=0$ and $J^{PC}=0^{++},2^{++}$ and show the way
of its generalization for  higher $J$.

\subsubsection{Amplitude with leading and daughter pion trajectory
exchanges}

The amplitude with $t$-channel pion trajectory exchanges can be
written as follows:
\be \label{Tk4}
 A_{\pi p\to\pi\pi n}^{(\pi-{\rm trajectories})}\!=\!
\sum\limits_{R(\pi_j)} \!A\bigg(\pi R(\pi_j)\to\pi\pi\bigg)
R_{\pi_j}(s_{\pi N},q^2) \left(\varphi_n^+(\vec \sigma \vec
q_\perp)\varphi_p \right)
 g^{(\pi_j)}_{pn}(t)~~~
\ee
 The summation is carried out over the leading and daughter
trajectories. Here $A(\pi R(\pi_j)\to\pi\pi)$ is the transition
amplitude for meson block in Fig. \ref{T6km-20},
$g^{(\pi_j)}_{pn}$ is the reggeon--$NN$  coupling and
$R_{\pi_j}(s_{\pi N},q^2)$ is the reggeon propagator:
\be
\label{Tk5}\hspace{-6mm} R_{\pi_j}(s_{\pi N},q^2)\! = \! \exp{\left
(\!-i\frac{\pi}{2}\alpha^{(j)}_\pi (q^2)\!\right )}
\frac{\left(s_{\pi N}/s_{\pi N0}\right )^{\alpha^{(j)}_\pi (q^2)}}
{\sin \left(\frac{\pi}{2}\alpha^{(j)}_\pi (q^2)\right) \Gamma \left
(\frac{1}{2} \alpha^{(j)}_\pi (q^2) +1\right ) }.
\ee
The $\pi$--reggeon has a positive signature, $\xi_\pi =+1$.
Following \cite{book3,Tbook,Tufn,Tsyst}, we use for pion trajectories:
\be
\label{Tpi6}
\alpha_{\pi}^{ (leading)} (q^2)\simeq -0.015 +0.72 q^2 , \quad
\alpha_{\pi }^{( daughter-1)} (q^2)\simeq -1.10 +0.72 q^2,~~~
\ee
where the slope parameters are given in (GeV/c)$^{-2}$ units. The
normalization parameter $ s_{\pi N0}$ is of the order of 2--20
GeV$^2$. To eliminate the poles at $q^2<0$ we introduce
Gamma-functions in the reggeon propagators (recall that
$1/\Gamma (x) =0 $ at $x=0,-1,-2, \ldots$).

For the nucleon--reggeon vertex $\hat G^{(\pi)}_{pn}$ we use in the
infinite momentum frame the two-component spinors $\varphi_p$ and
$\varphi_n$ (see, for example, \cite{book3,Tbook,Tkaidalov}):
\be \label{Tpi7}
g_\pi(\bar\psi (k_3)\gamma_5\psi(p_2)) \longrightarrow
\left(\varphi_n^+(\vec \sigma \vec q_\perp)\varphi_p \right)
 g^{(\pi)}_{pn}(t)\,.
\ee
 As to the meson--reggeon vertex,  we use the covariant representation
\cite{book3,Tbook,Toper}. For the production of two pseudoscalar
particles
 (let it be $\pi\pi$ in the considered case), it reads:
\bea
\label{Tpi8}
A\bigg(\pi R(\pi_j)\to\pi\pi\bigg)\!=\!\sum\limits_{J} A^{(J)}_{\pi
R(\pi_j)\to\pi\pi}(s) X^{(J)}_{\mu_1\ldots\mu_J}(p^{\perp}) \,
(-1)^J O^{\mu_1\ldots\mu_J}_{\nu_1\ldots\nu_J} (\perp P)
X^{(J)}_{\nu_1\ldots\nu_J}(k^{\perp})\,\xi_J\,,\nn \\
\xi_J=\frac{16\pi (2J+1)}{\alpha_J}, \qquad
\alpha_J=\prod\limits^J_{n=1}\frac{2n-1}{n}
.~~~~~~~~~~~~~~~~~~~~~~~~~~~~~~~~~~~
\eea
 The angular momentum operators are constructed of momenta $p^{\perp}$
and $k^{\perp}$ which are orthogonal to the momentum of the two-pion
system $P=k_1+k_2$:
\be
g_{\mu\nu}^\perp=g_{\mu\nu}-\frac{P_\mu P_\nu}{P^2},\qquad
k^\perp_\mu=\frac 12 (k_1-k_2)_\nu g^\perp_{\mu\nu} \qquad
p^\perp_\mu=\frac 12 (p_1+q)_\nu g^\perp_{\mu\nu}\,.
\ee
The coefficient $\xi_J$ normalizes the angular momentum operators,
so that the unitarity condition appears in a simple form (for
details see Appendix A).

\subsubsection{The $t$-channel $\pi_2$-exchange}

The $R(\pi_j)$-exchanges dominate the spin flip amplitudes, and the
amplitudes with $ m=1$  are here suppressed, see  (\ref{Tperiph4}).
However, their contributions are visible in the differential cross
sections and should be taken into account. The effects appear owing
to the interference in the two-meson production amplitude  because
of the reggeized $\pi_2$ exchange in the $t$-channel. The
corresponding amplitude is written as:
\bea
\label{Tamp_pi2} &&\sum\limits_a A_{\alpha\beta}\bigg(\pi
R(\pi_2)\to\pi\pi\bigg) \varepsilon^{(a)}_{\alpha\beta}
R_{\pi_2}(s_{\pi N},q^2)
\frac{\varepsilon^{(a)+}_{\alpha'\beta'}}{s^2_{\pi N}}
X^{(2)}_{\alpha'\beta'}(k_3^{\perp q}) \left(\varphi_n^+(\vec\sigma
\vec q_\perp)\varphi_p \right)
 g^{(\pi_2)}_{pn}(t)\ ,
\eea
where $A_{\alpha\beta}\bigg(\pi R(\pi_2)\to\pi\pi\bigg)$ is the
meson block of the amplitude related to the  $\pi_2$-reggeized
$t$-channel transition, $g^{(\pi_2)}_{pn}$ is the reggeon--$pn$
vertex, $R_{\pi_2}(s_{\pi N},q^2)$ is the reggeon propagator, and
$\varepsilon^{(a)}_{\alpha\beta}$ is the polarization tensor for the
$2^{-+}$ state. Let us remind that $k_3$ is the momentum of the
outgoing nucleon.
\be
k_{3\mu}^{\perp q}=g^{\perp q}_{\mu\nu}k_{3\nu}\qquad  g^{\perp
q}_{\mu\nu}=g_{\mu\nu}-\frac{q_\mu q_\nu}{q^2}\,.
\ee

The $\pi_2$ particles are located on the pion trajectories and are
described by a similar reggeized propagator. But in the meson block,
the $2^{-+}$ state exchange leads to vertices different from those
in the $0^{-+}$-exchange, so it is convenient to single out these
contributions. Therefore, we use for $R_{\pi_2}(s_{\pi N},q^2)$ the
propagator given by (\ref{Tk5}) but with eliminated
$\pi(0^{-+})$-contribution:
\be \label{Tamp_pi3}
R_{\pi_2}(s_{\pi N},q^2) = \exp{\left
(-i\frac{\pi}{2}\alpha^{(leading)}_\pi (q^2)\right )}
\frac{\left(s_{\pi N}/s_{\pi N0}\right )^{\alpha^{(leading)}_\pi
(q^2)}} {\sin \left(\frac{\pi}{2}\alpha^{(leading)}_\pi (q^2)\right)
\Gamma \left (\frac{1}{2} \alpha^{(leading)}_\pi (q^2) \right ) }\,.
\ee
Taking into account that
\be \label{Tamp_pi4}
\sum\limits_{a=1}^{5}
\varepsilon^{(a)}_{\alpha\beta}\varepsilon^{(a)+}_{\alpha'\beta'}=
\frac 12 \left ( g^{\perp q}_{\alpha\alpha'}g^{\perp
q}_{\beta\beta'}+ g^{\perp q}_{\beta\alpha'}g^{\perp
q}_{\alpha\beta'}-\frac 23 g^{\perp q}_{\alpha\beta}g^{\perp
q}_{\alpha'\beta'}\right )\,,
\ee
one obtains:
\be
\frac {X^{(2)}_{\alpha'\beta'}(k_3^{\perp q})}{2s^2_{\pi N}} \left (
g^{\perp q}_{\alpha\alpha'}g^{\perp q}_{\beta\beta'}+ g^{\perp
q}_{\beta\alpha'}g^{\perp q}_{\alpha\beta'}-\frac 23 g^{\perp
q}_{\alpha\beta}g^{\perp q}_{\alpha'\beta'}\right )= \frac 32
\frac{k^{\perp q}_{3\alpha}k^{\perp q}_{3\beta}}{s^2_{\pi N}}
-\frac{4m_N^2-q^2}{8s^2_{\pi N}} \left
(g_{\alpha\beta}-\frac{q_\alpha q_\beta}{q^2}\right )\ .
\label{Tpi_2_exch}
\ee
 In the large momentum limit of the initial pion,
the second term in  (\ref{Tpi_2_exch}) is always small and can be
neglected, while the convolution of
$k^{\perp q}_{3\alpha}k^{\perp  q}_{3\beta}$
with the momenta of the meson block results in the term
$\sim s^2_{\pi N}$. Hence, the amplitude for $\pi_2$-exchange can be
rewritten as follows:
\be
\label{Tamp_pi2_f}
 A_{\pi p\to\pi\pi n}^{(\pi_2-{\rm exchange})}=\frac{3}{2}
 A_{\alpha\beta}(\pi R(\pi_2)\to\pi\pi)
\frac{k^{\perp q}_{3\alpha}k^{\perp q}_{3\beta}}{s^2_{\pi N} }
R_{\pi_2}(s_{\pi N},q^2) \left(\varphi_n^+(\vec\sigma \vec
q_\perp)\varphi_p \right) g^{(\pi_2)}_{pn}\,.
\ee
 A resonance with spin $J$ and fixed parity can be produced owing to
the $\pi_2$-exchange with three angular momenta $L=J-2$, $L=J$ and
$L=J+2$, so we have:
\bea
&&A_{\alpha\beta}(\pi R(\pi_2)\to\pi\pi)= \sum\limits_J
A^{(J)}_{+2}(s)
X^{(J+2)}_{\alpha\beta\mu_1\ldots\mu_J}(p^{\perp})(-1)^J
O^{\mu_1\ldots\mu_J}_{\nu_1\ldots\nu_J}(\perp P)
X^{(J)}_{\nu_1\ldots\nu_J}(k^{\perp})\xi_J
\nn \\
&& +\sum\limits_J A^{(J)}_0(s) O_{\chi\tau}^{\alpha\beta}(\perp q)
X^{(J)}_{\chi\mu_2\ldots\mu_J}(p^{\perp})(-1)^J
O^{\tau\mu_2\ldots\mu_J}_{\nu_1\nu_2\ldots\nu_J}(\perp P)
X^{(J)}_{\nu_1\ldots\nu_J}(k^{\perp})\xi_J
\nn \\
&& +\sum\limits_J A^{(J)}_{-2}(s)
X^{(J-2)}_{\mu_3\ldots\mu_J}(p^{\perp})(-1)^J
O^{\alpha\beta\mu_3\ldots\mu_J}_{\nu_1\nu_2\nu_3\ldots\nu_J}(\perp
P) X^{(J)}_{\nu_1\ldots\nu_J}(k^{\perp})\xi_J \ .
\label{Tpi_2_3}
\eea
 The sum of the two terms presented in  (\ref{Tk4}) and (\ref{Tamp_pi2_f})
gives us an amplitude with a full set of the $\pi_j$-meson
exchanges.

Let us emphasize an important point: in the $K$-matrix
representation the amplitudes $ A^{(J)}_{\pi R(\pi_j)\to\pi\pi}(s)$
(Eq. (\ref{Tpi8}), $j=leading, daughter$-1) and $A^{(J)}_{+2}(s)$, $
A^{(J)}_0(s)$, $A^{(J)}_{-2}(s)$ (Eq. (\ref{Tpi_2_3})) differ only
due to the prompt-production $K$-matrix block (the term $\widehat
K_{\pi R(t)}(s)$ in (\ref{Tperiph})) while the final state
interaction factor ($[1-\hat\rho(s)\widehat K(s)]^{-1}$ in
(\ref{Tperiph})) is the same for each $J$.

\subsection{ Amplitudes with $a_J$-trajectory exchanges}

 Here we present formulae for
for leading and daughter $a_1$-trajectories and
leading  $a_2$-trajectory.

\subsubsection{ Amplitudes with  $a_1$-trajectory exchanges}

The amplitude with $t$-channel $a_1$-exchanges is a sum of leading
and daughter trajectories:
\be \label{Ta1-4}
 A_{\pi p\to\pi\pi n}^{(a_1{\rm-trajectories})}=
\sum\limits_{a_1^{(j)}} A\left(\pi R(a_1^{(j)})\to\pi\pi\right)
R_{a^{(j)}_1}(s_{\pi N},q^2) i\left(\varphi_n^+(\vec \sigma \vec
n_z)\varphi_p \right)
 g^{(a_{1j})}_{pn}(t)\ ,
\ee
where $g^{(a_{1j})}_{pn}$ is the reggeon--NN coupling and the
reggeon propagator $R_{a_1^{(j)}}(s_{\pi N},q^2)$ has the form:
\be
 R_{a_1^{(j)}}(s_{\pi N},q^2) =i
\exp{\left (-i\frac{\pi}{2}\alpha^{(j)}_{a_1} (q^2)\right )}
\frac{\left(s_{\pi N}/s_{\pi N0}\right )^{\alpha^{(j)}_{a_1} (q^2)}}
{\cos \left(\frac{\pi}{2}\alpha^{(j)}_{a_1} (q^2)\right) \Gamma \left
(\frac{1}{2} \alpha^{(j)}_{a_1} (q^2) +\frac 12 \right )
}\,  .\nn \\
\ee
 Recall that the $a_1$ trajectories have a negative signature, $\xi_\pi
=-1$. Here we take into account the leading and first daughter
trajectories which are linear and have a universal slope parameter
\cite{Tbook,Tufn,Tsyst}:
\be \label{Ta1-6}
\hspace{-0.8cm} \alpha_{a_1}^{ (leading) } (q^2) \simeq   -0.10
+0.72 q^2, \quad \alpha_{a_1}^{( daughter-1)} (q^2) \simeq   -1.10
+0.72 q^2 .
\ee
 As previously, the normalization parameter $ s_{\pi N0}$ is of the
order of 2--20 GeV$^2$, and the Gamma-functions in the reggeon
propagators are introduced in order to eliminate the poles at
$q^2<0$.

For the nucleon--reggeon vertex we use two-component spinors in the
infinite momentum frame, $\varphi_p$ and $\varphi_n$, so
the vertex reads $\left(\varphi_n^+ i(\vec \sigma \vec
n_z)\varphi_p \right) g^{(a_1)}_{pn}$ where $\vec n_z $ is the unit
vector directed along the nucleon momentum in the c.m. frame of
colliding particles.

At fixed partial wave $J^{PC}=J^{++}$, the $\pi R(a_1^j)$ channel
($j=leading, \, daughter$-1) is characterized by two angular momenta
$L=J+1,\, L=J-1$, therefore we have two amplitudes for each $J$:
 \bea \label{Ta1-8}
A\bigg(\pi R(a_1^{(j)})\to\pi\pi\bigg)&=&\sum\limits_J
\epsilon^{(-)}_{\beta} \left [ A^{(J+)}_{\pi
a_1^{(j)}\!\to\!\pi\pi}(s)
X^{(J\!+\!1)}_{\beta\mu_1\ldots\mu_J}(p^{\perp})+ A^{(J-)}_{\pi
a_1^{(j)}\!\to\!\pi\pi}(s)
Z_{\mu_1\ldots\mu_J,\beta}(p^{\perp})\right ]\nn\\
&&\times (-1)^J O^{\mu_1\ldots\mu_J}_{\nu_1\ldots\nu_J}(\perp P)
X^{(J)}_{\nu_1\ldots\nu_J}(k^{\perp})\,,
\eea
 where the polarisation vector
 $\epsilon^{(-)}_{\beta} \sim  n^{(-)}_{\beta}$; the GLF-vectors
 \cite{TGLF} defined in the c.m. system of the colliding particles
  as follows:
\be \label{GLF}
n^{(-)}_{\beta}=(1,0,0,-1)/2p_z,\quad n^{(+)}_{\beta}=(1,0,0,1)/2p_z
\ee
with $p_z\to \infty$.

The products of $Z$ and $X$ operators can be expressed through vectors
$V^{(J+)}_{\beta}$ and $V^{(J-)}_{\beta}$:
\bea \label{Ta1-11}
&&X^{(J+1)}_{\beta\mu_1\ldots \mu_J}(p^\perp) (-1)^J X_{\mu_1\ldots
\mu_J}(k^\perp)= \alpha_J
(\sqrt{-p^2_\perp})^{J+1}(\sqrt{-k^2_\perp})^J
V^{(J+)}_{\beta}\, , \nn \\
&&V^{(J+)}_{\beta}=\frac{1}{J\!+\!1} \left[P'_{J+1}(z)
\frac{p^\perp_\beta}{\sqrt{-p_\perp^2}} -P'_J(z)
\frac{k^\perp_\beta}{\sqrt{-k_\perp^2}}
\right ]\,, \nn\\
&&Z_{\mu_1\ldots\mu_J,\beta}(p^\perp) (-1)^J X^{(J)}_{\mu_1\ldots
\mu_J}(k^\perp)=\alpha_J
(\sqrt{-p^2_\perp})^{J-1}(\sqrt{-k^2_\perp})^{J} V^{(J-)}_{\beta}
\,, \nn\\
&&V^{(J-)}_{\beta}=\frac{1}{J}\left[P'_{J-1}(z)
\frac{p^\perp_\beta}{\sqrt{-p_\perp^2}} -P'_{J}(z)
\frac{k^\perp_\beta}{\sqrt{-k_\perp^2}}\right ]\, .
\eea
Here, $k_\perp^2$, $p_\perp^2$ and $z$ are defined as:
$k_\perp^2=(k^\perp k^\perp)$, $ p_\perp^2=(p^\perp p^\perp)$,
$ z=\bigg(-(k^\perp p^\perp)\bigg)/\bigg(\sqrt{-k_\perp^2}\sqrt{-p_\perp^2}\bigg)$.

\subsubsection{ The amplitude with $a_2$-trajectory exchange}

 The amplitude with $t$-channel
$a_2$-trajectory exchange reads:
\be
\label{Ta2-4}
 A_{\pi p\to\pi\pi n}^{(a_2)}&=&
\sum\limits_{a} A_{\alpha\beta}\left(\pi R(a_2)\to\pi\pi\right)
\varepsilon^{(a)}_{\alpha\beta} R_{a_2}(s_{\pi N},q^2) \frac
{\varepsilon^{(a)+}_{\alpha'\beta'}}{s^2_{\pi N}}
\times\nn \\
&&X^{(2)}_{\alpha '
\beta '}(k_3^{\perp q}) \bigg(\bar\psi (k_3)\psi (p_2) \bigg)
 g^{(a_{2})}_{pn}( q^2)\ ,
\ee
where $g^{(a_2)}_{pn}$ is the reggeon--NN coupling and the
reggeon propagator $R_{a_2}(s_{\pi N},q^2)$ has the form:
\be
 R_{a_2}(s_{\pi N},q^2) =
\exp{\left (-i\frac{\pi}{2}\alpha_{a_2} (q^2)\right )}
\frac{\left(s_{\pi N}/s_{\pi N0}\right )^{\alpha_{a_2} (q^2)}}
{\sin \left(\frac{\pi}{2}\alpha_{a_2} (q^2)\right) \Gamma \left
(\frac{1}{2} \alpha_{a_2} (q^2)  \right )
}\,  .
\ee
 Recall that the leading $a_2$ trajectory has a positive signature,
  $\xi_\pi =+1$, it is linear with the following slope parameter
\cite{Tbook,Tufn,Tsyst}:
\be \label{Ta2-6}
\hspace{-0.8cm} \alpha_{a_2} (q^2) =   0.45 \pm 0.05
+(0.72\pm 0.05) q^2\, .
\ee
 As previously, the normalization parameter $ s_{\pi N0}$ is of the
order of 2--20 GeV$^2$, and the Gamma-function in the reggeon
propagator is introduced in order to eliminate the poles at
$q^2<0$.

Using Eqs. (\ref{Tamp_pi4}), (\ref{Tpi_2_exch}), we obtain:
\be
\label{Ta2-tf}
A_{\pi p\to\pi\pi n}^{(a_2)}\!=\frac 32 A_{\alpha\beta}\left(\pi
R(a_2)\!\to\!\pi\pi\right) \frac {k^{\perp q}_{3\alpha}k^{\perp
q}_{3\beta}}{s^2_{\pi N}} R_{a_2}(s_{\pi N},q^2)\bigg(\bar\psi
(k_3)\psi (p_2) \bigg)
 g^{(a_{2})}_{pn}(q^2)\ .
\ee

Due to $a_2$ exchange, the resonance with spin $J$ can be produced
 from orbital momentum either $J-1$ or $J+1$. Thus,
\be
A_{\alpha\beta}(\pi R(a_2)\to\pi\pi)= \sum\limits_J \Big (
A^{(J)}_{-1}(s)T_{\alpha\beta}^{(J\!-\!1)}+A^{(J)}_{+1}(s)T_{\alpha\beta}^{(J\!+\!1)}
\Big )\,,
\ee
where
\be
T_{\alpha\beta}^{(J\!-\!1)}&=&\varepsilon_{\xi\alpha\tau
\eta}\frac{P_\eta}{\sqrt{s}}
X^{(J-1)}_{\xi\mu_3\ldots\mu_J}(p^\perp)
O^{\tau\beta\mu_3\ldots\mu_J}_{\nu_1\ldots\nu_J}(\perp P) (-1)^J
X^{(J)}_{\nu_1\ldots\nu_J}(k^\perp),
\nn \\
T_{\alpha\beta}^{(J\!+\!1)}&=&\varepsilon_{\xi\alpha\tau
\eta}\frac{P_\eta}{\sqrt{s}} X^{(J+1)}_{\xi\beta
\mu_2\ldots\mu_J}(p^\perp)
O^{\tau\mu_2\ldots\mu_J}_{\nu_1\ldots\nu_J}(\perp P)(-1)^J
X^{(J)}_{\nu_1\ldots\nu_J}(k^\perp).
\label{Ta2_a}
\ee
Taking into account that the tensors $T_{\alpha\beta}^{(J\!\pm\!1)}$
convolute with symmetrical tensor
$k^{\perp q}_{3\alpha}k^{\perp q}_{3\beta}$, we obtain:
\be
T_{\alpha\beta}^{(J\!-\!1)}k^{\perp q}_{3\alpha}k^{\perp
q}_{3\beta}= \frac{\varepsilon_{p\alpha k P}}{\sqrt{s}}
\frac{\alpha_{J-1}}{J(J-1)} \frac{\big(\sqrt{p_\perp^2
k_\perp^2}\big)^{J-1}}{\sqrt{-p_\perp^2}}\left (
P''_J(z)\frac{k^\perp_{\beta}}{\sqrt{-k_\perp^2}}-P''_{J-1}
\frac{p^\perp_\beta}{\sqrt{-p_\perp^2}}\right )k^{\perp
q}_{3\alpha}k^{\perp
q}_{3\beta}, \nn\\
T_{\alpha\beta}^{(J\!+\!1)}k^{\perp q}_{3\alpha}k^{\perp q}_{3\beta}
=-\frac{\alpha_{J+1}}{J}\varepsilon_{p\alpha k P}
\frac{\big(\sqrt{p_\perp^2 k_\perp^2}\big)^{J-1}}{\sqrt{s}}
p^\perp_{\beta} P'_J(z)k^{\perp q}_{3\alpha}k^{\perp q}_{3\beta} -
\frac{p_\perp^2(J\!-\!1)\alpha_{J}}{(J\!+\!1)\alpha_{J-1}}
T_{\alpha\beta}^{(J\!-\!1)} k^{\perp q}_{3\alpha}k^{\perp
q}_{3\beta}\, .
\ee

\subsubsection{Calculations in the Godfrey--Jackson system}

In the c.m. system of the produced mesons, which is used for the calculation of
the meson block (the GJ system), we
write:
\be \label{Ta1-9}
\epsilon^{(-)}_{\beta} = \frac{1}{s_{\pi N}}\left
(k_{3\mu}-\frac{q_\mu}{2}\right).
\ee
 In this system the momenta are as follows:
\bea \label{Ta1-10}
&&p_1^{\perp P}\equiv p_{\perp }= (0,0,0, p),\quad
 p^2=\frac{(s+m^2_\pi-t)^2}{4s}-m^2_\pi\, ,\quad k^2=\frac{s}{4}-m^2_\pi\, ,   \\
&&k_1^{\perp P}\equiv k_{\perp }=(0, k\sin \Theta\cos\varphi,
             k\sin \Theta\sin\varphi,  k\cos \Theta),
             \nn \\
&& q=(q_0,0,0, p), \qquad q_0=(s-m^2_\pi+t)/(2\sqrt {s}) \,
,\nn \\
&&k_3=(k_{30},k_{3x},0,k_{3z}),\quad k_{30}=(s_{\pi
N}-s-m_n^2)/(2\sqrt s),\quad k_{3z}=(2k_{30}q_0-t)/(2p)\,. \nn
\eea
 Recall that we use the notation $A=(A_0,A_x,A_y,A_z)$ and
 $\cos\Theta\equiv z=-(k^\perp p^\perp)/(\sqrt{-k_\perp^2}\sqrt{-p_\perp^2})$.

For the $a_1$-exchange the convolutions
$V^{(J+)}_{\beta}\left(k_{3\beta}-q_\beta/2\right)$,
$V^{(J-)}_{\beta}\left(k_{3\beta}-q_\beta/2\right)$ give us the
amplitude for the transition  $\pi R(a_1^{(j)})$ into two pions (in
a GJ-system the momentum $\vec k_3$  is usually situated in the
$(xz)$-plane). We write the amplitude in the form
\be \label{Ta1-12}
A\bigg(\pi R(a_1^{(j)})\to\pi\pi\bigg)&=& \sum\limits_{J} \alpha_{J}
  p^{J-1} k^J \left (W^{(J)}_0(s) Y^0_J(\Theta,\varphi)+ W^{(J)}_1(s) Re
Y^1_J(\Theta,\varphi\right )\ , \nn
\ee
 where the coefficients $W^{(J)}_0(s)$, $W^{(J)}_1(s)$ are easily
calculated:
\be
W^{(J)}_{0}&=&\sum\limits_i-N_{J0}\left(k_{3z}-\frac{|\vec
p|}{2}\right ) \left (|\vec p|^2 A^{(J+)}_{\pi
a_1^{(i)}\!\to\pi\pi}- A^{(J-)}_{\pi
a_1^{(i)}\!\to\pi\pi} \right ), \nn \\
W^{(J)}_{1}&=&\sum\limits_i-\frac{N_{J1}}{J(J\!+\!1)}k_{3x} \left
(|\vec p|^2 J\,A^{(J+)}_{\pi a_1^{(i)}\!\to\pi\pi}+
(J\!+\!1)A^{(J-)}_{\pi a_1^{(i)}\!\to\pi\pi}\right ).
\ee

For $a_2$-exchange, one has:
\be
T_{\alpha\beta}^{(J\!-\!1)}k^{\perp q}_{3\alpha}k^{\perp
q}_{3\beta}= \frac{\alpha_{J-1}}{J}p^{J\!-\!1}k^{J}k_{3x} \big
[(k_{3z}\!-\!\frac{p}{2})N_{1J}\,Im\,Y^1_J(\Theta,\varphi)-
\frac{k_{3x}}{2}\frac{N_{2J}}{J\!-\!1}Im\,Y^2_J(\Theta,\varphi) \big ]
\ee

For the amplitude with orbital momentum $J+1$, we write:
\be
T_{\alpha\beta}^{(J\!+\!1)} k^{\perp q}_{3\alpha}k^{\perp
q}_{3\beta} =-\alpha_{J+1} p^{J\!+\!1}k^J \big
(k_{3z}\!-\!\frac{p}{2}\big)
\frac{N_{1J}}{J}Im\,Y^1_J(\Theta,\varphi) -
\frac{p_\perp^2(J\!-\!1)\alpha_{J}}{(J\!+\!1)\alpha_{J-1}}
T_{\alpha\beta}^{(J\!-\!1)} k^{\perp q}_{3\alpha}k^{\perp
q}_{3\beta}\, .
\ee
The final expression for the $a_2$-exchange amplitude can be written
as follows:
\be
\label{Ta2-fin}
A_{\pi p\to\pi\pi n}^{(a_2)}&=&\frac {3k_{3x}}{2s^2_{\pi N}}
\sum\limits_J p^{J\!-\!1}k^J\big [
W_{a_2}^{1J}Im\,Y^1_J(\Theta,\varphi)+
W_{a_2}^{2J}Im\,Y^2_J(\Theta,\varphi)\big ]\times
\nn \\
&&R_{a_2}(s_{\pi N},q^2)\bigg(\bar\psi (k_3)\psi (p_2) \bigg)
 g^{(a_{2})}_{pn}(q^2)\ ,
\ee
where
\be
W_{a_2}^{1J}&=&\frac{N_{1J}}{J}\big(k_{3z}-\frac{p}{2}\big)\Big
[-p^2\alpha_{J\!+\!1}
A^{(J)}_{+1}+\left(\frac{\alpha_{J\!-\!1}}{J\!-\!1}A^{(J)}_{-1}+
\frac{p^2\alpha_{J}}{J\!+\!1}A^{(J)}_{+1}\right )(J-1)\Big ]
\nn \\
W_{a_2}^{2J}&=&-\,\frac{N_{2J}}{J}\,\frac{k_{3x}}{2}\Big
[\frac{\alpha_{J\!-\!1}}{J\!-\!1}A^{(J)}_{-1}+
\frac{p^2\alpha_{J}}{J\!+\!1}A^{(J)}_{+1}\Big ].
\ee

For the unpolarized cross section, the amplitude related to $a_2$
exchange does not interfere  with either $\pi$, $\pi_2$ or $a_1$
exchange amplitudes. If the highest moments are  small in the cross
section, one can assume that the combination in front of $Y^2_n$ is
close to 0. Then
\be
W_{a_2}^{1J}&=&-N_{1J}\big(k_{3z}-\frac{p}{2}\big)
p^2\alpha_{J\!+\!1} A^{(J)}_{+1} ,
\nn \\
W_{a_2}^{2J}&=&0,
\ee
and, as a result, we have:
\be
A_{\pi p\to\pi\pi n}^{(a_2)}&=&-\frac {3k_{3x}}{2s^2_{\pi N}}
\sum\limits_J\frac{\xi_J}{J} p^{J\!+\!1}k^J\big [
N_{1J}\big(k_{3z}-\frac{p}{2}\big) \alpha_{J\!+\!1}
A^{(J)}_{+1}Im\,Y^1_J(\Theta,\varphi)\big ]
\nn \\
&\times&R_{a_2}(s_{\pi N},q^2)\bigg(\bar\psi (k_3)\psi (p_2) \bigg)
 g^{(a_{2})}_{pn}(q^2)\, .
\ee

\subsubsection{Partial wave decomposition}

The partial wave amplitude $\pi R(a^{(j)}_1)\to \pi\pi$
with fixed $J^{++}$ is presented in the $K$-matrix form:
\be
\label{Ta1-13}
A^{(L=J\pm 1,J^{++})}_{\pi R(a^{(j)}_1),\pi\pi}(s) =\sum\limits_b
K_{\pi R(a^{(j)}_1),\, b}^{(L=J\pm 1,J^{++})}(s,q^2) \left [\frac
{\hat I}{\hat I-i\hat{\rho}(s) \hat K^{(J^{++})}(s)}\right
]_{b,\pi\pi},
\nn \\
\ee
where $K_{\pi R(a^{(j)}_1),b}^{(L=J\pm 1,J^{++})}(s,q^2)$ is the
following vector ($b=\pi\pi$, $K\bar K$, $\eta\eta$, $\eta\eta'$,
$\pi\pi\pi\pi$):
\be
K_{\pi R(a^{(j)}_1),\, b}^{(L=J\pm 1,J^{++})}(s,q^2) &=&\bigg(
\sum_\alpha \frac{G^{(L=J\pm 1,J^{++},\,
 \alpha)}_{\pi R(a^{(j)}_1)}(q^2)
g^{(J^{++},\, \alpha)}_b} {M^2_\alpha-s}\bigg.
\nn \\
&+&\bigg.F^{(J^{L=J\pm 1,++})}_{\pi R(a^{(j)}_1),\, b}(q^2)
\frac{1\;\mbox{GeV}^2+s_{R0}}{s+s_{R0}} \bigg)\;
\frac{s-s_A}{s+s_{A0}}\ .~~~~~~
\label{Ta1-14}
\ee
Here $G^{(L=J\pm 1,J^{++},\, \alpha)}_{\pi R(a^{(j)}_1)}(q^2)$ and
$F^{(J^{L=J\pm 1,++})}_{\pi R(a^{(j)}_1),\, b}(q^2)$ are the
$q^2$-dependent reggeon form factors.

\subsection{$\pi^- p \to K\bar K \, n$ reaction with $K\bar K
$-exchange by $\rho$-meson trajectories}

In the case of the production of a $K\bar K$ system the resonance in
this channel can have isospins $I=0$ and $I=1$, with even spin
(production of states of the types $\phi$ and $a_0$). Such processes
are described by $\rho$-exchanges.

\subsubsection{Amplitude with exchanges by $\rho$-meson
trajectories}

The amplitude with $t$-channel $\rho$-meson exchanges is written as
follows:
\be \label{Tr-k4}
 A_{\pi p\to K\bar K n}^{({\rm \rho - trajectories})}
= \sum\limits_{\rho_j}
 A\bigg(\pi R(\rho_j)\to K\bar K \bigg)
 R_{\rho_j}(s_{\pi N},q^2)
\hat g^{(\rho_j)}_{pn},
\ee
 where the reggeon propagator $R_{\rho_j}(s_{\pi N},q^2)$ and
the reggeon--nucleon vertex $\hat g^{(\rho_j)}_{pn}$ read,
respectively:
\bea \label{Tr-k5}
&& R_{\rho_j}(s_{\pi N},q^2) = \exp{\left
(-i\frac{\pi}{2}\alpha^{(j)}_\rho (q^2)\right )} \frac{\left(s_{\pi
N}/s_{\pi N0}\right )^{\alpha^{(j)}_\rho (q^2)}} {\sin
\left(\frac{\pi}{2}\alpha^{(j)}_\rho (q^2)\right) \Gamma \left
(\frac{1}{2} \alpha^{(j)}_\rho (q^2) +1\right ) }\,  ,\nn \\
 && \hat
g^{(\rho_j)}_{pn}=
 g^{(\rho_j)}_{pn}(1)(\varphi_n^+\varphi_p ) + g^{(\rho_j)}_{pn}(2)
 \left(\varphi_n^+
\frac{i}{2m_N}(\vec q_\perp [\vec n_z ,\vec\sigma ])
 \varphi_p \right) .
\eea
 The $\rho_j$-reggeons have positive signatures,
$\xi_\rho =+1$, being determined by linear trajectories
 \cite{Tbook,Tufn,Tsyst}:
\be \label{Tr-pi6}\hspace{-0.5cm}
\alpha_{\rho}^{ (leading)} (q^2)\simeq 0.50 +0.83 q^2 , \alpha_{\rho
}^{( daughter-1)} (q^2)\simeq -0.75 +0.83 q^2  .
 \ee
The slope parameters are in (GeV/c)$^{-2}$ units,
$ s_{\pi N0}\sim 2-20$ GeV$^2$.
Two vertices in $\hat g^{(\rho_j)}_{pn}$ correspond to
charge- and magnetic-type interactions (they are written in the
infinite momentum frame of the colliding particles).

The meson--reggeon amplitude can be written as
\be \label{Tr-pi8}
 A\bigg(\pi R(\rho_j)\to K\bar K \bigg)
 = \sum\limits_{J}
\varepsilon_{\beta \epsilon^{(-)} p P} Z_{\mu_1\mu_2
\ldots\mu_J,\beta}(p^\perp) A^{(J^{++})}_{\pi R_\rho(q^2),K\bar K
}(s)X^{(J)}_{\mu_1\mu_2 ...\mu_J}(k^\perp)(-1)^J\ ,
\ee
 where the polarisation vector $\epsilon^{(-)}_{\beta}$ was introduced
in (\ref{Ta1-9}).

We use the convolution of the $Z$ and $X$ operators in the GJ-system
(see notations in (\ref{Ta1-10}):
\be
Z_{\mu_1\ldots\mu_J,\beta}(p^\perp) (-1)^J X^{(J)}_{\mu_1\ldots
\mu_J}(k^\perp)= \frac{\alpha_J}{J}
(\sqrt{-p^2_\perp})^{J-1}(\sqrt{-k^2_\perp})^{J}  \left[P'_{J-1}(z)
\frac{p^\perp_\beta}{\sqrt{-p_\perp^2}} -P'_{J}(z)
\frac{k^\perp_\beta}{\sqrt{-k_\perp^2}}\right ]\!.
\ee
 The convolution of the spin--momentum operators in (\ref{Tr-pi8})
gives:
\be
\hspace{-8mm}A(\pi \rho_j\to\pi\pi)= \sum\limits_{J}
\frac{\alpha_{J}}{J}
  p^J k^J k_{3x}\sqrt{s} N_{j1} {\rm Im}\,Y^1_J(\Theta,\varphi)
 A^{(J^{++})}_{\pi R_\rho(q^2),K\bar K}(s).
\ee
 Let us remind that in the GJ-system the vector $\vec k_3$ is situated in
the $(xz)$-plane.

\subsubsection{Partial wave decomposition}

The  amplitude for the transition $\pi R_{\rho_j}(q^2)\to K\bar K$ in the
$K$-matrix representation reads:
\be
\hspace{-9mm}A^{(J^{++})}_{\pi R(\rho_j),K\bar K}(s) =\sum\limits_b
K_{\pi R(\rho_j),\, b}^{(J^{++})}(s,q^2) \left [\frac {\hat I}{\hat
I-i\hat{\rho}(s) \hat K^{(J^{++})}(s)}\right ]_{b,K\bar K},
\label{Tr-13}
\ee
where $K_{\pi R(\rho_j),b}^{(J^{++})}(s,q^2)$ is the following
vector ($b=\pi\pi$, $K\bar K$, $\eta\eta$, $\eta\eta'$,
$\pi\pi\pi\pi$):
\be
K_{\pi R(\rho_j),\, b}^{(J^{++})}(s,q^2) &=&\bigg( \sum_\alpha
\frac{G^{(J^{++},\, \alpha)}_{\pi R(\rho_j)}(q^2) g^{(J^{++},\,
\alpha)}_b}
{M^2_\alpha-s}\bigg. \nn \\
&+&\bigg.F^{(J^{++})}_{\pi R(\rho_j),\, b}(q^2)
\frac{1\;\mbox{GeV}^2+s_{R0}}{s+s_{R0}} \bigg)\;
\frac{s-s_A}{s+s_{A0}}\;\; .
\label{Tr-14}
\ee
Here $G^{(J^{++},\, \alpha)}_{\pi R(\rho_j)}(q^2)$ and
$F^{(J^{++})}_{\pi R(\rho_j),\, b}(q^2)$ are the reggeon
$q^2$-dependent form factors.

\section{Low-energy three-meson production in the
$K$-matrix approach}

Here we present elements of the $K$-matrix technique for the
low-energy reactions $p\bar p \to \pi\pi\pi,\pi\eta\eta$, $\pi K\bar
K$. The $K$-matrix technique provides a compact and, hence, a
convenient way for studying resonances in multiparticle processes of
such a type. However, we have to pay a price for the simplifications
the $K$-matrix technique gives us: we cannot take into account in a
full scale the partial wave left singularities as well as the
singularities related to the rescattering of all particles (for
example, the singularities of the triangle type diagrams)

The use of the $K$-matrix approach to the combined
analysis of different processes is based on
the fact that the denominator of the $K$-matrix two-particle
amplitude, $[1-\hat \rho \hat K]^{-1}$ is common for all processes,
depending only on quantum numbers of the considered two-meson system.

Let us illustrate this statement using as an example the amplitude
of the $p\bar p$ annihilation from the $^1S_0$ level: $p\bar
p(^1S_0)\to three\, mesons$. In the $K$-matrix approach, the
production amplitude for the resonance with the spin $J=0$ in the
channel ($1+2$) reads:
\beq
A_3(s_{12})_{ca}=\sum\limits_b\biggl(
K^{(prompt)}_3(s_{12})\biggr)_{cb}\biggl(\frac{1}{1-i\hat\rho_{12}
\widehat K_{12}(s_{12})}\biggr)_{ba},
\label{Tsubsub1}
\eeq
where
 $c=p\bar p(^1S_0)$ and $a,b\in \pi\pi,\eta\eta, K\bar K$. The
denominator $[1-i\hat \rho_{12}\widehat K_{12}(s_{12})]^{-1}$
depends on the invariant energy squared of mesons 1 and 2 and it
coincides with the denominator of the two-particle scattering amplitude.
The factor $\widehat K^{(prompt)}_3(s_{12})$ stands for the prompt
production of particles and resonances in this channel:
\beq
\biggl (K^{(prompt)}_3(s_{12}) \biggr )_{cb}  =\ \sum \limits_n
\frac{\Lambda^{(n)}_c g^{(n)}_b}{\mu^2_n-s_{12}}+
\varphi_{cb}(s_{12})\ ,
\label{Tsubsub2}
\eeq
where $\Lambda^{(n)}_c$ and $\varphi_{cb}$ are the parameters of the
prompt-production amplitude, and $g_b^{(n)}$ and $\mu_n$  are the
same as in the two-meson scattering amplitude.

The whole amplitude for the production of the $(J=0)$-resonances is
defined by the sum of  contributions from  all channels:
\beq
A_3(s_{12})+A_2(s_{13})+A_1(s_{23}).
\label{Tsubsub3}
\eeq
 The amplitudes $A_2(s_{13})$ and $A_1(s_{23})$ are given by formulae
similar to (\ref{Tsubsub1}), (\ref{Tsubsub2}) but with different
sets of final and intermediate states.

To take into account the resonances with non-zero spins $J$, one has
to substitute in (\ref{Tsubsub1})
\be
A_3(s_{12}) \to \sum\limits_J A_3^{(J)}(s_{12})
X_{\mu_1\mu_2...\mu_J}^{(J)}(k_{12}^{\perp p_{12}})
X_{\mu_1\mu_2...\mu_J}^{(J)}(k_{3}^{\perp P}),
\label{Tsubsub4}
\ee
where the $K$-matrix amplitude $A^{(J)}_3(s_{12})$ is determined by
an expression similar to (\ref{Tsubsub1}).

  The analysis performed in
\cite{Tf1500,Tf1500PR} showed that in the reactions $p\bar p({\rm
at\, rest})\to \pi^0\pi^0\pi^0$, $\pi^0\pi^0\eta$, $\pi^0\eta\eta$
the determination of  parameters of resonances produced in the
two-meson channels does not require the explicit consideration of
the triangle diagram singularities --- it is important to take into
account only the complexity of parameters $\Lambda^{(n)}_a $ and
$\varphi_{ab}$ in (\ref{Tsubsub2}) which are due to multiparticle
final-state interactions. Note that this is not a universal rule for
the meson production processes in the $p\bar p$ annihilation -- for
example, in the reaction $p\bar p \to \eta\pi^+\pi^-\pi^+\pi^-$
\cite{Tnana}, the triangle singularity contribution is important.

\section{Fitting procedure}

Here we present results of the combined fit for low energy
annihilation reactions  $p\bar p\to \pi\pi\pi$, $\pi\pi\eta$,
$\pi\eta\eta$  and high energy peripheral production  $\pi^- p\to
\pi^0\pi^0+n$.

\subsection{
The $K$-matrix fit of annihilation reactions at rest
 $p\bar p$ into
$ \pi\pi\pi$, $\pi\pi\eta$, $\pi\eta\eta$ }

We  have included into the fit procedure the following data sets for
the production of three mesons in $p\bar p$
annihilation:\\
 (1)  Crystal Barrel data on
 $p\bar p({\rm at\, rest,\, from\, liquid\,H_2})\to \pi^0\pi^0\pi^0$,
 $\pi^0\pi^0\eta$, $\pi^0\eta\eta$
\cite{Tcbc} and \\
(2) the data in gas
$p\bar p({\rm at\, rest,\, from\, gaseous\, H_2})\to \pi^0\pi^0\pi^0$,
$\pi^0\pi^0\eta$ \cite{Tcb,Tcbc_new}.

The considered $K$-matrix amplitude takes into account
$\pi\pi\pi\pi$, $ K\bar K$ and $\eta\eta'$ channels as well --
parameters for these channels are taken from \cite{Tepja}.

First, we present the formulae for the reactions $p\bar p\to
\pi^0\pi^0\pi^0$, $\pi^0\pi^0\eta$, $\pi^0\eta\eta$ from the liquid
$H_2$, when annihilation occurs from the $^1S_0p\bar p$ state and
scalar resonances, $f_0$ and $a_0$, are formed in the final state.
This is a case which represents well the applied technique of the
three-meson production reactions. A full set of amplitude terms
taken into account in the analysis \cite{Tepja} (production of
vector and tensor resonances, $p\bar p$ annihilation from the
$P$-wave states $^3P_1$, $^3P_2$, $^1P_1$) is constructed in an
analogous way.

{\centerline{\bf (i) Production  of the S-wave resonances.}}
For the
transition $p\bar p\; (^1S_0)\to \pi^0\pi^0\pi^0$ with the
production of two pions in a $(00^{++})$-state, we use the following
amplitude:
\bea \label{T15pi}
 &&A_{p\bar p\; (1^1S_0)\to
\pi^0\pi^0\pi^0}= \left (\bar\psi(-q_2)\frac{i\gamma_5}{2\sqrt 2
m_N} \psi(q_1)\right )
\\
&& \times \left [A_{p\bar p\; (1^1S_0)\pi^0,\pi^0\pi^0}(s_{23})
\!+\!A_{p\bar p\; (1^1S_0)\pi^0,\pi^0\pi^0}(s_{13}) \!+\!A_{p\bar
p\; (1^1S_0)\pi^0,\pi^0\pi^0}(s_{12})\right ] . \nn
\eea
 The four-spinors $\bar\psi(-p_2)$ and $ \psi(p_1)$ refer to the initial
antiproton and proton in the $I^{(2S+1)}L_J=1^1S_0$ state. For the
produced pseudoscalars we denote amplitudes in the left-hand side of
(\ref{T15pi}) as $A_{p\bar p\; (1^1S_0)P_\ell\, ,P_iP_j}(s_{ij})$.

The amplitudes for the transitions $p\bar p\; (0^1S_0)\to
\eta\pi^0\pi^0$, $p\bar p\; (1^1S_0)\to\pi^0\eta\eta$ have a similar
form:
\bea \label{T15eta}
&&A_{p\bar p\; (0^1S_0)\to \eta\pi^0\pi^0}= \left
(\bar\psi(-p_2)\frac{i\gamma_5}{2\sqrt 2 m_N} \psi(p_1)\right ) \\
&& \times\left [A_{p\bar p\; (0^1S_0)\eta,\pi^0\pi^0}(s_{23})
+A_{p\bar p\; (0^1S_0)\pi^0,\eta\pi^0}(s_{13}) +A_{p\bar p\;
(0^1S_0)\pi^0,\eta\pi^0}(s_{12})\right ] \; ,\nn
\eea
and
\bea \label{T15etaeta}
&&A_{p\bar p\; (1^1S_0)\to \pi^0\eta\eta}= \left
(\bar\psi(-p_2)\frac{i\gamma_5}{2\sqrt 2 m_N} \psi(p_1)\right )
\\
&& \times\left [A_{p\bar p\; (1^1S_0)\pi^0,\eta\eta}(s_{23})
+A_{p\bar p\; (1^1S_0)\eta,\eta\pi^0}(s_{13}) +A_{p\bar p\;
(1^1S_0)\eta,\eta\pi^0}(s_{12})\right ] \; . \nn
\eea
 For the description of the $S$-wave interaction of two mesons
in the scalar--isoscalar state (index $(00)$) the following
amplitudes are used in (\ref{T15pi}), (\ref{T15eta}) and
(\ref{T15etaeta}):
\be
A_{p\bar p\; (I^1S_0)\pi^0,b}(s_{ij}) =\sum\limits_{a}\widetilde
K_{p\bar p(I^1S_0)\pi^0,a}^{(00)} (s_{ij}) \left[\hat I-i
\hat\rho^{(0)}_{ij} (s_{ij})
\hat K^{(00)}(s_{23})\right]^{-1}_{ab}\ .\nn \\
\label{T16rest}
\ee
Here $b=\pi^0\pi^0$, $\eta\eta$ and $a=\pi^0\pi^0$, $\eta\eta$,
$K\bar K$, $\eta\eta'$,  $\pi^0\pi^0\pi^0\pi^0$. The $K$-matrix term
is responsible for the two-meson scattering.
The $\widetilde K$-matrix terms which describe the prompt
resonance and background meson production in the $p\bar p$
annihilation read:
\be
\widetilde K_{p\bar p(1^1S_0)\pi^0,a}^{(00)}(s_{23})&=&\bigg(
\sum_\alpha \frac{\Lambda^{(00,\alpha)}_{p\bar p(1^1S_0)\pi^0}
g^{(\alpha)}_a} {M^2_\alpha-s_{23}}
\bigg. \nn \\
&+&\bigg.\phi_{p\bar p(1^1S_0)\pi^0,a}^{(00)}\; \frac{1\;
\mbox{GeV}^2+\tilde s_0}{s_{23}+\tilde s_0} \bigg) \left
(\frac{s_{23}-\tilde s_A}{s_{23}+\tilde s_{A0}}\right ).~~~~~~~
\label{T18rest}
\ee
The parameters $\Lambda^{(00,\alpha)}_{p\bar p(1^1S_0)\pi^0,a}$ and
$\phi_{p\bar p(1^1S_0)\pi^0,a}^{(00)}$ are complex-valued, with
different phases due to
 three-particle interactions. Let us recall: the matter is that in
the final state interaction term  we take into account the leading
(pole) singularities only.  The next-to-leading singularities are
accounted for effectively, by considering the vertices $p\bar p\to
mesons$ as complex-valued factors.

{\bf (ii) Three-meson amplitudes with the production of
spin-non-zero resonances.}

In the three-meson production processes, the final-state two-meson
interactions in other states are taken into account in a way similar
to what was considered above.

The invariant part of the production amplitude $ A^{(I,tj)}_{p\bar
p\; (I\; ^{1}S_0,b)}(23)$ for the transition $p\bar p\; (I\;
^{1}S_0) \to 1+(2+3)_{tj}$, where the indices $tj$ refer to the
isospin and spin of the meson in the channel $b=2+3$, is as follows:
\bea
\label{T6A.6}
 A_{p\bar p\; (I\; ^{1}S_0)1,b} ^{(tj)}(23)&=&\sum\limits_{a}\widetilde
K_{p\bar p\; (I\; ^{1}S_0)1,a}^{(tj)} (s_{23})
 \left[\hat I-i \hat\rho^{(j)}_{23}
\hat K^{(tj)}(s_{23})\right]^{-1}_{ab}\; ,\nn \\
\widetilde K_{p\bar p\; (I\; ^{1}S_0)1,a} ^{(tj)}(s_{23})& =&\bigg(
\sum_\alpha \frac{\Lambda_{p\bar p\; (I\; ^{1}S_0)1}^{(tj,\alpha)}
g^{(\alpha)}_a} {M^2_\alpha-s_{23}}\bigg. \nn \\ &+&\bigg.
\phi_{p\bar p\; (I\; ^{1}S_0)1,a}^{(tj)} \frac{1\;
\mbox{GeV}^2+\tilde s_{tj0}}{s_{23}+\tilde s_{tj0}} \bigg)
D_a(s_{23})\, .
\eea
The parameters $\Lambda_{p\bar p\; (I\; ^{1}S_0)1}^{(tj,\alpha)} $,
$\phi_{p\bar p\; (I\; ^{1}S_0)1,a}^{(tj)}$ may be complex-valued,
with different phases due to three-particle interactions.

The $K$-matrix elements for the scattering amplitudes (which enter
the denominator of (\ref{T6A.6})) are determined in the partial
waves $02^{++}$, $10^{++}$, $12^{++}$ as follows:

{\it (1) Isoscalar--tensor, $02^{++}$, partial wave.}

The $D$-wave interaction in the isoscalar sector is parametrized by
the 4$\times $4 $K$-matrix where $1 = \pi\pi$, $2 = K\bar K$, $3 =
\eta\eta$ and $4 =  {\rm multi-meson\; states}$:
\be
\hspace{-7mm}K_{ab}^{(02)}(s)=D_a(s)\left ( \sum_\alpha
\frac{g^{(\alpha)}_a g^{(\alpha)}_b}{M^2_\alpha-s}
+f^{(02)}_{ab}\frac{1\,\mbox{GeV}^2+s_2}{s+s_2} \right ) D_b(s)\;.
\label{T6A.7}
\ee
Factor $D_a(s)$ stands for the $D$-wave
centrifugal barrier. We take this factor in the following form:
\be
D_a(s)=\frac{k_a^2}{k_a^2+3/r_a^2},\quad a=1,2,3\ ,
\label{T6A.8}
\ee
where $k_a=\sqrt{s/4-m_a^2} $ is the momentum of the decaying
meson in the c.m. frame  of the resonance. For the multi-meson
 decay the factor $D_4(s)$ is taken to be  1. The phase space
factors we use are the same as those for the isoscalar $S$-wave
channel.

{\it (2) Isovector--scalar, $10^{++}$, and isovector--tensor,
$12^{++}$, partial waves.}

For the amplitude in the isovector-scalar and isovector-tensor
channels we use the 4$\times $4 $K$-matrix with 1 = $\pi\eta$, 2 =
$K\bar K$, 3 = $\pi\eta'$ and 4 =  multi-meson states:
\be
\hspace{-7mm}K_{ab}^{(1j)}(s)=D_a(s)
 \left ( \sum_\alpha \frac{g^{(\alpha)}_a
g^{(\alpha)}_b}{M^2_\alpha-s}
 +f^{(1j)}_{ab}\frac{1.5\; \mbox{GeV}^2+s_1}{s+s_1} \right )D_b(s)\;.
\label{T6A.9}
\ee
Here $j=0,2$; the factors $D_a(s)$ are equal to 1 for the $10^{++}$
amplitude, while for the $D$-wave partial
 amplitude the factor $D_a(s)$ is taken in the form
\bea
 D_a(s)=\frac{k_a^2}{k_a^2+3/r_3^2}, \;\; a=1,2,3,  \qquad
 D_4(s)=1\; .
\label{T6A.10}
\eea

\begin{figure}[t]
\centerline{\epsfig{file=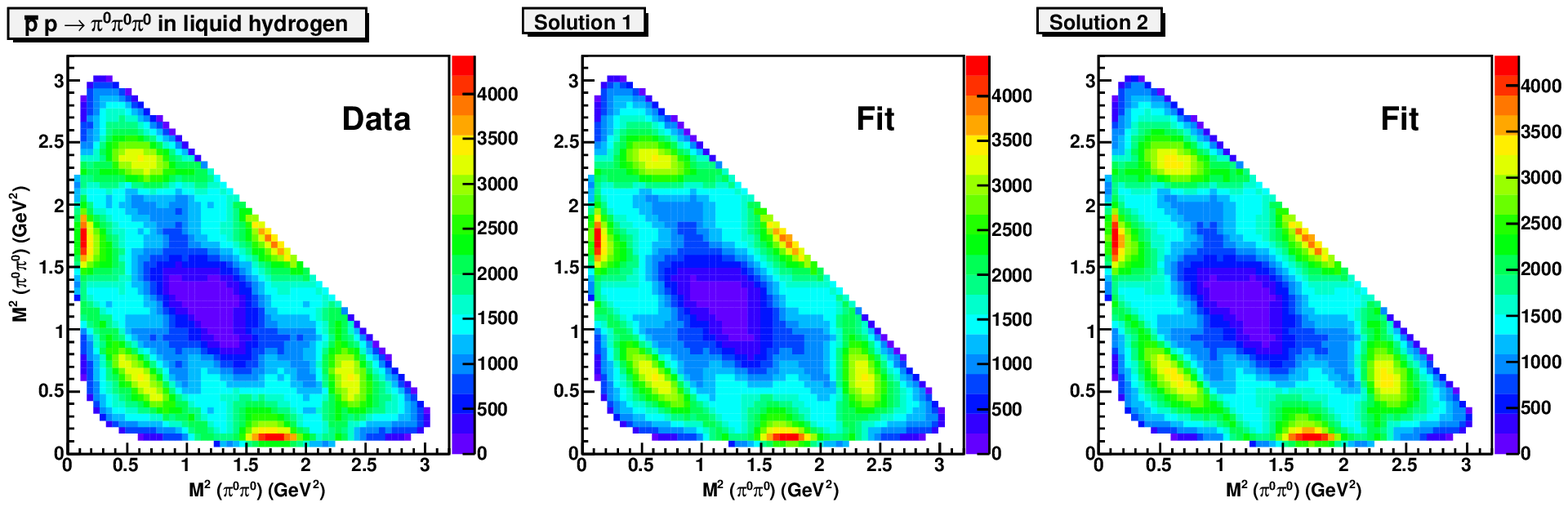,width=0.95\textwidth}}
%Fig.2
\caption{The acceptance-corrected Dalitz plot for the $p\bar p$
annihilation into $\pi^0\pi^0\pi^0$ in liquid $H_2$ and the result
of the two solutions.}
\label{T6km-9}
\vspace{-5mm}
%Fig.3
\centerline{\hspace*{0.7cm}\epsfig{file=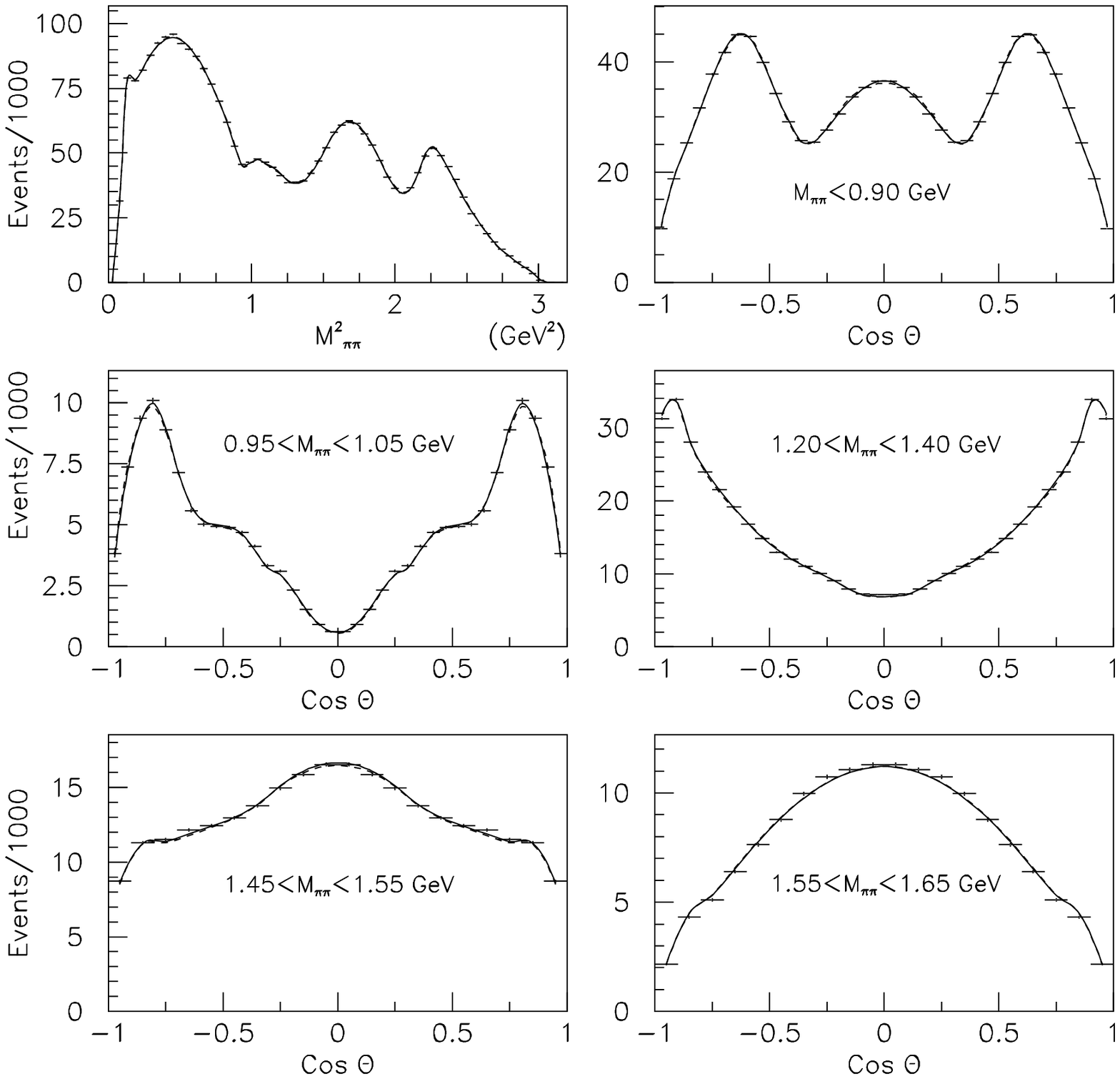,width=0.95\textwidth}}
\vspace{-5mm} \caption{Mass projection of the acceptance-corrected
Dalitz plot and angular distributions for specific mass slices. The
data are taken from the $p\bar p$ annihilation into $3\pi^0$ in
liquid $H_2$. Solid curves correspond to the solution 1 and dashed
curves to the solution 2. }
\label{T6km-9p}
\end{figure}

\begin{figure}
%Fig. 4
\centerline{\epsfig{file=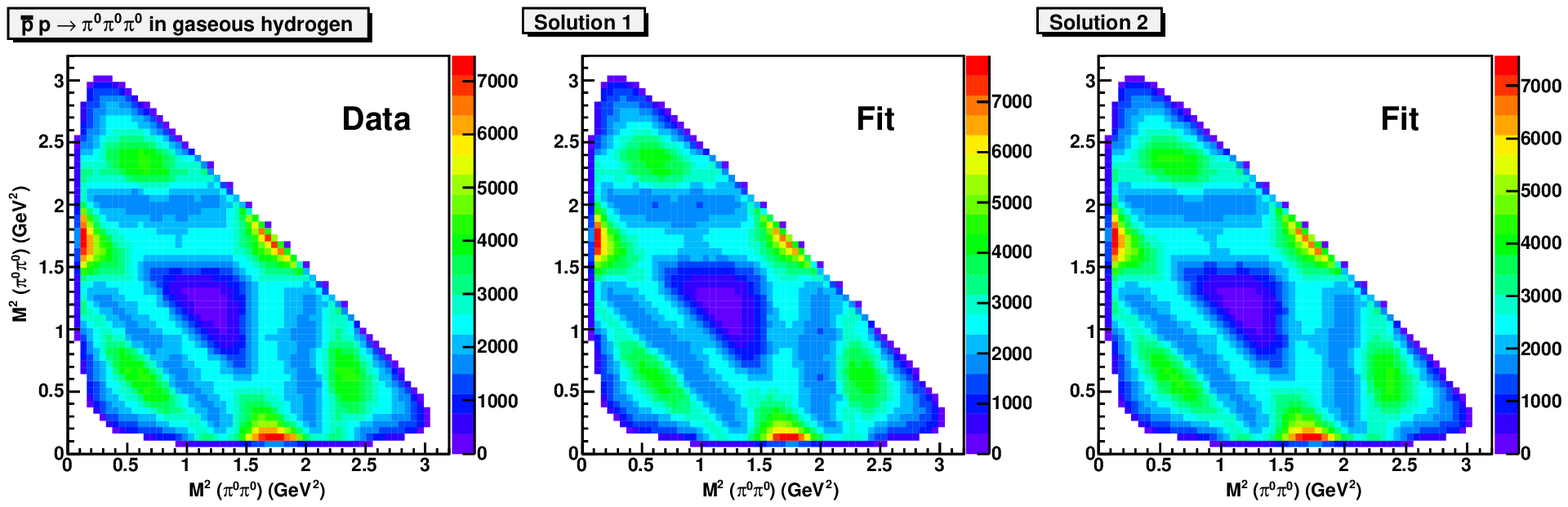,width=0.95\textwidth}} \caption{The
acceptance-corrected Dalitz plot for the $p\bar p$ annihilation into
$\pi^0\pi^0\pi^0$ in gaseous $H_2$ and the result of the two
solutions.}
\label{T6km-9g}
\vspace{-5mm}
%Fig.5
\centerline{\hspace*{0.7cm}\epsfig{file=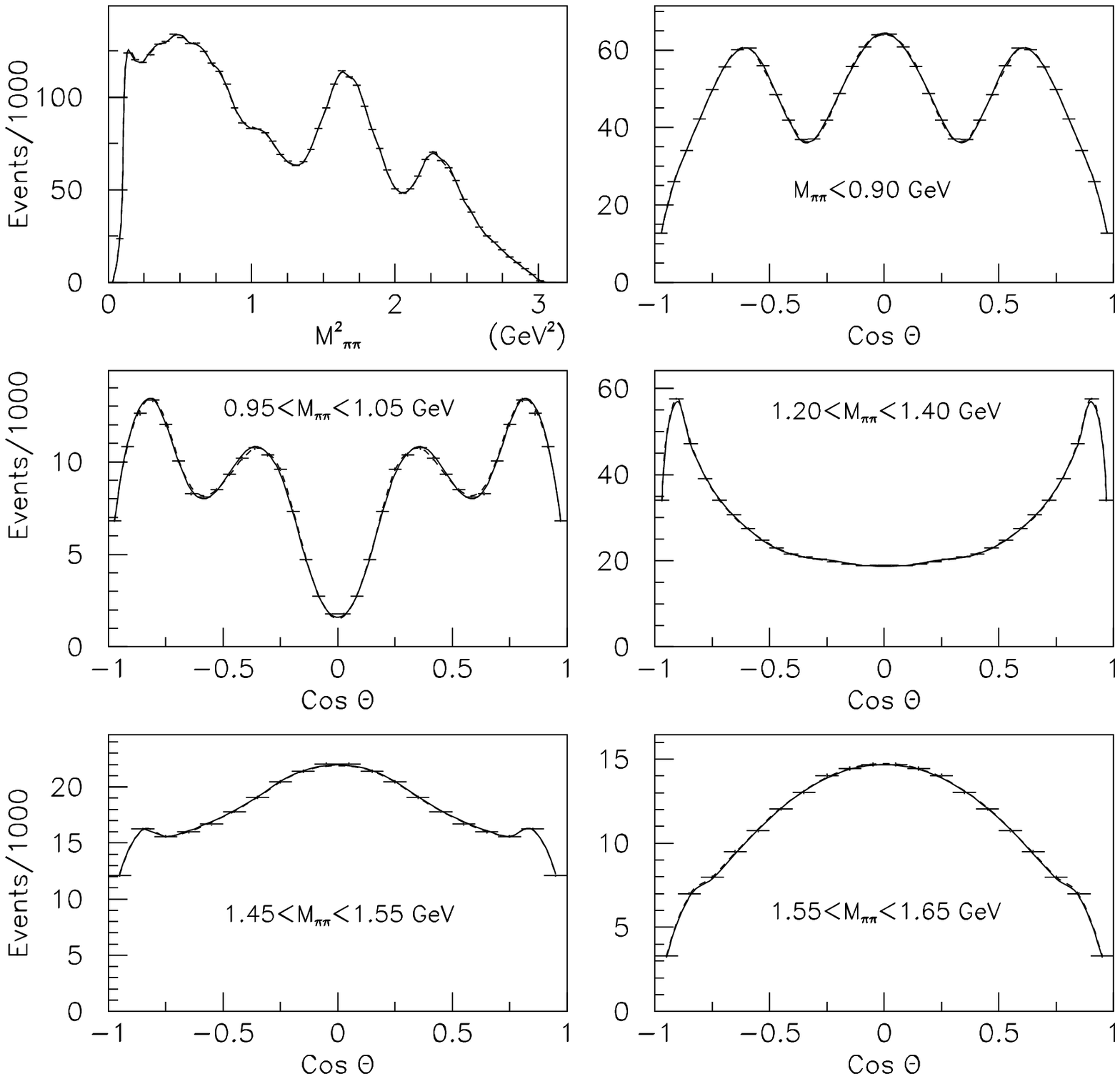,width=0.95\textwidth}}
\vspace{-5mm} \caption{Mass projection of the acceptance-corrected
Dalitz plot and angular distributions for specific mass slices. The
data are taken from the $p\bar p$ annihilation into
$\pi^0\pi^0\pi^0$ in gaseous  $H_2$. Solid curves correspond to the
solution 1 and dashed curves to the solution 2.}
\label{T6km-9gp}
\end{figure}

\begin{figure}
%Fig.6
\centerline{\epsfig{file=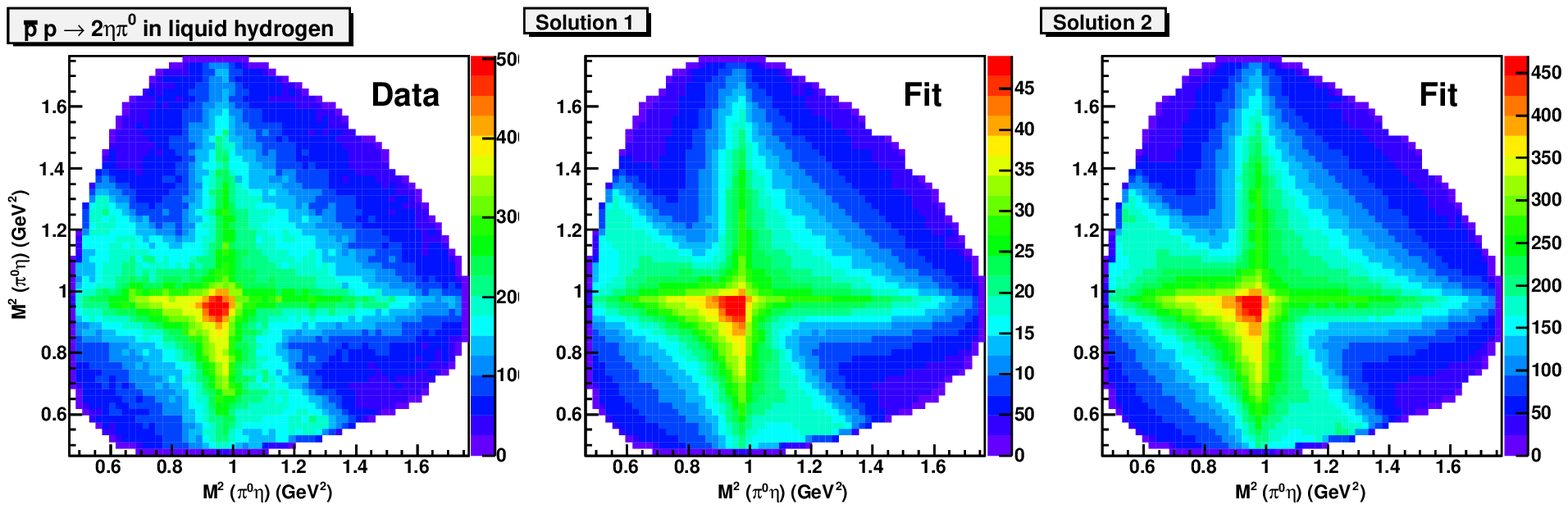,width=0.92\textwidth}} \caption{The
acceptance-corrected Dalitz plot for the $p\bar p$ annihilation into
$\eta\pi^0\eta$ in liquid $H_2$ and the result of the two
solutions.}
\label{T6km-10}
\vspace{-5mm}
%Fig. 7
\centerline{\hspace*{0.7cm}\epsfig{file=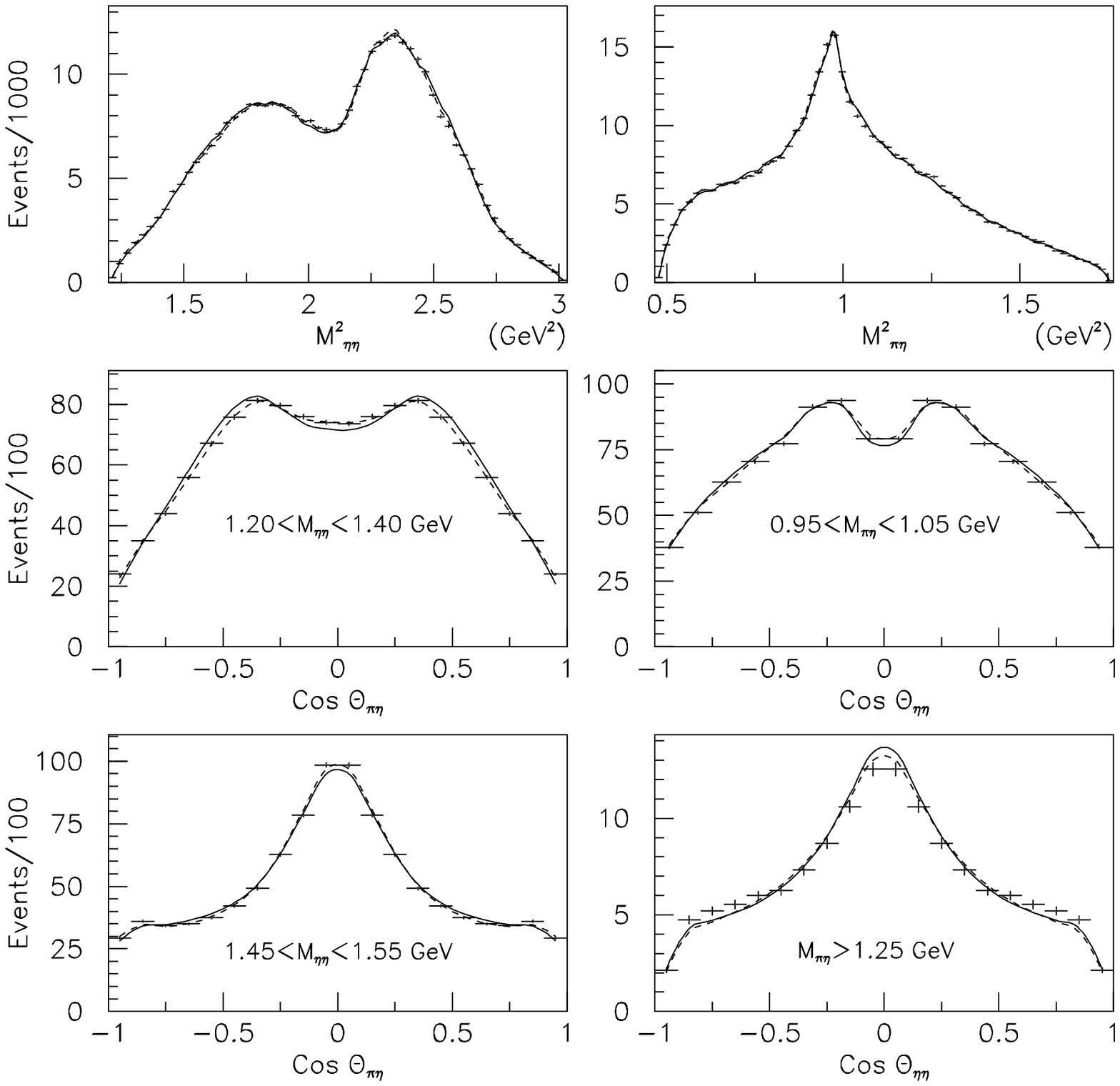,width=0.92\textwidth}}
\vspace{-5mm} \caption{Mass projections of the acceptance-corrected
Dalitz plot and angular distributions for specific mass slices. The
data are taken from the $p\bar p$ annihilation into $\eta\pi^0\eta$
in liquid $H_2$. Solid curves correspond to the solution 1 and
dashed curves to the solution 2.}
\label{T6km-10p}
\end{figure}

\begin{figure}
%Fig. 8
\centerline{\epsfig{file=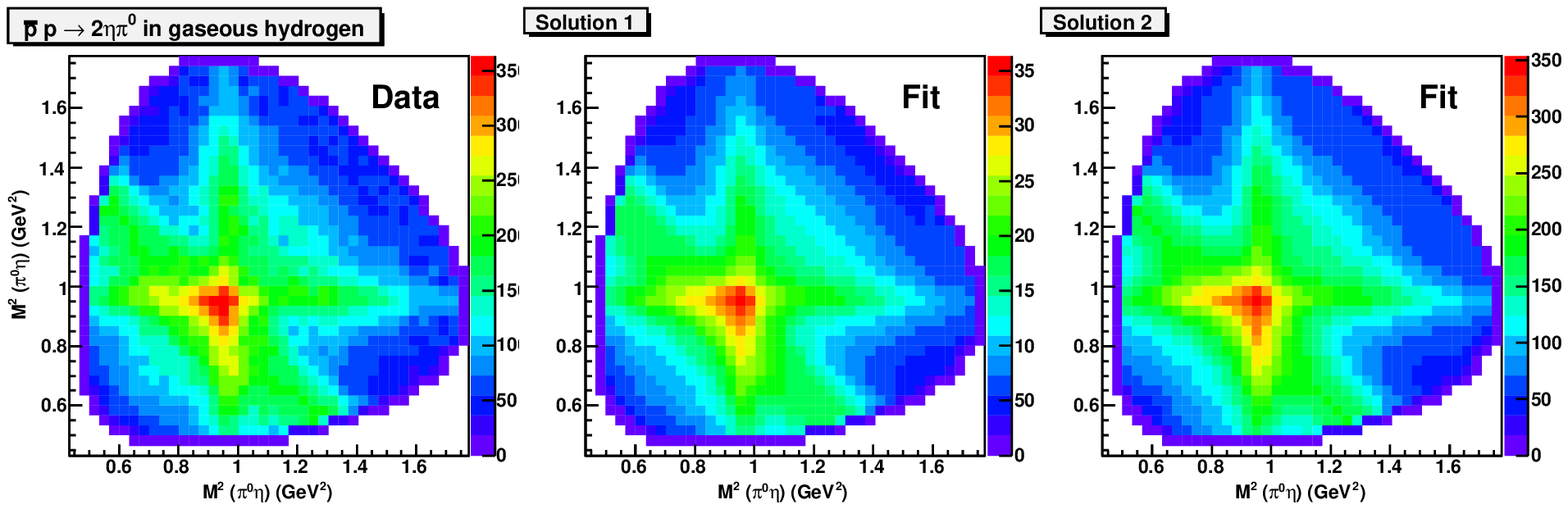,width=0.92\textwidth}} \caption{The
acceptance-corrected Dalitz plot for the $p\bar p$ annihilation into
$\eta\pi^0\eta$ in gaseous $H_2$ and the result of the two
solutions.}
\label{T6km-10g}
\vspace{-5mm}
%Fig. 9
\centerline{\hspace*{0.7cm}\epsfig{file=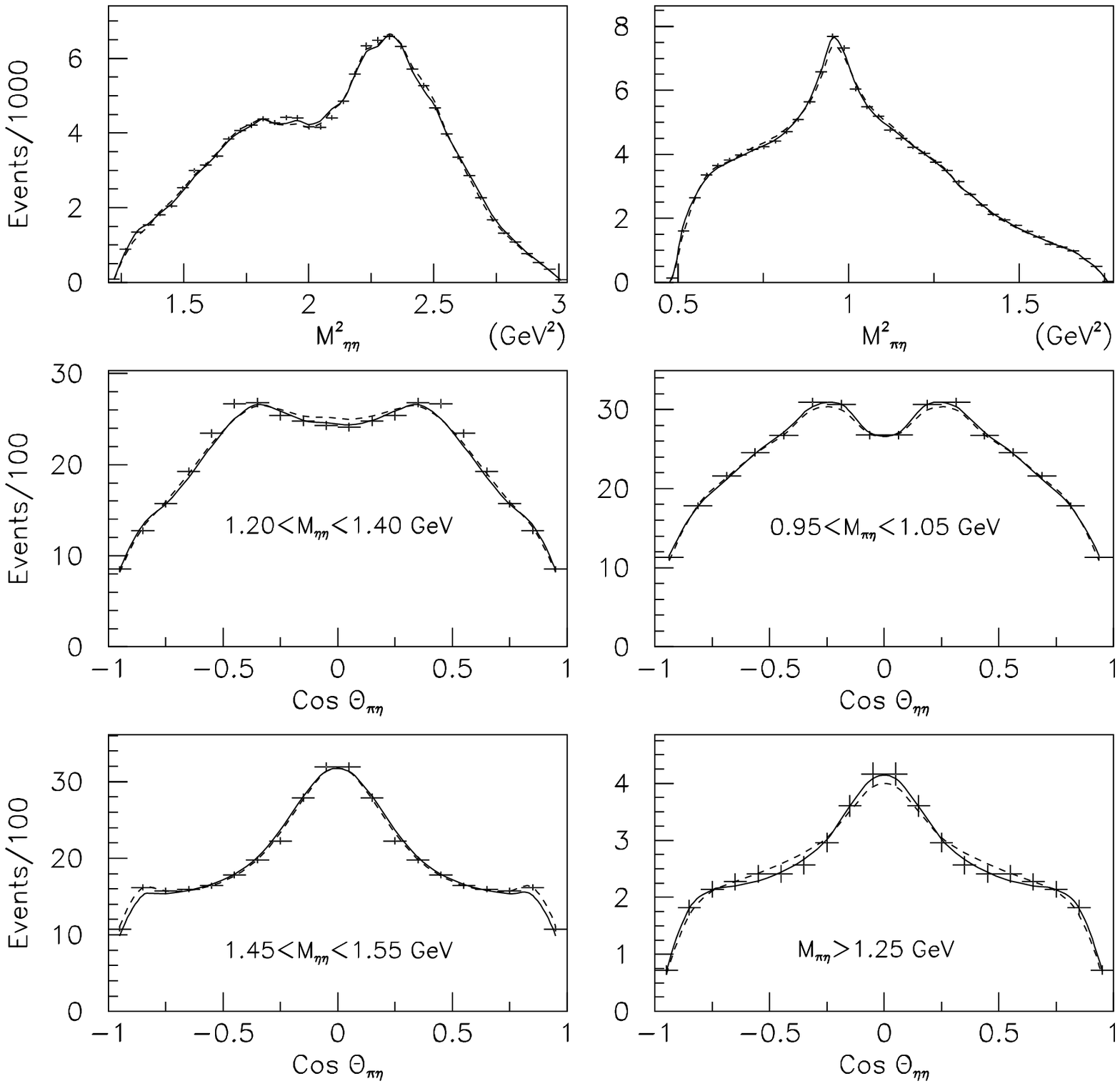,width=0.92\textwidth}}
\vspace{-5mm} \caption{Mass projections of the acceptance-corrected
Dalitz plot and angular distributions for specific mass slices. The
data are taken from the $p\bar p$ annihilation into $\eta\pi^0\eta$
in gaseous $H_2$. Solid curves correspond to the solution 1 and
dashed curves to the solution 2.}
\label{T6km-10gp}
\end{figure}

\begin{figure}
%Fig. 10
\centerline{\epsfig{file=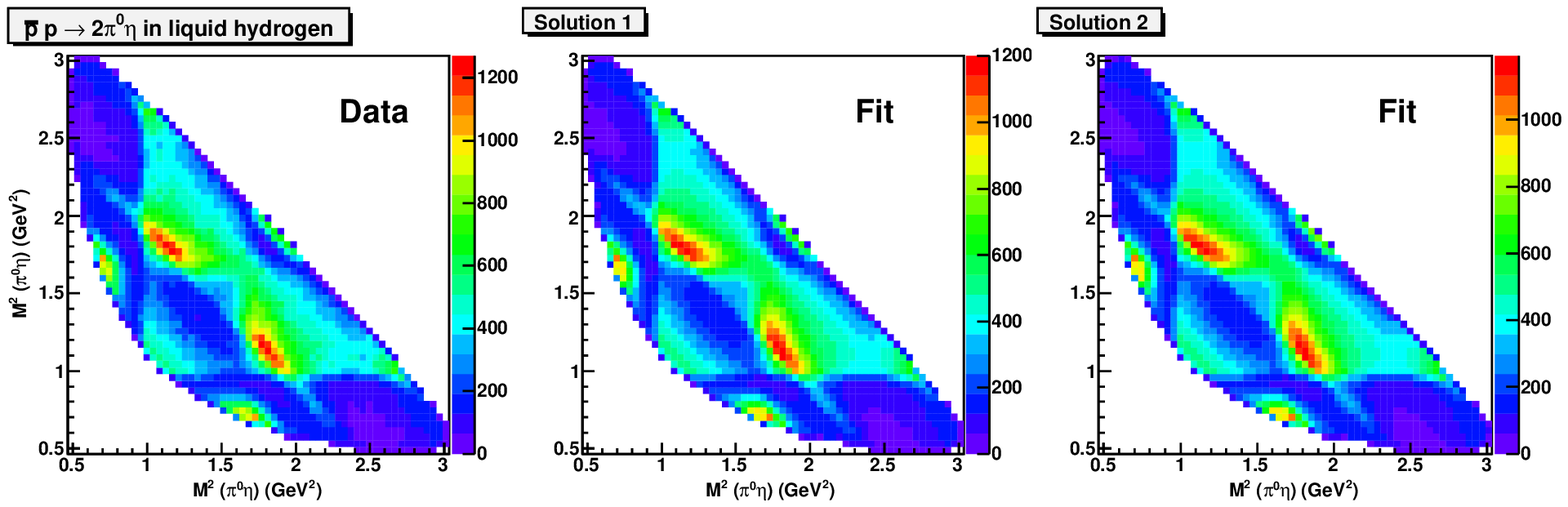,width=0.9\textwidth}} \caption{The
acceptance-corrected Dalitz plot for the $p\bar p$ annihilation into
$\pi^0\pi^0\eta$ in liquid $H_2$ and the result of the two
solutions.}
\label{T6km-11}
\vspace{-5mm}
%Fig. 11
\centerline{\hspace*{0.7cm}\epsfig{file=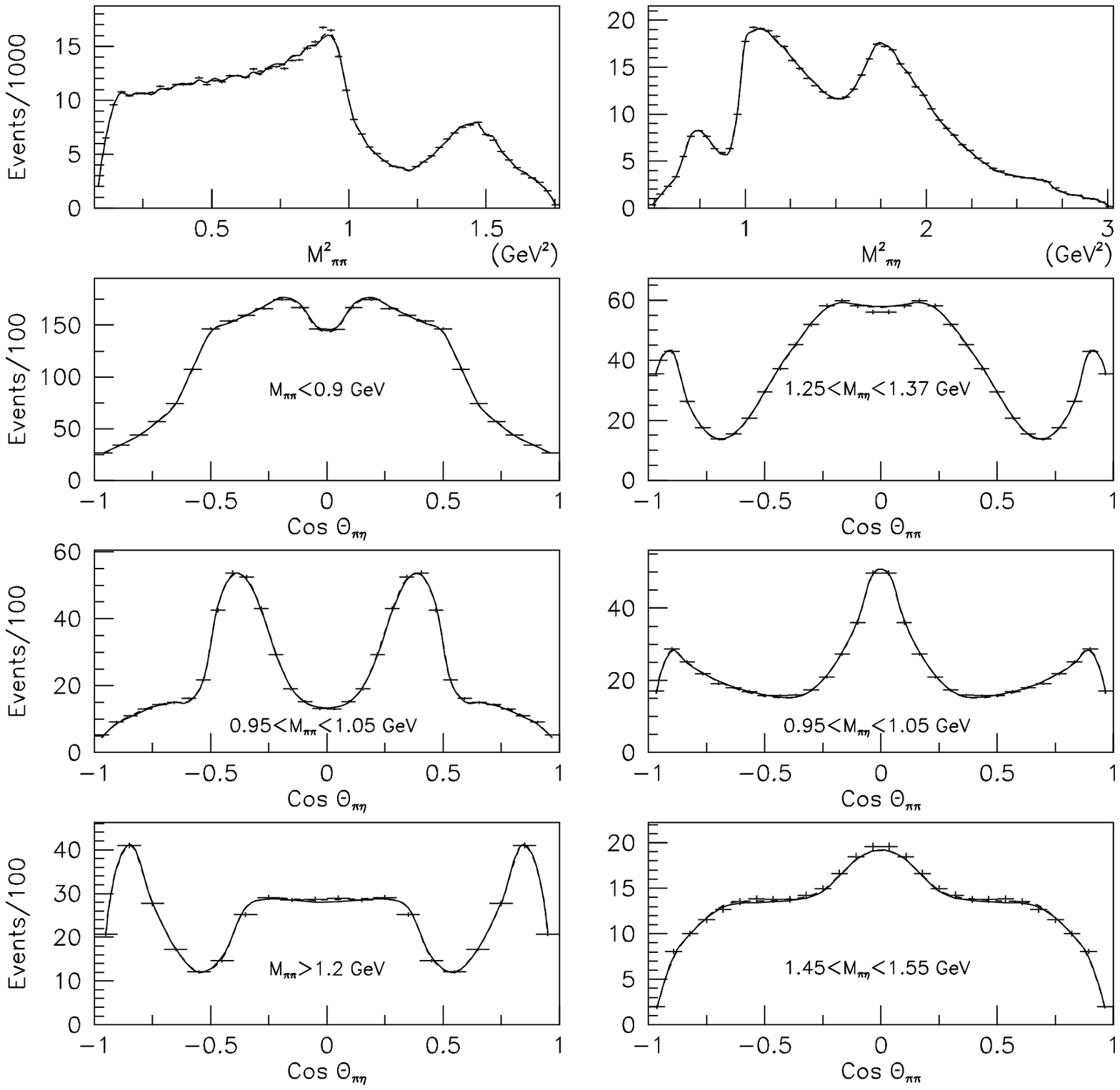,width=0.95\textwidth}}
\vspace{-5mm} \caption{Mass projections of the acceptance-corrected
Dalitz plot and angular distributions for specific mass slices. The
data are taken from the $p\bar p$ annihilation into $\pi^0\pi^0\eta$
in liquid $H_2$. Solid curves correspond to the solution 1 and
dashed curves to the solution 2.}
\label{T6km-11p}
\end{figure}

\begin{figure}
%Fig. 12
\centerline{\epsfig{file=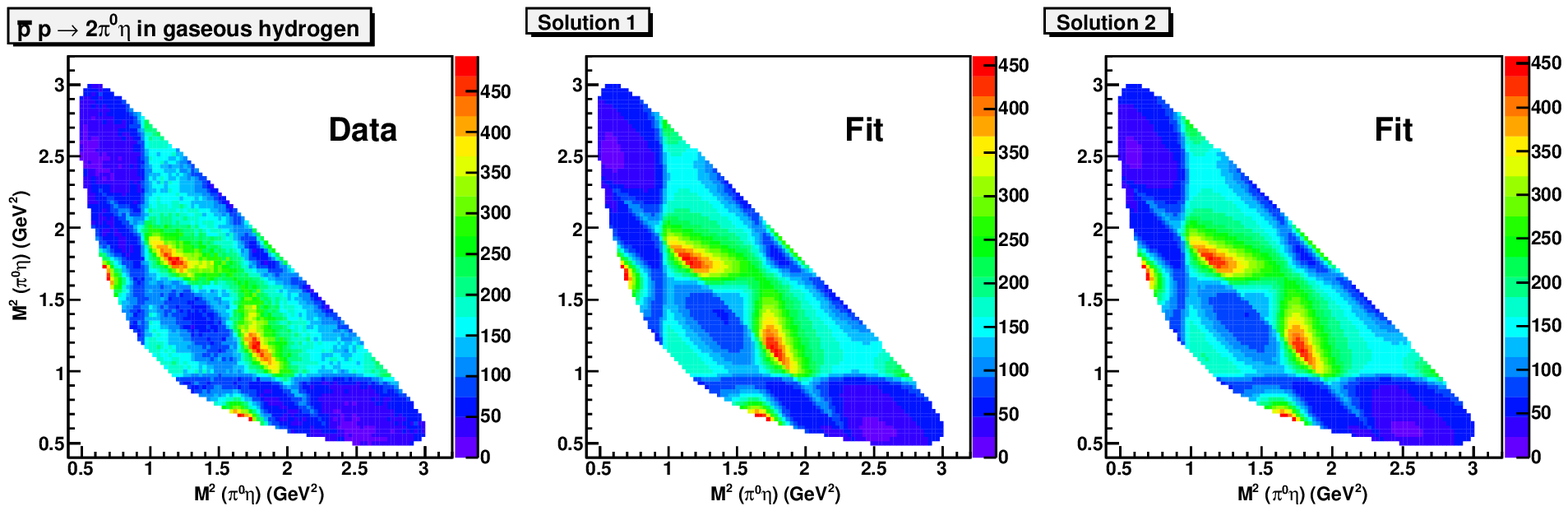,width=0.9\textwidth}} \caption{The
acceptance-corrected Dalitz plot for the $p\bar p$ annihilation into
$\pi^0\pi^0\eta$ in gaseous $H_2$ and two solutions.}
\label{T6km-11g}
%Fig. 13
\vspace{-5mm}
\centerline{\hspace*{0.7cm}\epsfig{file=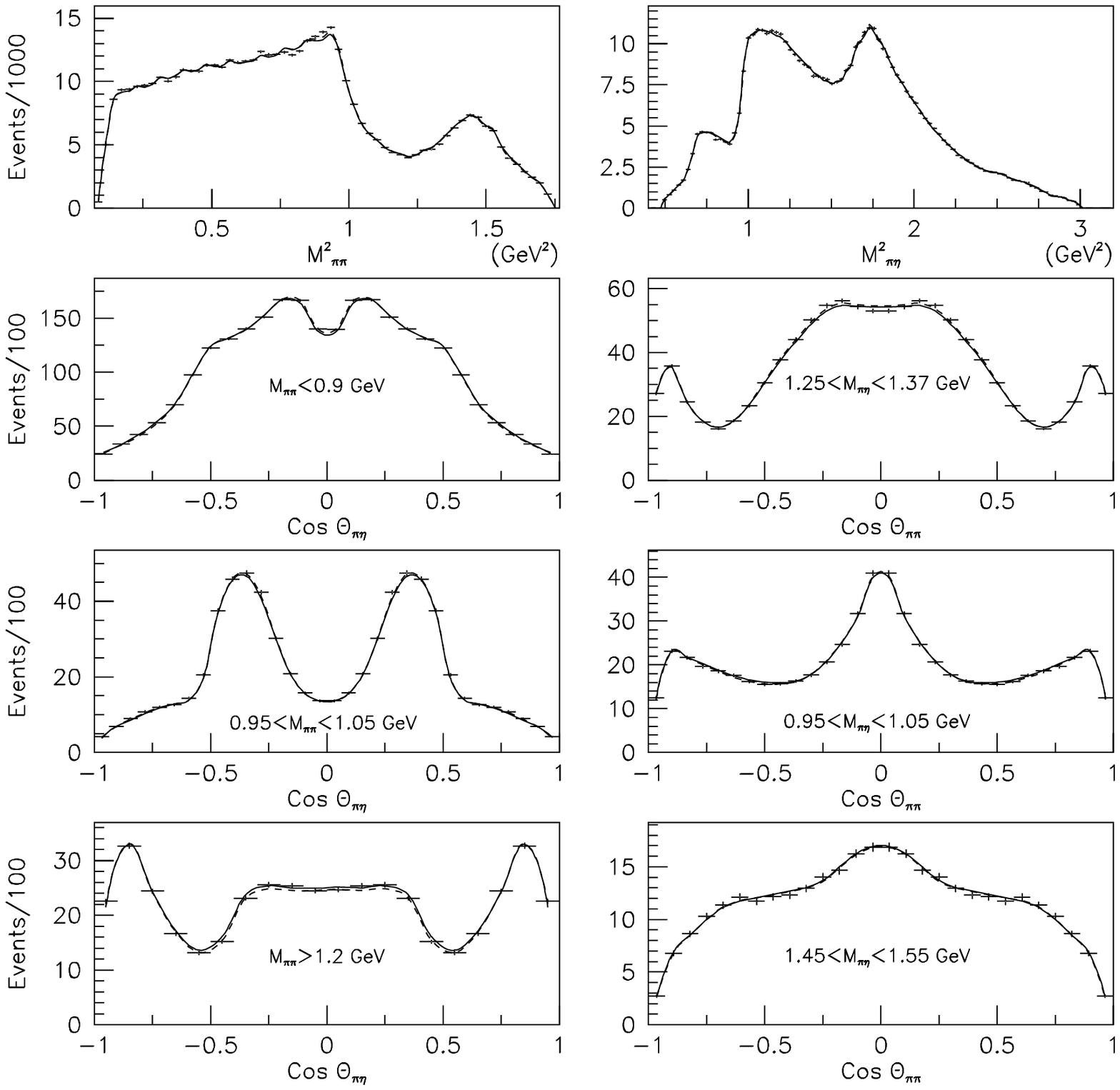,width=0.95\textwidth}}
\vspace{-5mm} \caption{Mass projections of the acceptance-corrected
Dalitz plot and angular distributions for specific mass slices. The
data are taken from the $p\bar p$ annihilation into $\pi^0\pi^0\eta$
in gaseous $H_2$. Solid curves correspond to the solution 1 and
dashed curves to the solution 2. }
\label{T6km-11gp}
\end{figure}

The results of the fit for the two solutions discussed below are
shown in Figs. \ref{T6km-9}--\ref{T6km-11gp}.

\begin{figure}[h!]
%Fig.14
\centerline{
\epsfig{file=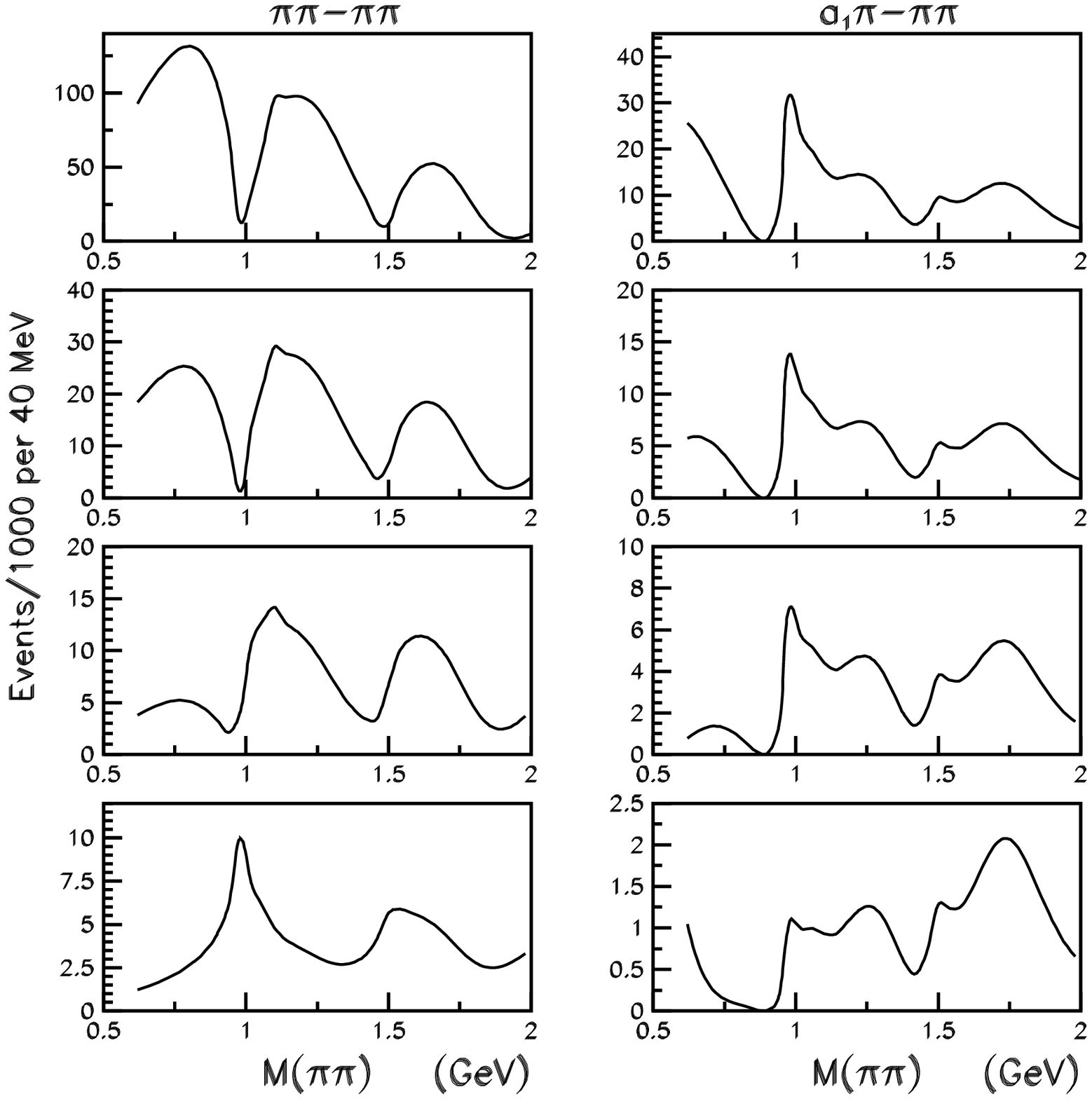,width=0.53\textwidth,clip=on}\hspace*{-0.5cm}
\epsfig{file=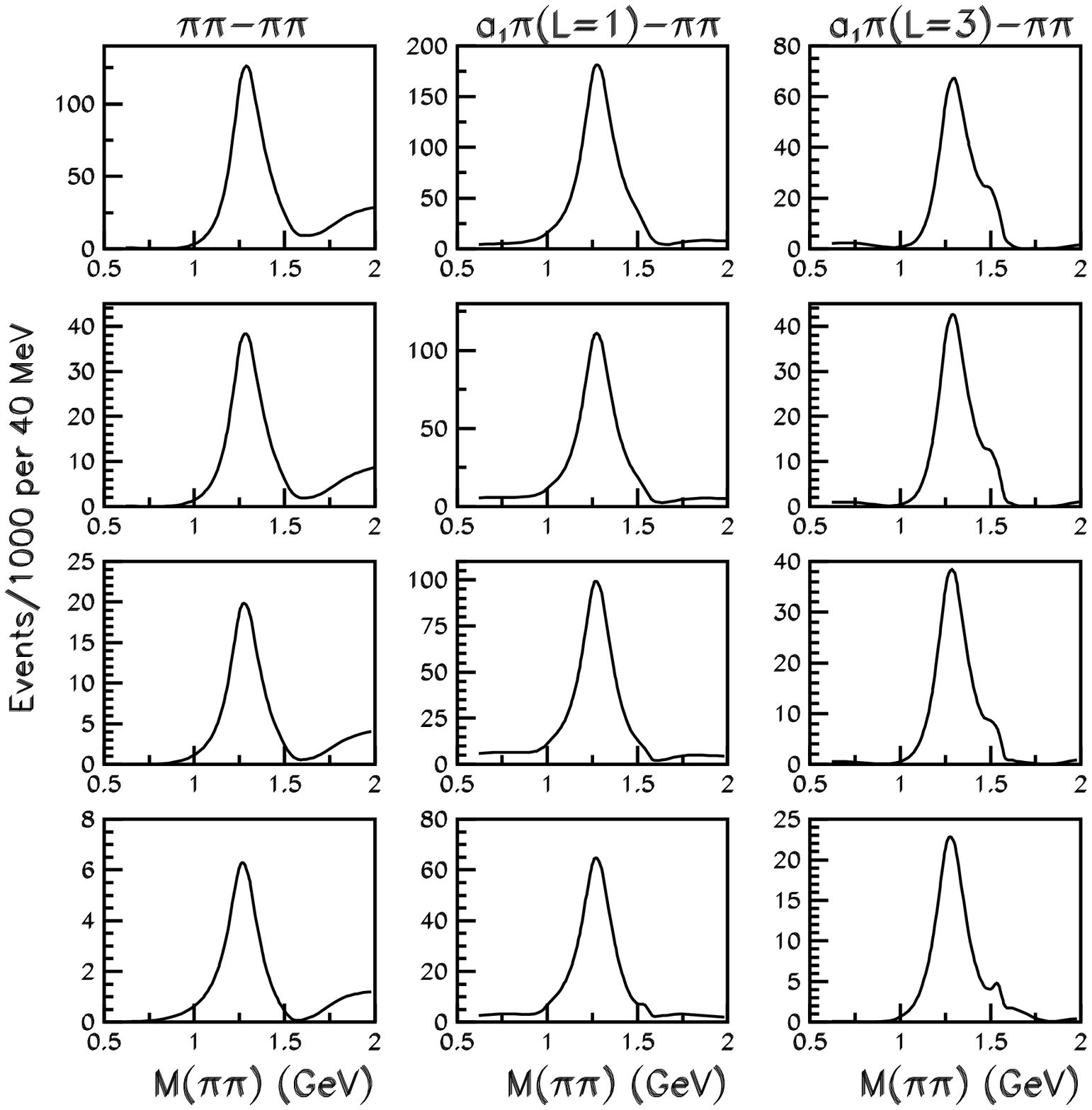,width=0.53\textwidth,clip=on} }
\caption{Solution 1. The contributions of S-wave (two left columns)
and D-wave (three right columns) to $Y_{00}$ moment integrated over
$t$ intervals. First line: $-0.1\!<t\!<\!-0.01$ GeV$^2$, second line:
$-0.2\!<\!t\!<\!-0.1$ GeV$^2$, third line:  $-0.4\!<\!t\!<\!-0.2$
GeV$^2$ and the bottom line: $-0.4\!<\!t\!<\!-1.5$ GeV$^2$.}
\label{Tsdwave_s1}
\end{figure}

\begin{figure}[h!]
%Fig.15
\centerline{
\epsfig{file=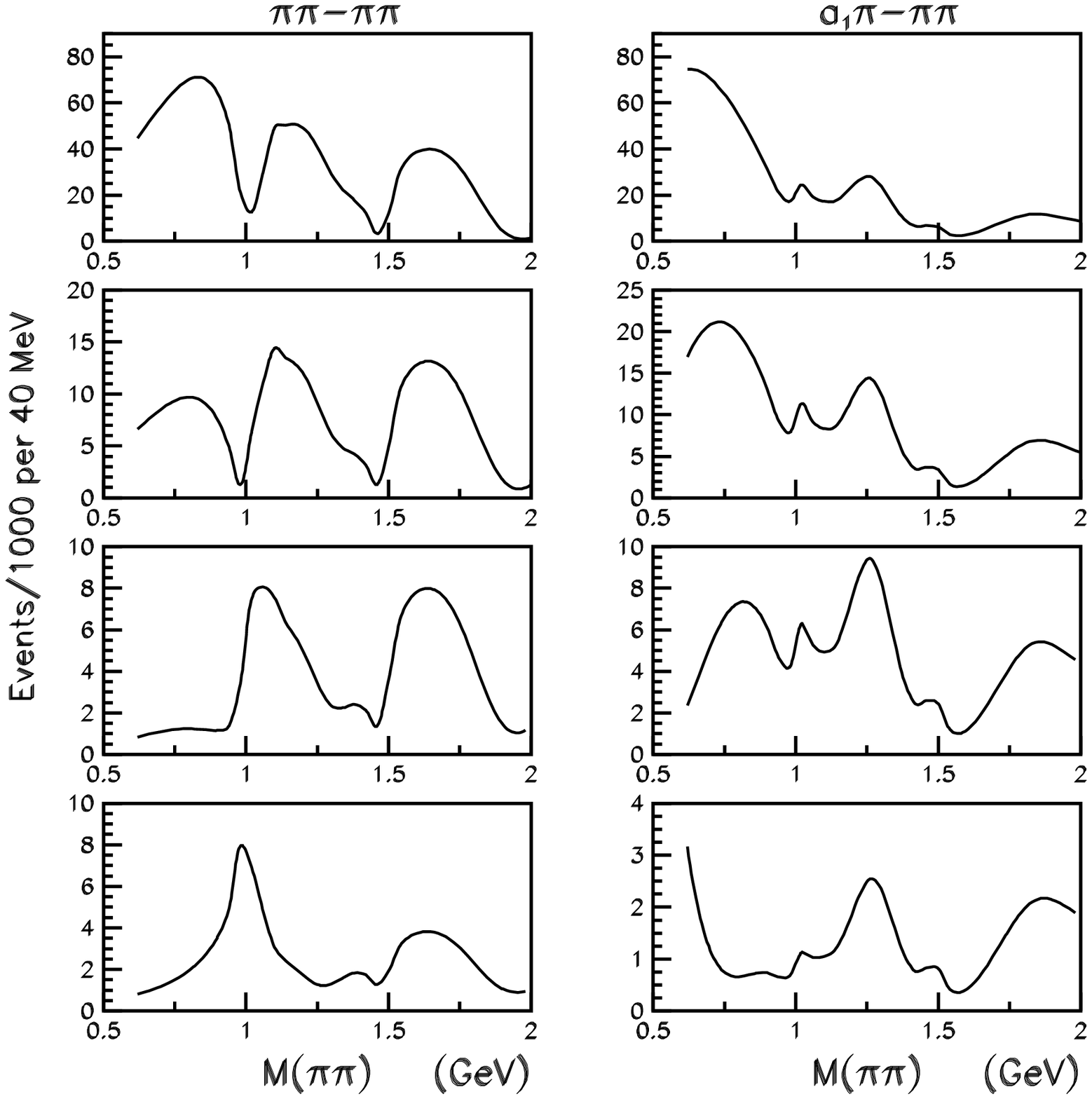,width=0.53\textwidth,clip=on}\hspace*{-0.5cm}
\epsfig{file=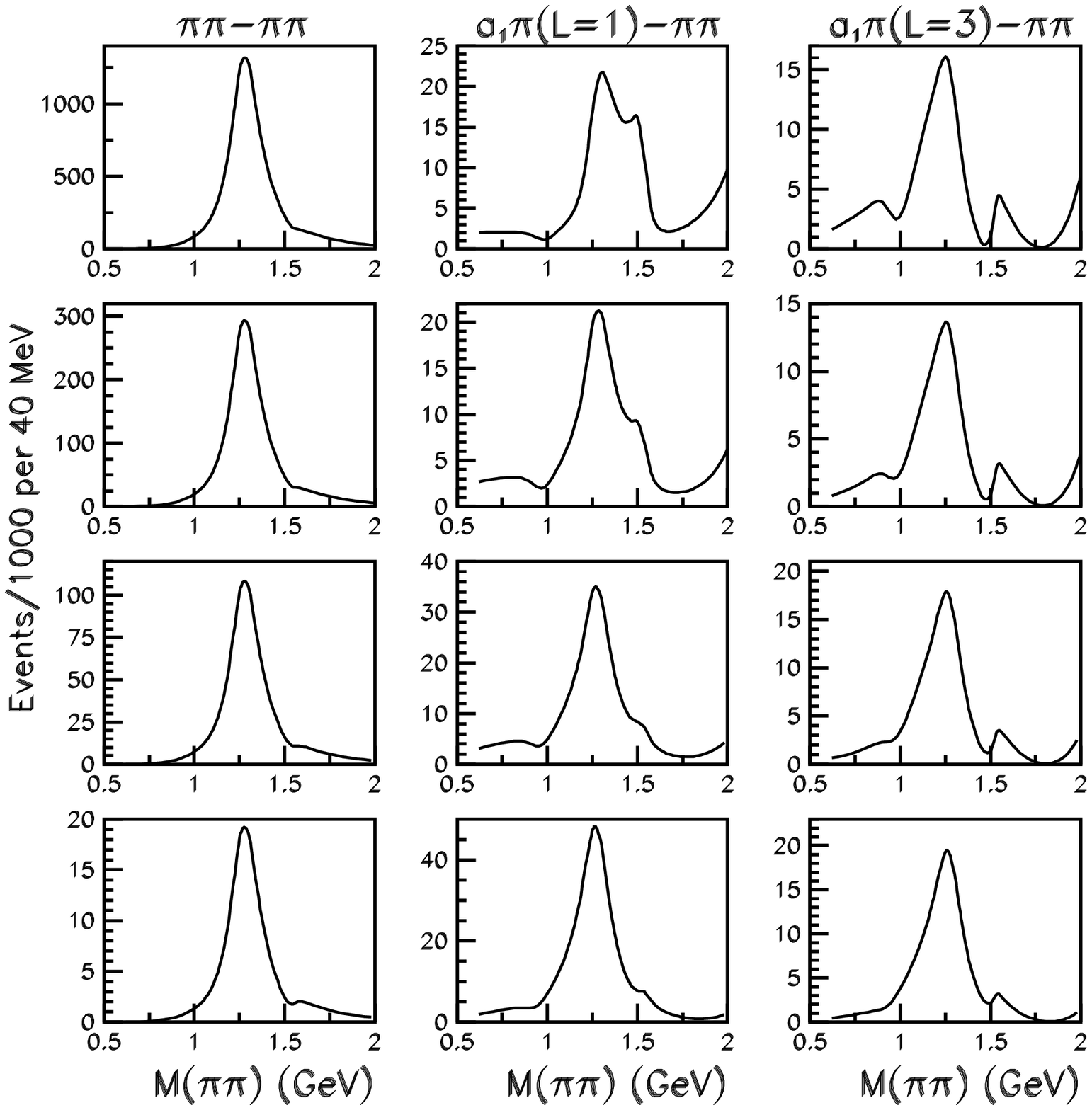,width=0.53\textwidth,clip=on} }
\caption{Solution 2. The contributions of S-wave (two left columns)
and D-wave (three right columns) to $Y_{00}$ moment integrated over
$t$ intervals. First line: $-0.1\!<\!t\!<\!-0.01$ GeV$^2$, second line:
$-0.2\!<\!t\!<\!-0.1$ GeV$^2$, third line:  $-0.4\!<\!t\!<\!-0.2$
GeV$^2$ and the bottom line: $-0.4\!<\!t\!<\!-1.5$ GeV$^2$.}
\label{Tsdwave_s2}
\end{figure}

\subsection{The $K$-matrix fit of high-energy meson
production:
 the $\pi$- and $a_1$-trajectory exchanges}

The leading contribution from the $\pi$-exchange trajectory can
contribute only to the moments with $m=0$, while the $a_1$-exchange
can contribute to the moments up to $m=2$. The characteristic
feature of the $a_1$ exchange is that moments with $m=2$ are
suppressed compared to moments with $m=1$ by the ratio
$k_{3x}/k_{3z}$ which is small for the system of two final mesons
propagating with a large momentum in the beam direction.

The amplitudes defined by the $\pi$ and $a_1$ exchanges are
orthogonal if the nucleon polarisation is not measured. This is due
to the fact that the pion trajectory states are defined by the
singlet combination of the nucleon spins while the $a_1$ trajectory
states are defined by the triplet combination. This effect is not
taken into account for the S-wave contribution in  (\ref{Tves})
which can lead to a misidentification of this wave at large momenta
transferred.

The $\pi_2$ particle is situated on the pion trajectory and
therefore should be described by the reggeized pion exchange.
However, the $\pi_2$-exchange has next-to-leading order
contributions with spherical functions at $m\ge 1$. The interference
of such amplitudes with the pion exchange can be important
(especially at small t) and is taken into account in the present
analysis.

\subsection{Results of the fit }

To reconstruct the total cross section of the reaction
$\pi^-p\to\pi^0\pi^0n$ which is not available to us we have used two
partial wave decompositions provided by the E852 collaboration
\cite{TE852}. The cross section was reconstructed by Eq. (\ref{Tves})
and decomposed over moments. The two partial wave decompositions
produced very close results for the moments and we included the
small differences between them as systematical errors.

The $\pi^- p\to \pi^0\pi^0 n$ moments can be described successfully
with only $\pi$, $a_1$ and $\pi_2$ leading trajectories taken into
account and a simple assumption about the  $t$-dependence of form
factor for all partial waves. Moreover, we have found two solutions
which differ by their contributions from these exchanges. Such an
ambiguity is likely to be connected with the lack of polarisation
data and can be resolved by data from future experiments.

The quality of the description of the Crystal Barrel data by both
solutions is shown in Figs. \ref{T6km-9}-\ref{T6km-11gp} and the
$\chi^2$  is given in Table \ref{chi2}. Here we also provide $\chi^2$
values for the $\pi\pi\to \eta\eta$ and $\pi\pi\to \eta\eta'$ S-wave
extracted by the GAMS collaboration \cite{Tgams1},\cite{Tgams2} from
the $\pi^-N$ data taken at small transfer energies. It is seen that
both solutions describe Crystal Barrel and GAMS data with the same
quality. The main reason that the K-matrix parameters for the S and
D-waves as well as P-vectors for the $\bar p p$ annihilation into
three mesons are very similar in the two solutions.  However the P-vectors
for description of the E852 data are different in solutions 1 and 2.

\begin{figure}[h!]
%Fig.14
\centerline{
\epsfig{file=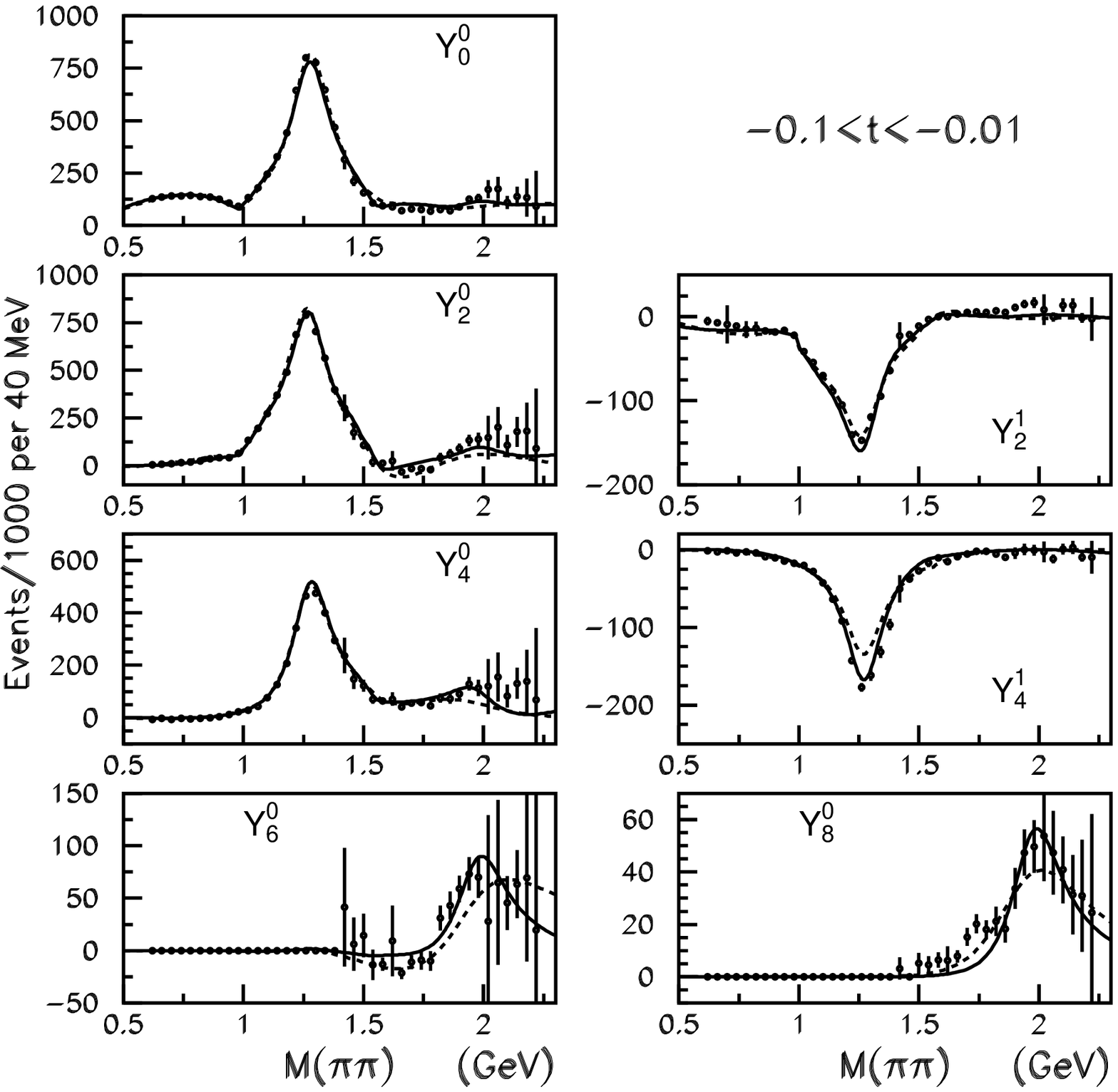,width=0.52\textwidth,clip=on}\hspace*{-0.3cm}
\epsfig{file=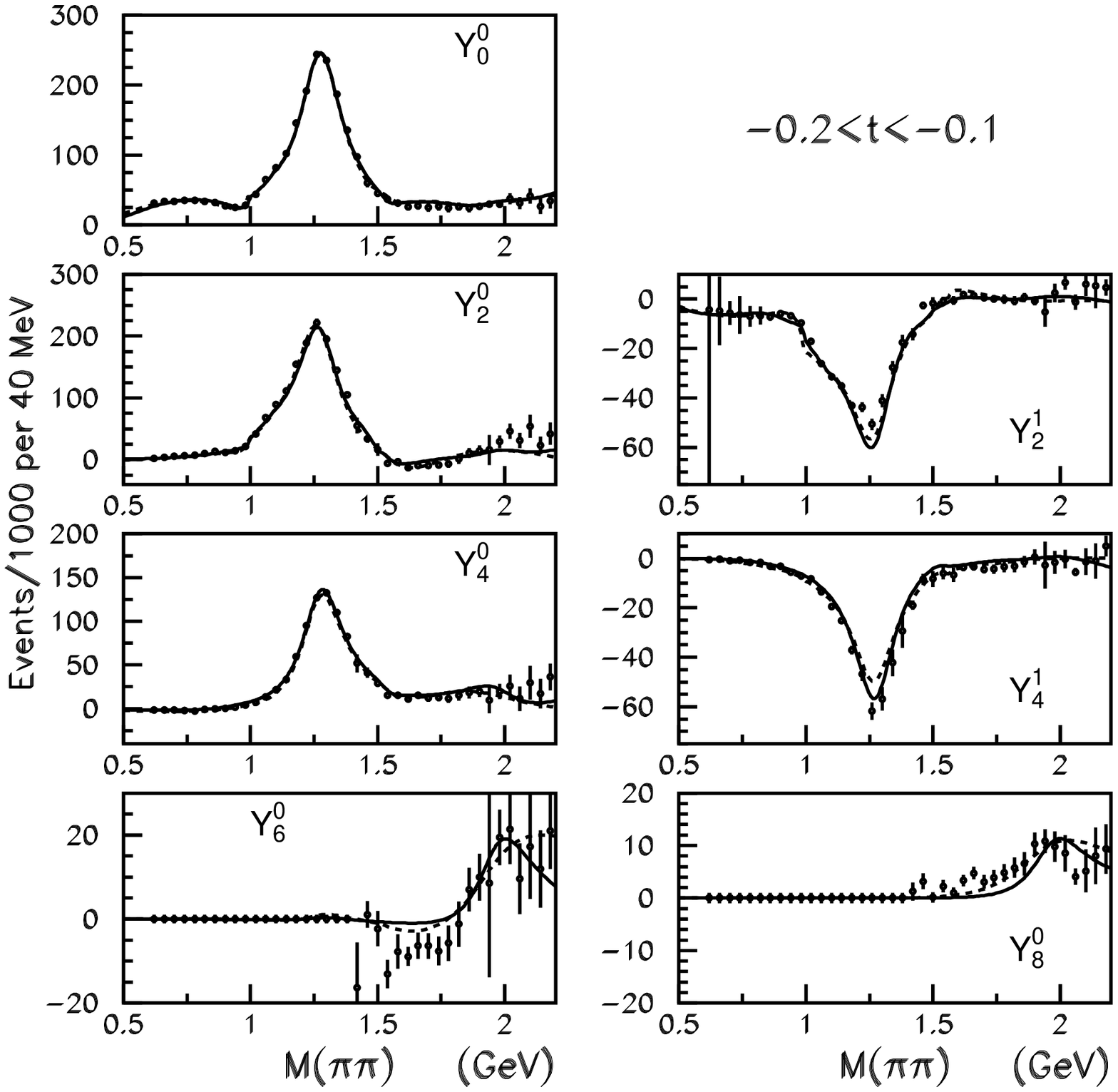,width=0.52\textwidth,clip=on} } \caption{The
description of the moments extracted at $-0.1\!<\!t\!<\!-0.01$
GeV$^2$ (the left two columns) and $-0.2\!<\!t\!<\!-0.1$ GeV$^2$
(the right two columns). Dashed curves correspond to the solution 1
and full curves to the solution 2.}
\label{Ty_sm_t}
\end{figure}

\begin{figure}[h!]
%Fig.15
\centerline{
\epsfig{file=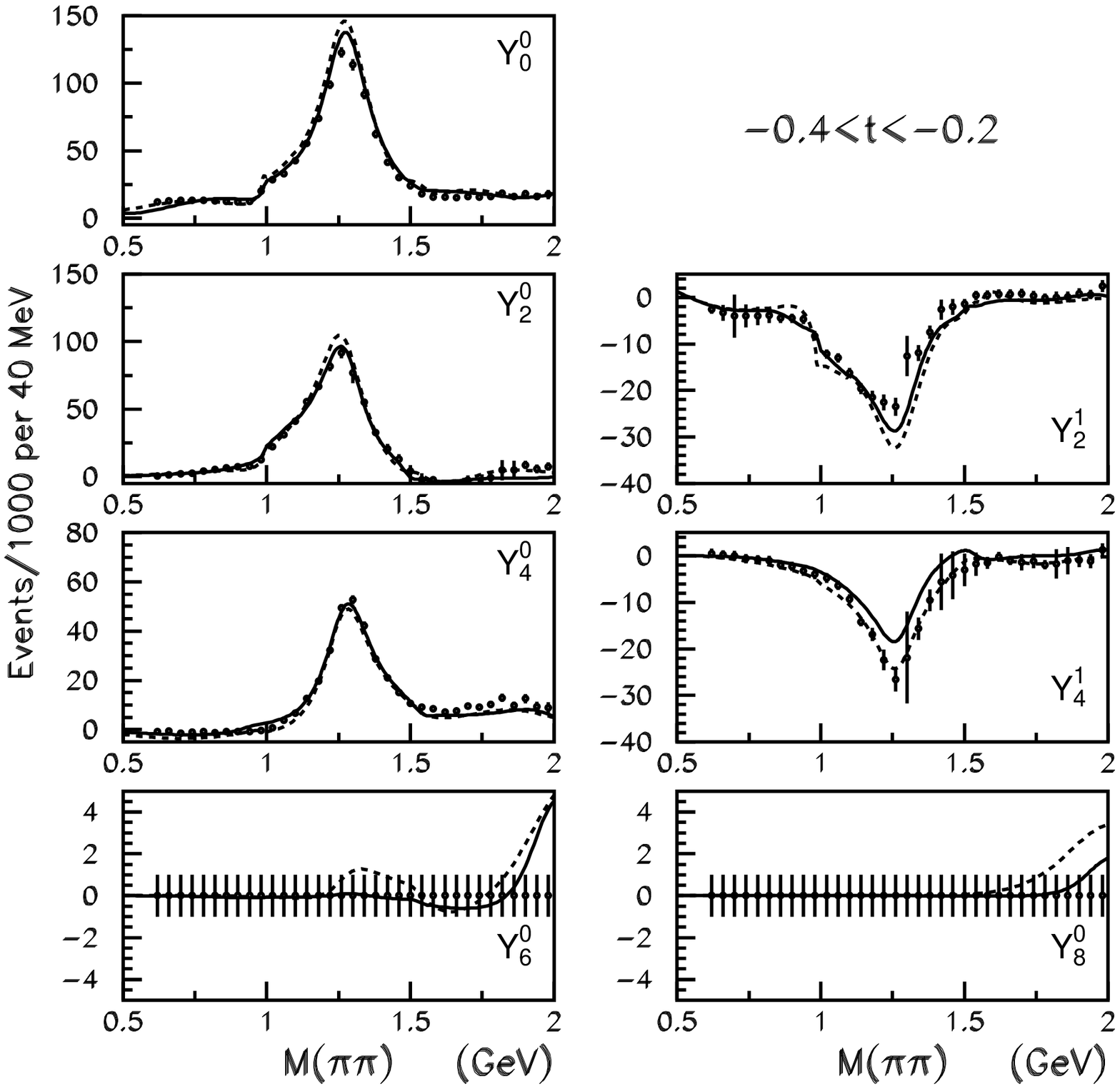,width=0.52\textwidth,clip=on}\hspace*{-0.3cm}
\epsfig{file=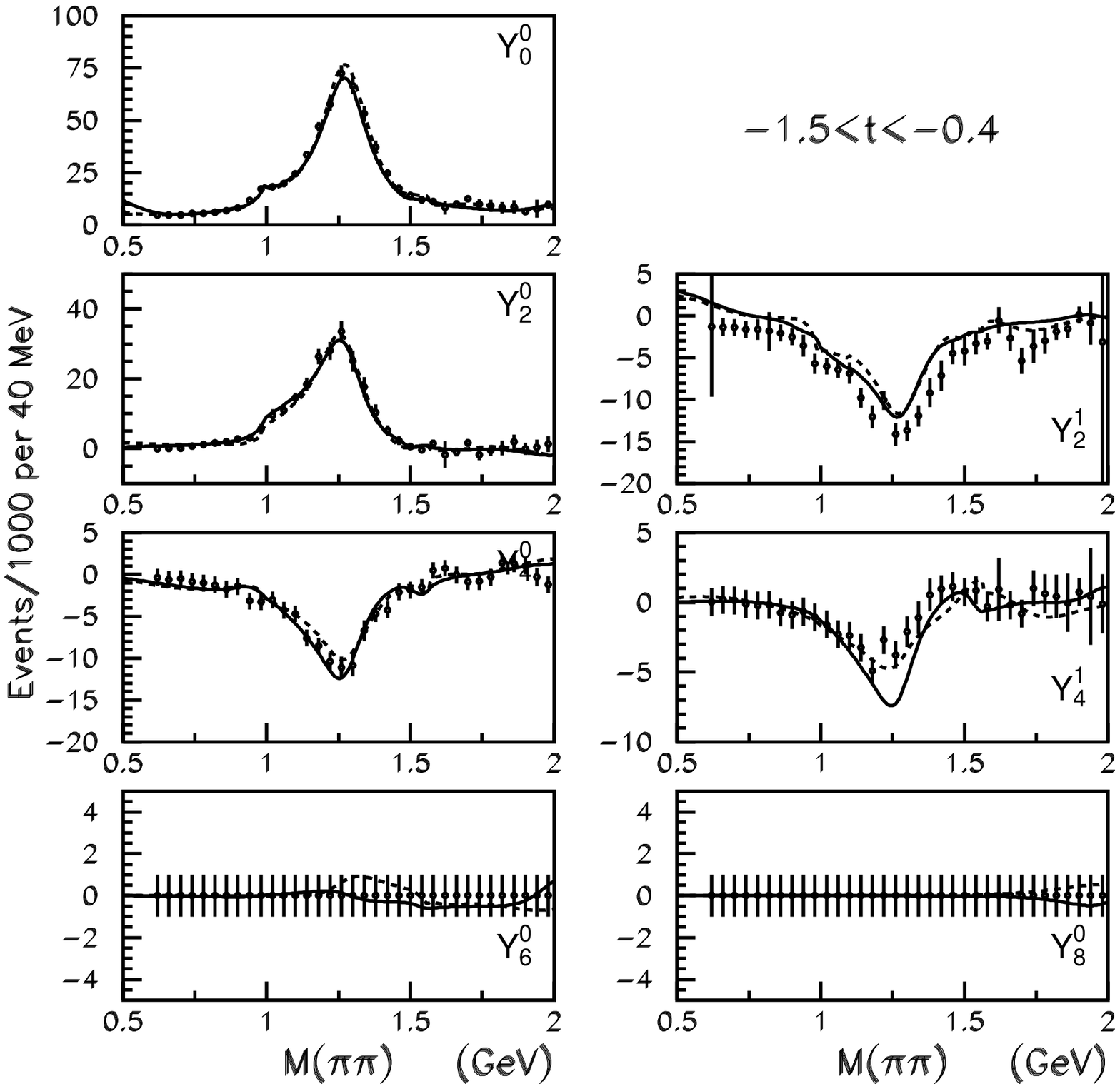,width=0.52\textwidth,clip=on} } \caption{The
description of the moments extracted at $-0.4\!<\!t\!<\!-0.2$
GeV$^2$ (two left  columns) and $-1.5\!<\!t\!<\!-0.4$ GeV$^2$ (two
right columns). Dashed curves correspond to the solution 1 and full
lines to the solution 2.}
\label{Ty_lg_t}
\end{figure}

 In a more detail:
these two solutions differ by the fraction of the $\pi$, $a_1$ and
$\pi_2$ exchanges already in the region of small energy transferred.
The first solution has a very large, practically dominant
contribution from the $a_1$ exchange to the $D$-wave (see
Fig.~\ref{Tsdwave_s1}). The contribution from the $a_1$ exchange to
the $S$-wave is small. In this solution there is no notable signal
from the $f_0(1300)$ state either at small or at large energy
transferred. If $f_0(1300)$ is excluded from this solution, only the
description of the Crystal Barrel and GAMS data is deteriorated
while the description of the E852 data has the same quality.

In the second solution the D-wave at small energies transferred is
dominantly produced from the $\pi$ exchange. The fraction $a_1$
exchange at $|t|<0.1$ is about $2.5-3\%$. At large energy transferred,
like in solution 1, the contribution from $a_1$ exchange becomes
comparable and even dominant. The S-wave has a well known structure
at small $|t|$. At intermediate energies the contribution from the
$a_1$ exchange becomes dominant and a signal from the $f_0(1300)$
state is well seen in this wave. At very large $|t|$ ($-1.5<t<-0.4$
GeV$^2$) the contribution from $a_1$ exchange is rather small. The
dominant contribution comes from the $f_0(980)$ state produced from
$\pi$ exchange. Here our analysis is in contradiction with the
result reported by the E852 collaboration which observed a strong
S-wave signal around 1300 MeV in this $t$-interval. However, the
contribution from $f_0(1300)$ at intermediate energies transferred
is important for the description of data
with this solution. If this state is excluded from the fit the
description is notably deteriorated,
see Table \ref{chi2}, solution 2(-). This subject is considered
in the following section in detail.

The Krakow group reported from the analysis of the polarized data
that at small $t$ the dominant contribution comes from the
$\pi$-exchange \cite{Kaminski:2001hv}. They point out that the
second solution is possibly a physical one. However, the final
conclusion can be made only after including these (yet unavailable
to us) data  in the present combined
analysis which uses reggeon exchanges.

The description of the moments at small and large $|t|$ for the two
solutions is shown in Figs. \ref{Ty_sm_t} and \ref{Ty_lg_t},
correspondingly. The second solution produces a systematically
better overall description except for the $Y^1_4$ moments at large
energies transferred.

The S-wave was fitted to 5 poles in the 5-channel $K$-matrix,
described in detail in  previous sections.
The parameters for the
first solution are very close to those for the second one, {\it e.g.}
the parametrization given in Table \ref{Tswave_par} describes
both solutions, and the given errors cover  a marginal change in
both descriptions.

\begin{figure}[h!]
%Fig.16
\centerline{\epsfig{file=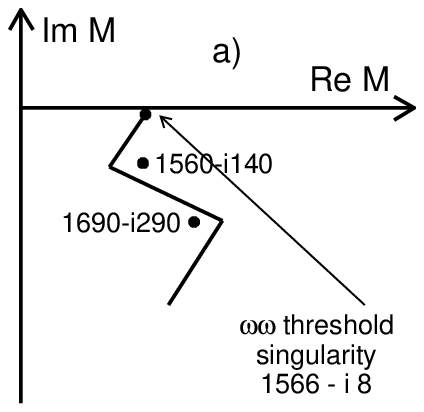,width=0.4\textwidth,clip=on}
\hspace*{1cm}
\epsfig{file=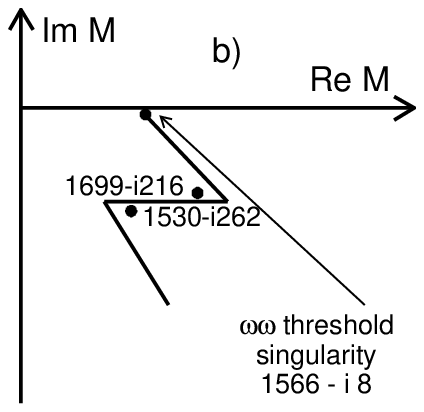,width=0.4\textwidth,clip=on} }
\vspace{-10mm} \caption{Pole structure of the 2$^{++}$-amplitude in
the region of the $\omega\omega$-threshold: the resonance
$f_2(1560)$: a) solution 1(i), b) solution 1(ii)}
\label{Tch6-poles}
\end{figure}

The D-wave was fitted to 4 poles in the 5-channel ($\pi\pi$, $K\bar
K$, $\eta\eta$, $\omega\omega$ and $4\pi$) $K$-matrix. The position
of the first two D-wave poles was found to be $1270-i97$ MeV and
$1530-i72$ MeV which corresponds to the well-known resonances
$f_2(1270)$ and $f_2(1525)$. The third state has a Flatt\'{e}-structure
near the $\omega\omega$ threshold and is defined by two poles on the
sheets defined by the $\omega\omega$ cut. Due to the fact that we do
not fit directly the $\omega\omega$ production data these positions
can not be defined unambiguously. For example, in the framework of
the solution 1 (dominant $a_1$-exchange in the D-wave) we found at
least two solutions for the pole structure in the region of 1560
MeV. In the first the pole is situated at $1565-i140$  MeV on the
sheet above the $\omega\omega$ threshold and $1690-i290$ MeV on the
sheet below the $\omega\omega$ threshold. In the other solution
the position of the pole is $1530-i262$ and $1699-i216$,
correspondingly.
The closest physical region is for both poles the beginning of the
$\omega\omega$ threshold $M\sim$1570 MeV, where they form a
relatively narrow (220--250 MeV) structure which is called the
$f_2(1560)$ state, see Fig. \ref{Tch6-poles}. A similar situation
was observed in the solution 2. The K-matrix D-wave parameters for
the solution 2 are given in Table \ref{Tdwave_par}.

The fourth  $D$-wave $K$-matrix pole, $f_2^{bare}(1980)$ cannot be
rigidly fixed  by the present data. The position of the
corresponding pole is also not stable: one can easily increase the
mass of the pole with the simultaneous increase of the width,
spoiling only slightly the description of data. Because of that we
consider this pole as some effective contribution of resonances
located above 1900 MeV.

\begin{figure}[h!]
%Fig.17
\centerline{ \epsfig{file=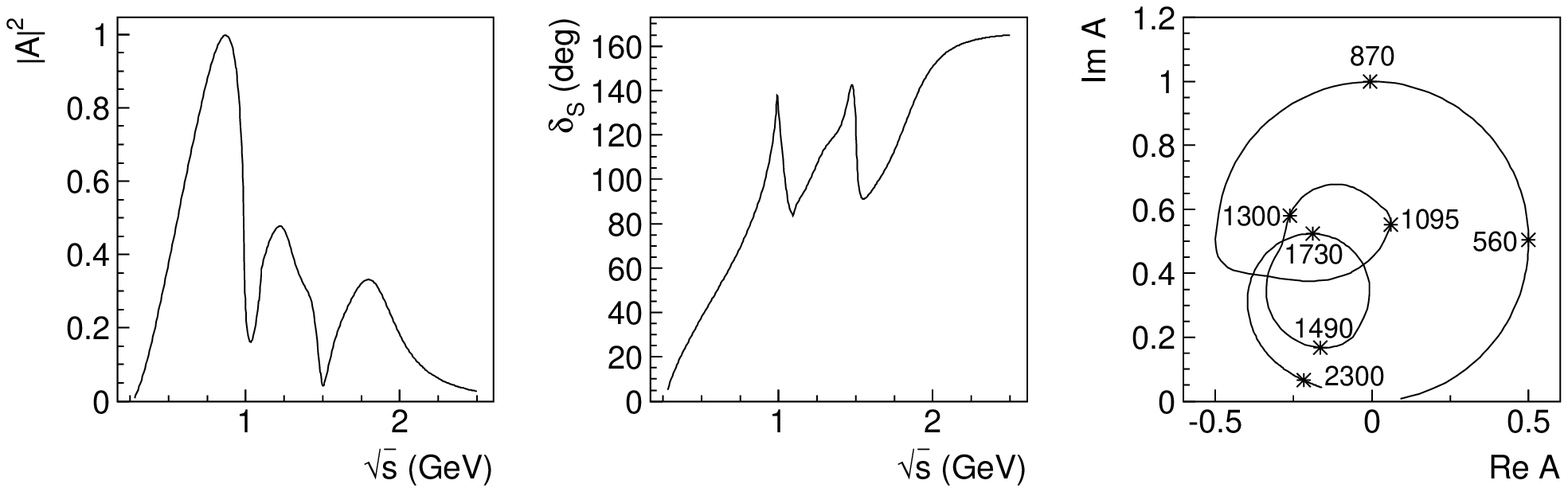,width=0.99\textwidth,clip=on} }
\caption{From left to right: a) The $\pi\pi\to\pi\pi$ S-wave
amplitude squared, b) the amplitude phase and c) the Argand diagram
for the S-wave amplitude $\pi\pi\to\pi\pi$.}
\label{Tsw_amp}
\end{figure}

The $\pi\pi\to \pi\pi$ S-wave elastic amplitude for the second
solution is shown in Fig.\ref{Tsw_amp}. The structure of the
amplitude is well known, it is defined by the destructive
interference of the broad component with $f_0(980)$ and $f_0(1500)$.
Neither $f_0(1300)$ nor $f_0(1750)$ provide a strong change of the
amplitudes. However, this is hardly a surprise: both these states
are relatively broad and dominantly inelastic.

\begin{table}
\caption{The $\chi^2$ per data point for the description of the Crystal
Barrel and GAMS data. Two solutions are given  as well as that
with $f_0(1300)$ excluded from the fit, solution 2(-).  For the E852 data
the $\chi^2$ is calculated for all moments in the given
$t$-interval} {
\begin{tabular}{|l|ccc|}
\hline
Data & Solution 1 & Solution 2 & Solution 2(-) (no $f_0(1300)$) \\
\hline
$\bar p p\to \pi^0\pi^0\pi^0$ (Liq) &  1.360 &   1.356 &   1.443  \\
$\bar p p\to \pi^0\pi^0\pi^0$ (Gas) &  1.238 &   1.242 &   1.496  \\
$\bar p p\to \eta \pi^0\eta$ (Liq)  &  1.350 &   1.442 &   1.446  \\
$\bar p p\to \eta \pi^0\eta$  (Gas) &  1.503 &   1.371 &   1.315  \\
$\bar p p\to \pi^0\eta \pi^0$ (Liq) &  1.210 &   1.236 &   1.412  \\
$\bar p p\to \pi^0\eta \pi^0 $(Gas) &  1.099 &   1.119 &   1.227  \\
$\pi\pi\to\eta\eta$ (S-wave)        &  1.08  &   1.19  &   1.38  \\
$\pi\pi\to \eta\eta'$ (S-wave)      &  0.26  &   0.41  &   0.45  \\
 \hline
\end{tabular}
\label{chi2}}
\end{table}

%----------------------------------------------------------------

\begin{table}
\caption{Masses and couplings (in GeV units)  for
the S-wave $K$-matrix poles
($f_0^{bare}$ states) as well as the
amplitude pole positions (given in MeV).
The II sheet is defined
under the  $\pi\pi$ and
$4\pi$ cuts, the IV sheet is under $\pi\pi$, $4\pi$, $K\bar K$ and
$\eta \eta$ cuts, and the V sheet is determined by
 $\pi\pi$, $4\pi$, $K\bar K$,
$\eta \eta$ and $\eta \eta'$ cuts.} {\small
{\begin{tabular}{|l|ccccc|} \hline ~ & $\alpha=1$ &$\alpha=2$ &
$\alpha=3$ & $\alpha=4$ & $\alpha=5$ \\ \hline ~         &~ & ~ & ~
& ~ & ~ \\
M              & $0.720^{+.50}_{-.080}$ &$1.220^{+.040}_{-.030}$ &
$1.210\pm 030$ & $1.550^{+.030}_{-.020}$ & $1.850\pm .040$ \\ ~ &~ &
~ & ~ & ~ & ~
\\
$g_0^{(\alpha)}$ &$0.760^{+.080}_{-.060}$ &$0.820\pm 0.090$
&$0.470\pm.050$
&$0.360\pm .050$ &$0.440\pm .050$\\
~         &~ & ~ & ~ & ~ & ~ \\
$g_{5}^{(\alpha)}$& 0 & 0 & $0.850\pm .100$ &
$0.570\pm.070$ & $-0.900\pm .070$  \\
~         &~ & ~ & ~ & ~ & ~ \\
$\varphi_\alpha $  & -($60\pm 12$) & $28\pm 12$ & $30\pm 14$
& $8\pm 15$ &-($52\pm 14$)\\
~         &~ & ~ & ~ & ~ & ~ \\
\hline
~ & $a=\pi\pi$ &$a=K\bar K$ & $a=\eta\eta$ & $a=\eta\eta'$ & $a=4\pi$ \\
\hline
~         &~ & ~ & ~ & ~ & ~ \\
$f_{1a} $ &$0.180\pm .120$ &$ 0.150\pm.100$ & $0.240\pm.100$
&$0.300\pm .100$ & $0.0\pm .060$ \\
~ &  ~ & $f_{ba}=0$ & $b=2,3,4,5$&~ & ~\\
~         &~ & ~ & ~ & ~ & ~ \\
\hline
~ & ~ &\multicolumn{3}{c}{Pole position}& ~  \\
II      &$1030^{+30}_{-10}$& ~&~&~&~ \\
  sheet           &$-i(35^{+10}_{-16})$& ~&~&~&~\\
\hline
III       &$850^{+80}_{-50}$& ~&~&~&~ \\
  sheet          &$-i(100\pm 25)$& ~&~&~&~\\
\hline IV    &~ &$1290\pm 50$ & $1486\pm 10$
              &$1510\pm 130$ & ~ \\
  sheet        & ~&$-i(170^{+20}_{-40})$ & $-i(57\pm 5)$
              &$-i(800^{+100}_{-150})$ & ~ \\
\hline
V    & ~ & ~ & ~ & ~     & $1800\pm 60$ \\
  sheet       & ~ & ~ & ~ & ~     & $-i(200\pm 30)$   \\
\hline
\end{tabular} \label{Tswave_par}}}
\end{table}

%----------------------------------------------------------------

\begin{table}
\caption{Masses and couplings (in GeV units)  for $D$-wave $K$-matrix
poles ($f_2^{bare}$ states) for the solution 2. The III sheet is defined by  $\pi\pi$
and $4\pi$ and $K\bar K$ cuts, the IV sheet by $\pi\pi$, $4\pi$,
$K\bar K$ and $\omega\omega$ cuts. The values marked by $*$ were
fixed in the fit.
} {\begin{tabular}{|l|cccc|} \hline
~ & $\alpha=1$ &$\alpha=2$ & $\alpha=3$ & $\alpha=4$  \\
\hline
M &$1.286\pm 0.025$ &$1.540\pm0.015$ & $1.560\pm 0.020$ & $2.200^{+0.300}_{-0.200}$ \\
$g_{\pi\pi}^{(\alpha)}      $ & $0.920\pm 0.020$ &$-0.05\pm 0.080$ &$0.280\pm 0.100$ &
$-0.30\pm 0.15$ \\
$g_{\eta\eta}^{(\alpha)}    $ & $0.420\pm 0.060$ &$ 0.27\pm 0.15$
&$0.400\pm 0.200$ & $1.2\pm 0.6^*$ \\ $g_{4\pi}^{(\alpha)}        $
&
$-0.150\pm 0.200$ &$0.370\pm 0.150$ &$1.170\pm 0.450$ & $1.0\pm 0.4$ \\
$g_{\omega\omega}^{(\alpha)}$ & $0^*$ &           $0^*$            &$0.540\pm 0.150$ &
$-0.05\pm 0.2$  \\
\hline
~ & $a=\pi\pi$ &$a=\eta\eta$ & $a=\omega\omega$ & $a=4\pi$ \\
\hline
$f_{1a} $ &$0.03\pm 0.15$ &$-0.11\pm0.10$ & $0^*$ &$0^*$  \\
$f_{2a} $ &$-0.11\pm 0.10$ &$-1.8\pm 0.60$ & $0^*$ &$0^*$  \\
~ &  ~ & $f_{ba}=0$ & $b=3,4,5$&~ \\
\hline
~ & ~ &\multicolumn{2}{c}{Pole position}& ~  \\
III sheet      &$1.270\pm0.008$& $1.530\pm0.012$&~&~ \\
  ~           &$-i\,0.097\pm 0.008$& -i$\,0.064\pm 0.010$&~&~\\
\hline
III sheet      &~ &  ~ & $1.690\pm 0.015$ &~ \\
  ~           &~ & ~ & $-i\,0.290\pm 0.020$&~\\
IV sheet     &~ &  ~ & $1.560\pm 0.015$ &~ \\
  ~           &~ & ~ & $-i\,0.140\pm 0.020$&~\\
\hline
\end{tabular}
\label{Tdwave_par}}
\end{table}

\begin{figure}[h!]
%Fig.20
\centerline{ \epsfig{file=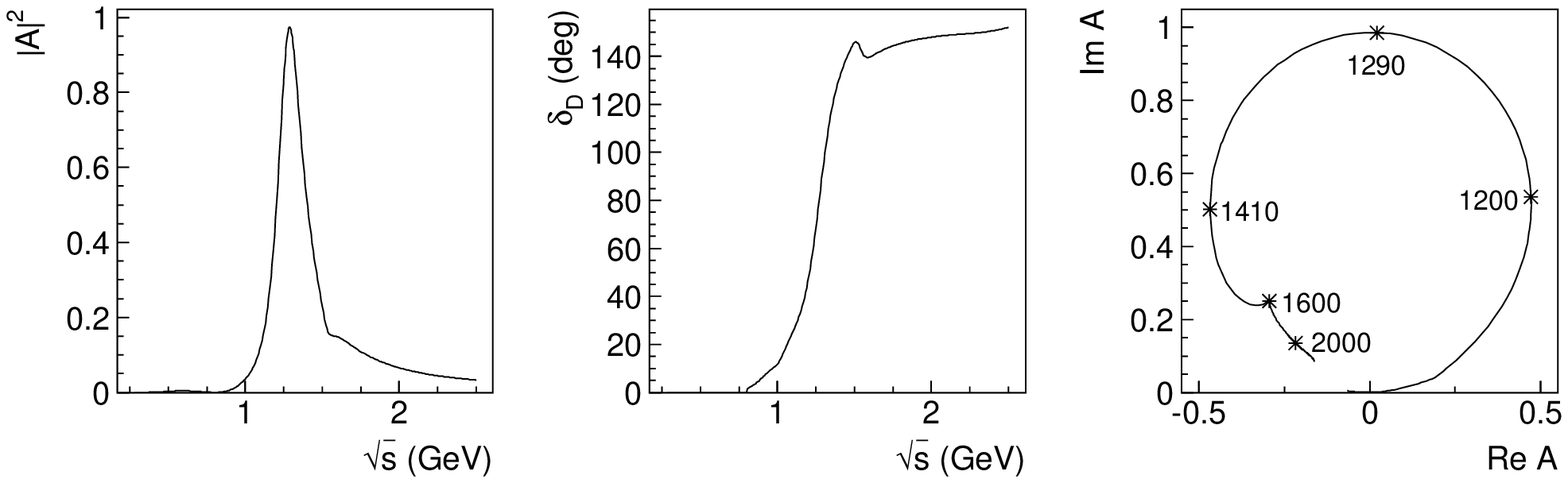,width=0.99\textwidth,clip=on}}
\caption{From left to right: The $\pi\pi\to\pi\pi$ D-wave amplitude
squared, the amplitude phase and the Argand diagram for the
amplitude.} \label{Tdw_amp} \end{figure}

The $\pi\pi\to \pi\pi$ D-wave elastic amplitude is shown in Fig.
\ref{Tdw_amp}. The amplitude squared is dominated by the $f_2(1270)$
state.
Neither of $f_2(1560)$ and $f_2(1510)$ (which are included into the
$K\bar K$ channel of the $K$-matrix) show a meaningful structure in
the amplitude squared. The $K$-matrix parameters found in the
solution are given in Table~\ref{Tdwave_par}.

\subsection{The $f_0(1300)$ state}

\begin{figure}
%Fig. 12
\centerline{\hspace*{0.7cm}\epsfig{file=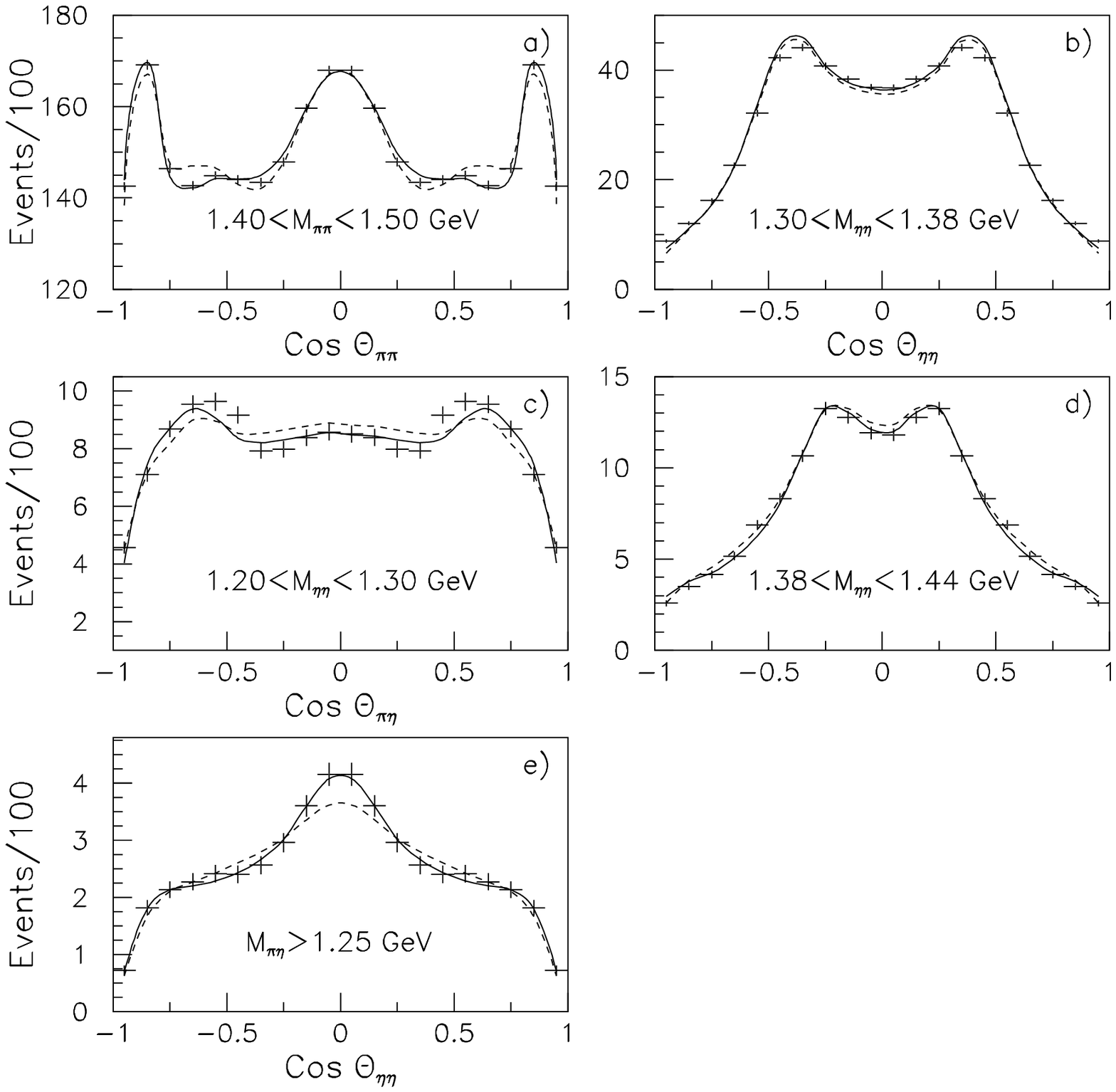,width=0.95\textwidth}}
\vspace{-5mm} \caption{Angular distributions for specific mass slices.
a) $\bar p p\to 3\pi^0$ (gas) b) $\bar p p\to \pi^0\eta\eta$ (liquid),
c,d,e) $\bar p p\to \pi^0\eta\eta$ (gas).
Solid curves correspond to the solution 2 and
dashed curves to the solution 2(-) with excluded $f_0(1300)$. }
\label{cb_no1300}
\end{figure}

In the solution 2 the fit of the E852 data shows a large
contribution from the $f_0(1300)$ state to $Y^0_0$ moment due to
$a_1$ exchange at $-0.2\!<\!t\!<\!-0.1$ and  $-0.4\!<\!t\!<\!-0.2$
GeV$^2$. At very small ($-0.1\!<\!t\!<\!-0.01$ GeV$^2$) and large
($-1.5\!<\!t\!<\!-0.4$ GeV$^2$) energy transferred the contribution
of this state to the $Y^0_0$ moment is less pronounced. If the
K-matrix pole which corresponds to the $f_0(1300)$ state is excluded
from the fit (all couplings are put to zero) the total $\chi^2$
changes rather appreciably. The corresponding values for the
description of the Crystal Barrel and GAMS data are given in the
last column of Table \ref{chi2} (solution 2(-)). The mass slices
made in the region
of the state show systematical discrepancies in this case (see
Fig.\ref{cb_no1300}).

\begin{figure}[h!]
%Fig.15
\centerline{
\epsfig{file=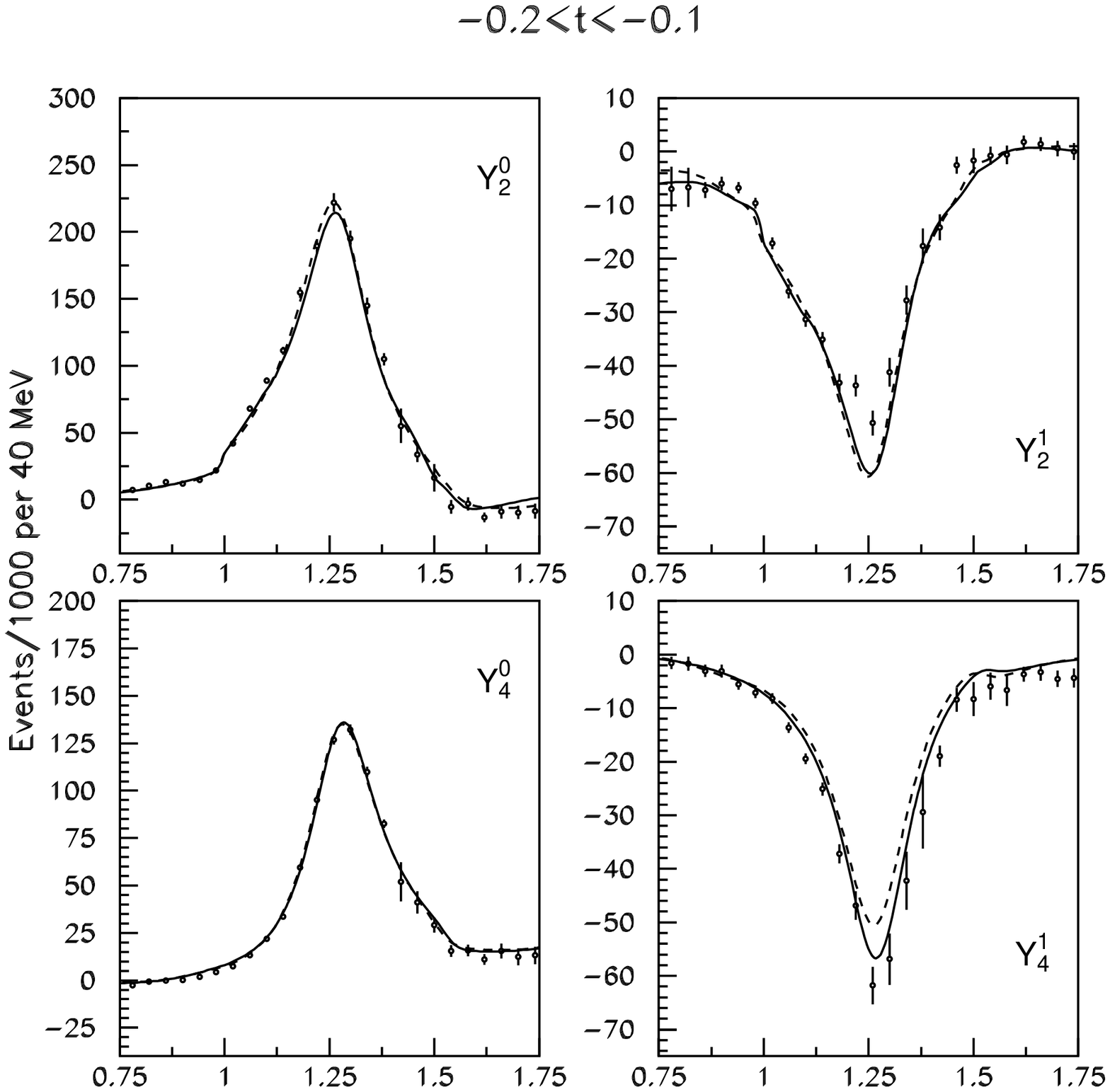,width=0.52\textwidth,clip=on}\hspace*{-0.3cm}
\epsfig{file=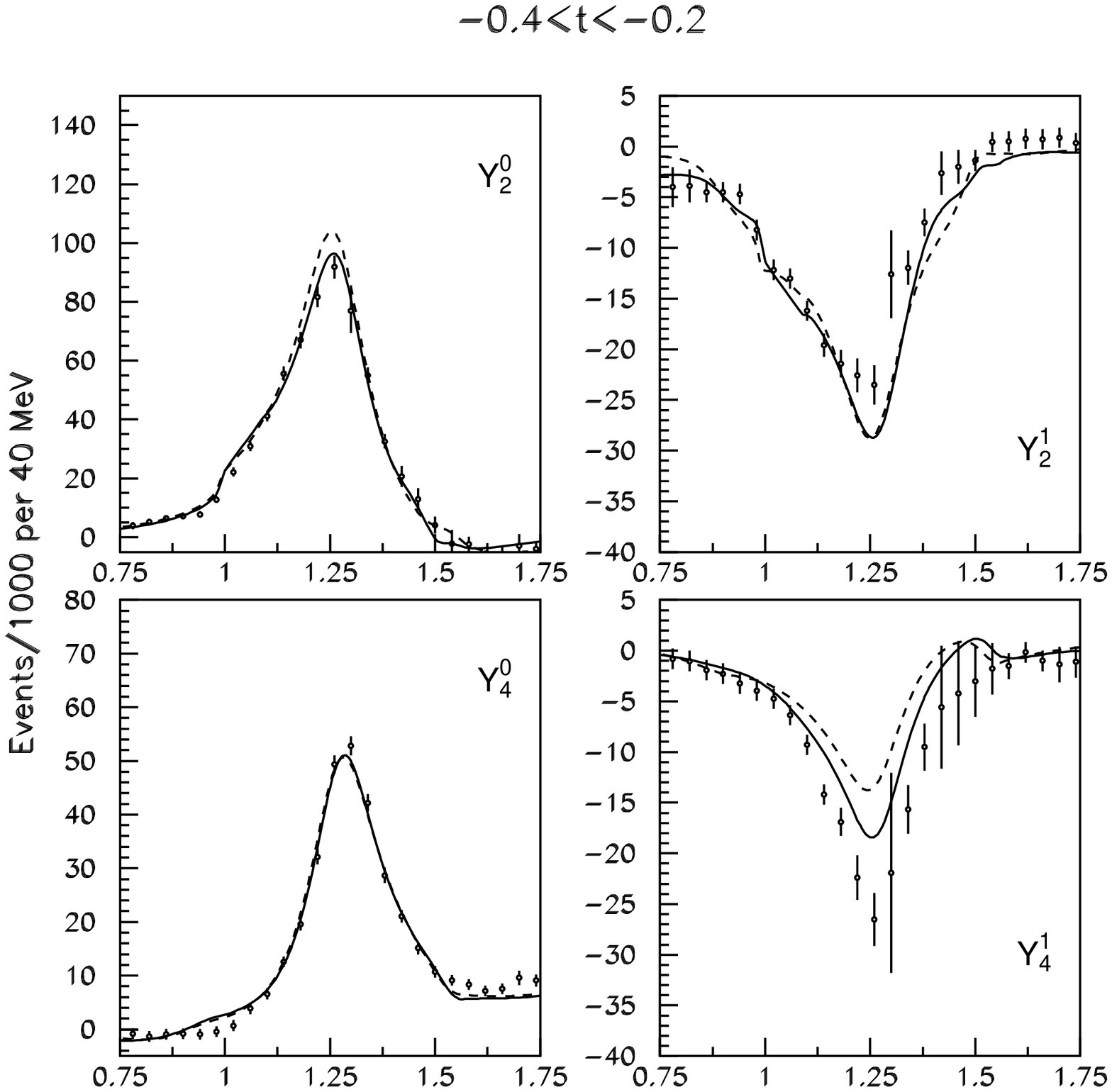,width=0.52\textwidth,clip=on} }
\caption{The description of the moments extracted at
$-0.1\!<\!t\!<\!-0.2$ GeV$^2$ (two left  columns) and
$-0.4\!<\!t\!<\!-0.2$ GeV$^2$ (two right columns). Solid curves
correspond to the solution 2 and dashed line to the solution 2(-) with
excluded $f_0(1300)$.}
\label{Ty_no1300}
\end{figure}

\begin{figure}[h!]
%Fig.18
\centerline{
\epsfig{file=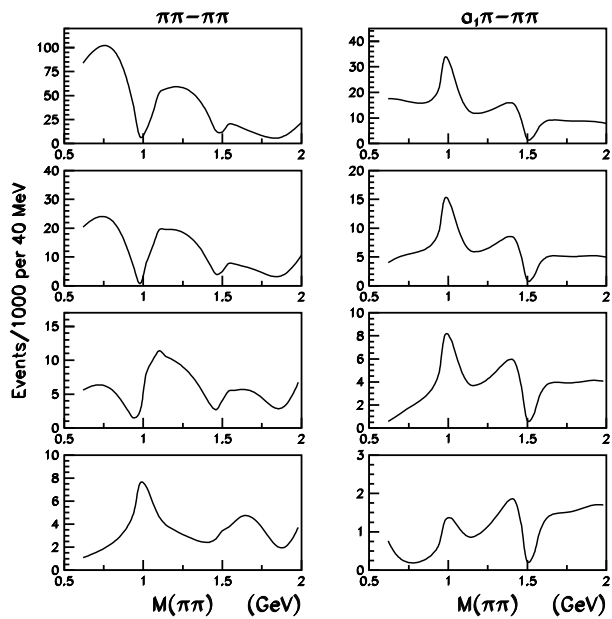,width=0.53\textwidth,clip=on}
}
\caption{Solution 2(-) with $f_0(1300)$ excluded from the fit.
The contributions of S-wave  to $Y_{00}$ moment integrated over
different $t$ intervals. First line: $t\!<\!-0.1$ GeV$^2$, second line:
$-0.1\!<\!t\!<\!-0.2$ GeV$^2$, third line:  $-0.2\!<\!t\!<\!-0.4$
GeV$^2$ and the bottom line: $-1.5\!<\!t\!<\!-0.4$ GeV$^2$.}
\label{Tsdwave_s2_nof0}
\end{figure}

For the description of the E852 data the main effect is seen, as
expected, for the second and third $t$ intervals. The comparison of
the solutions with and without $f_0(1300)$ for these $t$ regions is
shown in Fig.~\ref{Ty_no1300}. Here the description of the $Y^1_4$
moment is systematically worse for the fit where $f_0(1300)$ is
excluded. The $\chi^2$ per data points change for this moment from
1.84 to 3.63 for the $-0.2\!<\!t\!<\!-0.1$ GeV$^2$ interval and from
2.07 to 4.90 for the $-0.4\!<\!t\!<\!-0.2$ GeV$^2$ interval. The fit
without $f_0(1300)$ produces a worse description also for $Y^0_2$
and $Y_4^0$. At intervals of small and large $t$ the description has
the same quality and can hardly be distinguished on the pictures.
The contribution of the S-wave to the moment $Y^0_0$ from this
solution is shown in Fig. \ref{Tsdwave_s2_nof0}. It is seen that an
appreciable contribution from $a_1$ exchange at the mass region
1300-1500 MeV is needed and the fit tries to simulate it (although
not very successfully) by an interference between the broad
component and the $f_0(1500)$ state.

 Below we present the pole positions of the S-wave  amplitude (in MeV units) and
couplings calculated as pole residues (in GeV units):
$A_{a\to b}\simeq G_aG_b[(M-i\Gamma /2)^2-s]^{-1} + smooth\, terms$
with $a,b=\pi\pi,K\bar K,\eta\eta,\eta\eta',\pi\pi\pi\pi$;
the couplings are written
as $G_a=g_a\exp(i\varphi_a)$, the
 phases are given in degrees. For resonances
 $f_0(980)$, $f_0(1300)$,  $f_0(1500)$, $f_0(1200-1600)$, $f_0(1750)$ we
 obtain:
\begin{equation}
{\begin{tabular}{lcccccc}
&  $f_0(980)_{1st\,pole}$ & $f_0(980)_{2nd\,pole}$    & $f_0(1300)$
                     &  $f_0(1500)$      &  $f_0(1200-1600)$    & $f_0(1750)$     \\
M                    &$1030^{+30}_{-10}$&$850^{+80}_{-50}$ &$1290\!\pm\! 50$
                     &$1486\!\pm\! 10$      &$1510\!\pm\!130$ & $1800\!\pm\! 60$ \\
$\Gamma /2$             &$35^{+10}_{-16}$  &$100\!\pm\! 25$  &$170^{+20}_{-40}$
                     &$57\!\pm\! 5$& $800^{+100}_{-150}$ & $200\!\pm\! 30$ \\
Sheet                &     II       &     III     &     IV
                     &     IV       &   IV        &      V         \\
\hline
$g_{\pi\pi}$         &$0.42\!\pm\! 0.07$&$0.39\!\pm\!0.05$&$0.28\!\pm\!0.08$&$0.24\!\pm\!0.05$&$0.82\!\pm\!0.06$&$0.55\!\pm\!0.05$\\
$\varphi_{\pi\pi}$   &$-71\!\pm\! 8$    &$45\!\pm\! 7$    &$27\!\pm\!10$    &$65\!\pm\! 8$    &$10\!\pm\!12$    &$15^{+6}_{-15}$\\
$g_{K\bar K}$        &$0.62\!\pm\!0.06$ &$0.68\!\pm\!0.12$&$0.15\!\pm\!0.05$&$0.17\!\pm\!0.04$&$0.84\!\pm\!0.08$&$0.11\!\pm\!0.04$\\
$\varphi_{K\bar K}$  &$ 3 \!\pm\! 8$    &$155\!\pm\! 6$   &$ 35\!\pm\!15$   &$ 48\!\pm\! 8$   &$  2\!\pm\!10$   &$55\!\pm\! 20$   \\
$g_{\eta\eta}$       &$0.51\!\pm\!0.07$ &$0.58\!\pm\!0.10$&$0.14\!\pm\!0.06$&$0.10\!\pm\!0.03$&$0.40\!\pm\!0.06$&$0.18\!\pm\!0.05$\\
$\varphi_{\eta\eta}$ &$ 10\!\pm\!  8$   &$157\!\pm\! 10$  &$ 57\!\pm\! 8 $  &$ 96\!\pm\!  6$  &$ 16\!\pm\!  7$  &$ 40\!\pm\! 12$  \\
$g_{\eta\eta'}$      &$0.42\!\pm\!0.08$ &$0.46\!\pm\!0.12$&$0.17\!\pm\!0.07$&$0.18\!\pm\!0.06$&$0.14\!\pm\!0.05$&$0.35\!\pm\!0.07$\\
$\varphi_{\eta\eta'}$&$18 \!\pm\! 8$    &$160\!\pm\! 10$  &$ 75\!\pm\! 15$  &$143\!\pm\! 15$  &$ 80\!\pm\! 17$  &$18\!\pm\! 6$  \\
$g_{4\pi}$           &$0.16\!\pm\!0.05$ &$0.29\!\pm\!0.10$&$0.80\!\pm\!0.15$&$0.47\!\pm\!0.08$&$1.30\!\pm\!0.20$&$0.85\!\pm\!0.20$\\
$\varphi_{4\pi}$    &$25^{+8}_{-15}$&$155\!\pm\! 12$  &$205\!\pm\! 12$  &$156\!\pm\! 10$  &$  5\!\pm\! 12$  &$150\!\pm\! 14$  \\
\end{tabular} \label{TswaveG_par}}
\end{equation}

\subsection{Isovector scalar and tensor resonances below 1.7 GeV }

The isovector states are contributing strongly to the
$\bar p p\to \eta\eta\pi^0$ and $\bar p p\to \pi^0\pi^0\eta$
amplitude.

The
$a_0(980)$ is clearly seen on the $\eta\eta\pi^0$ Dalitz plot
and on the $\pi\eta$ mass projection. A successful description of
these data can be obtained with the Flatt\'{e} parametrization of this
resonance (pole parametrization with decays into channels $\pi\eta$
and $K\bar K$). Within this parametrization (for more detail see
\cite{Sarantsev:2004tn}) we obtain the following masses and
couplings for $a_0(980)$:
\begin{equation}
\begin{tabular}{ccc}
M (MeV) & $g^2_{\pi\eta}$ (GeV) & $R=g^2_{K\bar
K}/g^2_{\pi\eta}$ \\
$986\pm 4$ & $0.175\pm 0.015$ & $1.20\pm 0.15$
\\
\end{tabular}
\label{a0_par}
\end{equation}
These parameters result in the following two poles on the I-st sheet
(under the $\pi\eta$ cut) and the II-nd sheet (under the $\pi\eta$
and $K\bar K$ cuts) for $a_0(980)$:
\begin{equation}
\begin{tabular}{cc}
 I-st sheet & II-nd sheet \\
$1000\pm 6\!-\!i(35\pm 4)$ MeV, & $940\pm 20\!-\!i(85\pm 15)$ MeV\\
\end{tabular}
\label{a0_par-pole}
\end{equation}

The second isovector scalar state $a_0(1474)$ \cite{book3} (defined as
$a_0(1450)$ in PDG) is situated in the 1500 MeV region. We
found:
\begin{equation}
\label{a0-1450}
 {\rm \, Pole \, position\, of\,}a_0(1474):
 \qquad M=1515\pm 30-i(115\pm 18)\quad {\rm MeV}.
\end{equation}
 This state is highly inelastic with a branching ratio into the
$\pi\eta$ channel less than $10\%$. We have not found any
indications for an extra isovector scalar state in the mass region
between $a_0(980)$ and $a_0(1474)$.

The resonance $a_2(1320)$  contributes to the $\bar p p\to
\pi^0\pi^0\eta$ strongly, here we see also $a_2(1675)$-signal. We
parametrized these resonances in the  Breit-Wigner form and have
found the following amplitude poles:
\begin{equation}
\begin{tabular}{cc}
 $a_2(1320)$ & $a_2(1675)$ \\
$M=1309\pm 4-i(55\pm 2)$ MeV, & $M=(1675\pm 25)-i(135^{+25}_{-10})$ MeV\\
\end{tabular}
\label{a2_par-pole}
\end{equation}
For $a_2(1320)$ we have found a mass $1309\pm 4$ MeV which is lower
by 9 MeV than the average PDG value but corresponds very well to the
analysis of high statistical data performed by the VES collaboration
\cite{Amelin:1995gt}. The observed width of the state $111\pm 4$
corresponds well to other observations from the $\pi\eta$ decay
mode.

The $a_2(1675)$ state improves the fit; however, the mass and width
of this state can not be well defined from these data because the
resonance is situated on the phase volume boundary and is suppressed
by the D-wave centrifugal barrier. The obtained values are
compatible with previous findings of the Crystal Barrel
collaboration \cite{Amsler:2002qq}. Let us mention that the mass and
width of this state can be much better defined from the L3 data on
$\gamma\gamma$ interaction into $\pi^+\pi^-\pi^0$
\cite{Shchegelsky:2006es}.

\section{Summary for isoscalar resonances}

We develop a method for the analysis of the reactions
$\pi N\to two\, mesons+N$
at large energies of the initial pion. The approach
is based on the use of the reggeized exchanges that allow us to
analyze simultaneously the data obtained at small and large momentum
transfers. In the present article the method is applied to the
analysis of the $\pi^- N\to \pi^0\pi^0 N$ data measured by the E852
experiment. The inclusion of the Crystal Barrel data on the
proton-antiproton annihilation at rest into the $3\pi^0$,
$\pi^0\eta\eta$ and $\pi^0\pi^0\eta$ channels helps to reduce
ambiguities in the isoscalar sector and investigate the properties
of the isovector scalar and tensor states.

As the result of the analysis the K-matrix parameters of the
isoscalar-scalar and isoscalar-tensor
states was obtained up to the invariant mass 2 GeV and pole
positions of corresponding amplitudes are defined.

 \subsection{Isoscalar-scalar sector}

In the scalar sector the contribution of the $f_0(1300)$
is necessary to get a consistent description for the data set analyzed:
\begin{equation}
\label{f0-1300}
 {\rm \, Pole \, position\, of\,}f_0(1300):
 \qquad M=1290\pm50-i(170^{+20}_{-40})\quad {\rm MeV}.
\end{equation}
According to our fit, the strong signal in the $\pi\pi$ spectrum in
the region 1300 MeV is formed by two contributions, by $f_0(1300)$
(dominantly the $a_1$ reggeized exchange) and $f_2(1275)$ (the $\pi$
and $a_1$  reggeized exchanges).

The position of the $f_0(980)$ is defined very well.
 The resonance reveals a double pole structure around the $K\bar K$
threshold.
\bea \label{f0-980}
& {\rm \, Pole \, positions\, of\,}f_0(980):\\
& {\rm \, sheet\,  II \,
(under\,\pi\pi\, and \,\pi\pi\pi\pi\, cuts)}:
\qquad M=1030^{+30}_{-10} -i(35^{+10}_{-16})\quad {\rm MeV},\nn\\
& {\rm \,  sheet\,III \,(under\,\pi\pi,\, \pi\pi\pi\pi\, and \, K\bar K\, cuts)}:
\qquad M=850^{+80}_{-50}-i(200\pm 50)\quad {\rm MeV}.\nn
\eea
The $f_0(1500)$ is defined from the combined fit with a good accuracy:
\begin{equation}
\label{f0-1500}
 {\rm \, Pole \, position\, of\,}f_0(1500):
 \qquad M=1486\pm10-i(57\pm 5)\quad {\rm MeV}.
\end{equation}
The broad state $f_0(1200-1600)$
(the scalar glueball descendant)
gives contribution in $\pi\pi$ scattering
amplitudes in region up to 2 GeV;
the following pole position is found
\begin{equation}
\label{f0-broad}
 {\rm \, Pole \, position\, of\,}f_0(1200-1600):
 \qquad M=(1510\pm 130)-i(800^{+100}_{-150})\quad {\rm MeV}.
\end{equation}
The $f_0(1750)$
 is a dominantly $s\bar s $ state \cite{book3} and is needed to describe
 $\pi\pi\to \pi\pi$ and $\pi\pi\to
\eta\eta$ amplitudes above 1750 MeV.
\begin{equation}
\label{f0-1750}
 {\rm \, Pole \, position\, of\,}f_0(1750):
 \qquad M=1800\pm 60-i(200\pm 30)\quad {\rm MeV}.
\end{equation}
Parameters of this state differ from that observed by the BES
\cite{Bai:2000ss} and WA102 \cite{Barberis:2000cd} collaborations
(denoted as $f_0(1710)$); one should, however, have in mind that in
the case of strong interferences characteristics of a peak in the
data does not correspond to the resonance position. A combined fit
of the Crystal Barrel, CERN-Munich, E852, GAMS and BES data is
needed and it is one of our future objectives.

\subsection{Isoscalar-tensor  sector}

The D-wave reveals the resonances $f_2(1275)$, $f_2(1525)$,
$f_2(1565)$, and $f_2(1950)$ with the following pole positions:
\bea
\label{f2-all}
 f_2(1275)&:\qquad&  M=1270\pm 8-i(97\pm 8)\, {\rm MeV},\nn \\
 f_2(1525)&:\qquad&  M=1530\pm 12-i(64\pm 10)\, {\rm MeV},\nn \\
 f_2(1565)\;{\rm (2nd\, solution)}&:\qquad&   M_I=1690\pm 15-i(290\pm 20)\,
{\rm MeV},\nn\\
& \qquad& M_{II}=1560\pm 15-i(140\pm 20)\, {\rm
MeV},
\eea
 In the case of $f_2(1565)$ the K-matrix fit can be obtained only
with the large coupling of this state to $\omega\omega$ (and,
possibly, to $\rho\rho$) channel (note that this result is in a very
good agreement with the analysis of the proton-antiproton
annihilation into $\omega\omega\pi$ \cite{Baker:1999ac}). The large
coupling to $\omega\omega$ leads to the double pole structure of
$f_2(1565)$, see Fig. \ref{Tch6-poles}.

The state $f_2(1980)$ can not be identified unambiguously from the
present data due to its large inelasticity. It plays the role of
some broad contribution needed for the description of the $\pi N$
data

\subsection{Isoscalar  sector $J^{PC}=4^{++}$ }

\begin{figure}[h!]
%Fig.26
\centerline{ \epsfig{file=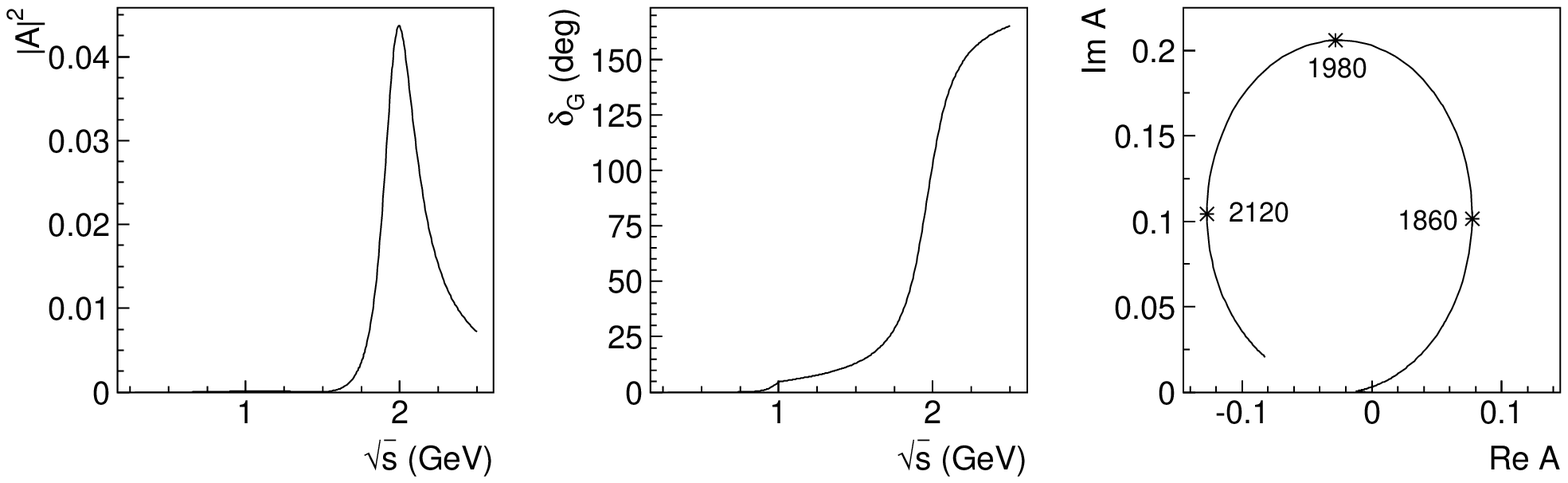,width=0.99\textwidth,clip=on}}
\caption{From left to right: The $\pi\pi\to\pi\pi$ G-wave amplitude
squared, the amplitude phase and the Argand diagram for the
amplitude.} \label{Tgw_amp}
\end{figure}

For the description of high moments in the $\pi N\to \pi^0\pi^0 N$
data a contribution from a $4^{++}$ state is needed. This state is
identified as $f_4(2025)$. Due to the lack of data at high masses
this state was fitted as a two channel ($\pi\pi$ and $4\pi$) one
pole K-matrix.
\begin{equation}
\begin{tabular}{cccc}
 M (GeV) &
$g_{\pi\pi}$ & $g_{4\pi}$ & $f_{\pi\pi\to \pi\pi}$ \\
 $1.970\pm
30$ & $0.550\pm 0.050$ &$ 0.490\pm 0.080$ & $-0.025\pm 0.050$
 \\
\end{tabular}
\label{Tgwave_par}
\end{equation}
Here, as previously, masses and couplings are in GeV units. The
position of the pole is equal to $(1966\pm 25)-i\,(130\pm 20)$. The
amplitude phase and the Argand diagram for the isoscalar $4^{++}$
state is shown in Fig.\ref{Tgw_amp}. The $\pi\pi\to\pi\pi$ $4^{++}$
amplitude has a peak at 1995 MeV and is slightly asymmetrical: the
half height is reached at the mass 1880 and 2165 MeV. The branching
ratio of the $\pi\pi$ channel at the pole position is $20\pm 3\%$
which is in agreement with the PDG value within the error.

\section*{Acknowledgments}

We thank A.V. Anisovich, L.G. Dakhno, J. Nyiri, V.A. Nikonov, M.A. Matveev
for helpful discussions. The paper was supported by the RFFI grant
07-02-01196-a.

\section{Appendix A. Angular Momentum Operators}

The angular-dependent part of the wave function of a composite state
is described by operators constructed for the relative momenta of
particles and the metric tensor. Such operators (we  denote them as
$X^{(L)}_{\mu_1\ldots\mu_L}$, where $L$ is the angular momentum) are
called angular momentum operators; they correspond to irreducible
representations of the Lorentz group.
They satisfy the following properties: \\
(i) Symmetry with respect to the permutation of any two indices:
\be
X^{(L)}_{\mu_1\ldots\mu_i\ldots\mu_j\ldots\mu_L}\; =\;
X^{(L)}_{\mu_1\ldots\mu_j\ldots\mu_i\ldots\mu_L}.
\label{Voth_b1}
\ee
(ii) Orthogonality to the total momentum of the system, $P=k_1+k_2$:
\be P_{\mu_i}X^{(L)}_{\mu_1\ldots\mu_i\ldots\mu_L}\ =\ 0.
\label{Voth_b2}
\ee
(iii) Tracelessness with respect to the summation
over  any two indices:
\be
g_{\mu_i\mu_j}X^{(L)}
_{\mu_1\ldots\mu_i\ldots\mu_j\ldots\mu_L}\ \ =\ 0. \label{Voth_b3}
\ee
Let us consider a one-loop diagram describing the decay of a
composite system into two spinless particles, which propagate and
then form again a composite system. The decay and formation
processes are described by angular momentum operators. Owing to the
quantum number conservation, this amplitude must vanish for initial
and final states with different spins. The S-wave operator is a
scalar and can be taken as a unit operator. The P-wave operator is a
vector. In the dispersion relation approach it is sufficient that
the imaginary part of the loop diagram, with S- and P-wave operators
as vertices, equals 0. In the case of spinless particles, this
requirement entails
\be
\int\frac{d\Omega}{4\pi} X^{(1)}_\mu =0\ ,
\ee
where the integral is taken over the solid angle of the relative
momentum. In general, the result of such an integration is
proportional to the total momentum  $P_\mu$ (the only external
vector):
\be
\int\frac{d\Omega}{4\pi} X^{(1)}_\mu =\lambda P_\mu\;.
\ee
Convoluting this expression with $P_\mu$ and demanding
$\lambda=0$, we obtain the orthogonality condition (\ref{Voth_b2}).
The orthogonality between the D- and S-waves is provided by the
tracelessness condition (\ref{Voth_b3}); equations (\ref{Voth_b2}),
(\ref{Voth_b3}) provide the orthogonality for all operators with
different angular momenta.

The orthogonality condition (\ref{Voth_b2}) is automatically fulfilled
if the operators are constructed from the relative momenta
$k^\perp_\mu$ and tensor $g^\perp_{\mu\nu}$. Both of them are
orthogonal to the total momentum of the system:
\be
k^\perp_\mu=\frac12
g^\perp_{\mu\nu}(k_1-k_2)_\nu \ ,  \qquad
g^\perp_{\mu\nu}=g_{\mu\nu}-\frac{P_\mu P_\nu}{s}\;.
\ee
 In the c.m. system, where $P=(P_0,\vec P)=(\sqrt s,0)$, the vector
$k^\perp$ is space-like: $k^\perp=(0,\vec k,0)$.

The operator for $L=0$ is a scalar (for example, a unit operator),
and the operator for $L=1$ is a vector, which can be constructed
from $k^\perp_\mu$ only. The orbital angular momentum operators for
$L =0 $ to 3 are:
\begin{eqnarray}
&&\hspace{-6mm}
X^{(0)}(k^\perp)=1 , \qquad X^{(1)}_\mu=k^\perp_\mu , \\
&&\hspace{-6mm}
X^{(2)}_{\mu_1 \mu_2}(k^\perp)=\frac32\left(k^\perp_{\mu_1}
k^\perp_{\mu_2}-\frac13\, k^2_\perp g^\perp_{\mu_1\mu_2}\right), \nn  \\
&&\hspace{-6mm}X^{(3)}_{\mu_1\mu_2\mu_3}(k^\perp)=
\frac52\Big[k^\perp_{\mu_1} k^\perp_{\mu_2 }
k^\perp_{\mu_3} -
\frac{k^2_\perp}5\left(g^\perp_{\mu_1\mu_2}k^\perp
_{\mu_3}+g^\perp_{\mu_1\mu_3}k^\perp_{\mu_2}+
g^\perp_{\mu_2\mu_3}k^\perp_{\mu_1}
\right)\Big] . \nonumber
\end{eqnarray}
The operators $X^{(L)}_{\mu_1\ldots\mu_L}$ for $L\ge 1$ can be
written in the form of a recurrency relation:
\be
X^{(L)}_{\mu_1\ldots\mu_L}(k^\perp)&=&k^\perp_\alpha
Z^{\alpha}_{\mu_1\ldots\mu_L}(k^\perp) \; ,
\nonumber\\
Z^{\alpha}_{\mu_1\ldots\mu_L}(k^\perp)&=&
\frac{2L-1}{L^2}\Big (
\sum^L_{i=1}X^{{(L-1)}}_{\mu_1\ldots\mu_{i-1}\mu_{i+1}\ldots\mu_L}(k^\perp)
g^\perp_{\mu_i\alpha}
\nonumber \\
 -\frac{2}{2L-1}  \sum^L_{i,j=1 \atop i<j}
&g^\perp_{\mu_i\mu_j}&
X^{{(L-1)}}_{\mu_1\ldots\mu_{i-1}\mu_{i+1}\ldots\mu_{j-1}\mu_{j+1}
\ldots\mu_L\alpha}(k^\perp) \Big ).
\label{Vz}
\ee
The convolution equality reads
\be
X^{(L)}_{\mu_1\ldots\mu_L}(k^\perp)k^\perp_{\mu_L}=k^2_\perp
X^{(L-1)}_{\mu_1\ldots\mu_{L-1}}(k^\perp).
\label{Vceq}
\ee
On the basis of Eq.(\ref{Vceq}) and
taking into account the tracelessness property of $X^{(L)}_{\mu_1\ldots\mu_L}$,
one can write down the orthogonali\-ty--normalisation condition for
orbital angular  operators
\begin{eqnarray}
&&\int\frac{d\Omega}{4\pi}
X^{(L)}_{\mu_1\ldots\mu_L}(k^\perp)X^{(L')}_{\mu_1\ldots\mu_L'}(k^\perp)
\ =\ \delta_{LL'}\alpha_L k^{2L}_\perp \; , \label{Vort-x}\nn
\\
&&\alpha_L\ =\ \prod^L_{l=1}\frac{2l-1}{l} \; .
\label{Valpha}
\end{eqnarray}
Iterating equation (\ref{Vz}), one obtains the
following expression for the operator $X^{(L)}_{\mu_1\ldots\mu_L}$:
\be
\label{Vx-direct}
&&X^{(L)}_{\mu_1\ldots\mu_L}(k^\perp)=
\alpha_L \bigg [
k^\perp_{\mu_1}k^\perp_{\mu_2}k^\perp_{\mu_3}k^\perp_{\mu_4}
\ldots k^\perp_{\mu_L}
\nn \\
&&-\frac{k^2_\perp}{2L-1}\bigg(
g^\perp_{\mu_1\mu_2}k^\perp_{\mu_3}k^\perp_{\mu_4}\ldots
k^\perp_{\mu_L}
+g^\perp_{\mu_1\mu_3}k^\perp_{\mu_2}k^\perp_{\mu_4}\ldots
k^\perp_{\mu_L} + \ldots \bigg)
\nn \\
&&+\frac{k^4_\perp}{(2L\!-\!1)(2L\!-\!3)}\bigg(
g^\perp_{\mu_1\mu_2}g^\perp_{\mu_3\mu_4}k^\perp_{\mu_5}
k^\perp_{\mu_6}\ldots k^\perp_{\mu_L}
\nn \\
&&+
g^\perp_{\mu_1\mu_2}g^\perp_{\mu_3\mu_5}k^\perp_{\mu_4}
k^\perp_{\mu_6}\ldots k^\perp_{\mu_L}+
\ldots\bigg)+\ldots\bigg ].
\ee

The projection operator $O^{\mu_1\ldots\mu_L}_{\nu_1\ldots \nu_L}$
(or  $O^{\mu_1\ldots\mu_L}_{\nu_1\ldots \nu_L}(\perp P)$)
is constructed of the metric tensors $g^\perp_{\mu\nu}$. It
has the  properties as follows:
\be
X^{(L)}_{\mu_1\ldots\mu_L}
O^{\mu_1\ldots\mu_L}_{\nu_1\ldots \nu_L}\
&=&\ X^{(L)}_{\nu_1\ldots \nu_L}\ , \nonumber \\
O^{\mu_1\ldots\mu_L}_{\alpha_1\ldots\alpha_L} \
O^{\alpha_1\ldots\alpha_L}_{\nu_1\ldots \nu_L}\ &=&
O^{\mu_1\ldots\mu_L}_{\nu_1\ldots \nu_L}\ .
\label{Vproj_op}
\ee
Taking into account the definition of  projection operators
(\ref{Vproj_op}) and the properties of the $X$-operators
(\ref{Vx-direct}), we obtain
\be
k_{\mu_1}\ldots k_{\mu_L}
O^{\mu_1\ldots\mu_L}_{\nu_1\ldots \nu_L}\ = \frac{1}{\alpha_L}
X^{(L)}_{\nu_1\ldots\nu_L}(k^\perp).
\label{V19}
\ee
 This equation is the basic property of the projection operator:
it projects any operator with $L$ indices onto the partial wave
operator with angular momentum $L$.

For the lowest states,
\be
&&\hspace{-6mm}O= 1\ ,\qquad O^\mu_\nu=g_{\mu\nu}^\perp \, ,\nn \\
&&\hspace{-6mm}O^{\mu_1\mu_2}_{\nu_1\nu_2}=
\frac 12 \left (
g_{\mu_1\nu_1}^\perp  g_{\mu_2\nu_2}^\perp \!+\!
g_{\mu_1\nu_2}^\perp  g_{\mu_2\nu_1}^\perp  \!- \!\frac 23
g_{\mu_1\mu_2}^\perp  g_{\nu_1\nu_2}^\perp \right ).\;\;
\ee
For higher states, the operator can be calculated using the
recurrent expression:
\be
&&O^{\mu_1\ldots\mu_L}_{\nu_1\ldots\nu_L}=
\frac{1}{L^2} \bigg (
\sum\limits_{i,j=1}^{L}g^\perp_{\mu_i\nu_j}
O^{\mu_1\ldots\mu_{i-1}\mu_{i+1}\ldots\mu_L}_{\nu_1\ldots
\nu_{j-1}\nu_{j+1}\ldots\nu_L}
 \\
&&- \frac{4}{(2L-1)(2L-3)}
\times
\sum\limits_{i<j\atop k<m}^{L}
g^\perp_{\mu_i\mu_j}g^\perp_{\nu_k\nu_m}
O^{\mu_1\ldots\mu_{i-1}\mu_{i+1}\ldots\mu_{j-1}\mu_{j+1}\ldots\mu_L}_
{\nu_1\ldots\nu_{k-1}\nu_{k+1}\ldots\nu_{m-1}\nu_{m+1}\ldots\nu_L}
\bigg ). \nn
\ee
The product of two $X$-operators integrated over a solid angle (that
is equivalent to the integration over internal momenta) depends only
on the external momenta and the metric tensor. Therefore, it must be
proportional to the projection operator. After straightforward
calculations we obtain
\be
\!\!\!\int\!\frac{d\Omega }{4\pi}\!
X^{(L)}_{\mu_1\ldots\mu_L}(k^\perp)
X^{(L)}_{\nu_1\ldots\nu_L}(k^\perp)\!=\! \frac{\alpha_L\,
k^{2L}_\perp}{2L\!+\!1} O^{\mu_1\ldots\mu_L}_{\nu_1\ldots \nu_L}\ .
\label{Vx-prod}
\ee
 Let us introduce the positive valued $|\vec k|^2$:
\be
|\vec k|^2\!=\!-k_\perp^2\!=\!
\frac{[s\!-\!(m_1\!+\!m_2)^2][s\!-\!(m_1\!-\!m_2)^2]}{4s}\ .
\label{Vk2_rel}
\ee
In the c.m.s. of the reaction, $\vec k$ is the
momentum of a particle. In other systems we use this definition only
in the sense of $|\vec k|\equiv \sqrt{-k_\perp^2}$; clearly, $|\vec
k|^2$ is a relativistically invariant positive value. If so,
equation (\ref{Vx-prod}) can be written as
\be \hspace{-1cm}
\int\!\frac{d\Omega
}{4\pi}\! X^{(L)}_{\mu_1\ldots\mu_L}(k^\perp)
X^{(L)}_{\nu_1\ldots\nu_L}(k^\perp)\!=\! \frac{\alpha_L\,|\vec
k|^{2L}}{2L\!+\!1}(-1)^L
O^{\mu_1\ldots\mu_L}_{\nu_1\ldots \nu_L}.
\ee
The tensor part of the numerator of the boson propagator is
defined by the projection operator. Let us write it as follows:
\be
F^{\mu_1\ldots\mu_L}_{\nu_1\ldots\nu_L}=
(-1)^L\,O^{\mu_1\ldots\mu_L}_{\nu_1\ldots \nu_L}\ ,
\label{Vboson_prop}
\ee
with the definition of the propagator
\be
\frac{F^{\mu_1\ldots\mu_L}_{\nu_1\ldots\nu_L}
 }
{M^2-s} .
\label{Vboson_prop1}
\ee
 This definition guarantees that the width of a resonance
(calculated using the decay vertices) is positive.

 \end{document}